\documentclass[11pt,letterpaper]{article}
\usepackage[margin=1in]{geometry}
\usepackage[T1]{fontenc}
\usepackage{authblk}
\usepackage{threeparttable}
\usepackage{indentfirst}
\usepackage[sort&compress,numbers]{natbib}

\usepackage[font=footnotesize,labelfont=bf]{caption}
\usepackage{xurl}
\usepackage{hyperref}
\providecommand{\burl}[1]{\url{#1}}

\providecommand{\keywords}[1]{\noindent\textbf{\textit{Keywords: }}#1}
\hypersetup{colorlinks=true,linkcolor=blue,citecolor=blue,urlcolor=blue,
  pdftitle={High-fidelity Three-Dimensional Flow Field Reconstruction Driven by Turbulent Fluid Dynamics},
  pdfauthor={Hanlin Kong, Jianan Wei, Yuntian Chen, Yujia Wang, Wenguan Wang}}

\usepackage{graphicx}%
\usepackage{overpic}
\usepackage{multirow}%
\usepackage{amsmath,amssymb,amsfonts}%
\usepackage{amsthm}%
\usepackage{mathrsfs}%
\usepackage{xcolor}%
\usepackage{textcomp}%
\usepackage{manyfoot}%
\usepackage{booktabs}%
\usepackage{algorithm}%
\usepackage{algorithmicx}%
\usepackage{algpseudocode}%
\usepackage{listings}%
\usepackage{xeCJK}
\usepackage{enumitem}
\usepackage{booktabs}
\usepackage{amsmath}
\usepackage{longtable}
\usepackage{array}
\usepackage{siunitx}
\usepackage{makecell}
\usepackage{bm}
\usepackage{placeins}
\hypersetup{hypertexnames=false} %

\newcommand{\ie}{\textit{i}.\textit{e}.}
\newcommand{\eg}{\textit{e}.\textit{g}.}

\theoremstyle{plain}%

\theoremstyle{remark}%

\theoremstyle{definition}%

\makeatletter
\newenvironment{supplementaryinformation}{%
  \clearpage
  \section*{Supplementary Information}
  \setcounter{section}{0}%
  \setcounter{subsection}{0}%
  \setcounter{figure}{0}%
  \setcounter{table}{0}%
  \setcounter{equation}{0}%
  \def\@seccntformat##1{\csname the##1\endcsname\quad}%
}{\clearpage}
\makeatother

\begin{document}
\title{\textbf{High-fidelity Three-Dimensional Flow Field Reconstruction Driven by Turbulent Fluid Dynamics}}
\author[1,2,$\dagger$]{Hanlin Kong}
\author[1,$\dagger$]{Jianan Wei}
\author[3,$\dagger$]{Yuntian Chen}
\author[4,*]{Yujia Wang}
\author[1,*]{Wenguan Wang}
\affil[1]{\small State Key Lab of Brain-Machine Intelligence, Zhejiang University, Hangzhou, Zhejiang, China}
\affil[2]{\small Shanghai Innovation Institute, Shanghai, China}
\affil[3]{\small Zhejiang Key Laboratory of Industrial Intelligence and Digital Twin, Eastern Institute of Technology, Ningbo, Zhejiang, China}
\affil[4]{\small Zhejiang Sci-Tech University, Hangzhou, Zhejiang, China}
\affil[$\dagger$]{\small These authors contributed equally to this work.}
\affil[*]{\small Corresponding authors.}
\date{}
\maketitle
\normalsize
\vspace{-18pt}
\begin{abstract}
\small

High-fidelity three-dimensional turbulent flow fields are essential for scientific and engineering analyses, yet obtaining them through direct numerical simulation or high-resolution measurements remains expensive. 
Super-resolution reconstruction from coarse-grained data therefore offers a promising alternative, 
but remains fundamentally challenging because fine-scale structures are intrinsically three-dimensional and dynamically coupled across scales. 
Existing data-driven approaches still struggle to recover multiscale volumetric structures and their coherent temporal evolution.
To address these limitations, we propose a turbulence-aware framework for time-resolved three-dimensional flow reconstruction. The framework couples a multiscale neural operator, designed to recover volumetric structures, with a state space model to promote temporal consistency in the reconstructed dynamics. In addition, Navier--Stokes residuals are incorporated as soft constraints to regularize the reconstruction toward physically plausible dynamics. A mixture-of-experts mechanism is also introduced to 
support unsupervised adaptation to unknown coarse-graining.
Across evaluations on canonical turbulent systems and atmospheric reanalysis data, the proposed framework reduces reconstruction errors relative to learning-based and classical baselines, while better preserving multiscale spatial structure and coherent temporal evolution. 
These results validate the effectiveness of the proposed framework in reconstructing high-fidelity three-dimensional flow fields from coarse-grained data.
\end{abstract}


\vspace{12pt}

\section{Introduction}\label{sec:introduction}

Turbulent flows exhibit multiscale structures that evolve over time~\cite{bratanov2015new}. 
High-fidelity three-dimensional flow fields are essential for both scientific research and engineering applications, \eg, atmospheric science \cite{guo2024direct}, phytoplankton ecology \cite{durham2013turbulence}, and scene imaging \cite{jin2021neutralizing}. In practice, however, obtaining such data remains costly: direct numerical simulation (DNS) resolves all relevant scales but is computationally prohibitive~\cite{luther2025nonlinear}, while more affordable approaches such as large-eddy simulation (LES) or Reynolds-averaged Navier--Stokes methods involve simplifying modeling assumptions that reduce fidelity. 
Consequently, time-resolved volumetric flow data are often available only at limited spatial resolution.

This motivates super-resolution reconstruction, which aims to recover high-resolution (HR) flow fields from their low-resolution (LR) counterparts.
Early attempts leverage signal processing and reduced-order representations~\cite{venturi2004gappy,willcox2006unsteady}, and recent progress in learning-based super-resolution improves the recovery of missing fine-scale content from data~\cite{li2024learning}.
A representative study by Fukami \textit{et al.} \cite{fukami2019super} shows that convolutional neural networks (CNNs) are able to achieve a ``data-driven interpolation'' that is more accurate than low-order polynomials. Subsequent studies explored more expressive architectures, such as transformer-based models~\cite{xu2023super,zeng2025swin}, generative adversarial networks~\cite{nista2024influence,kim2021unsupervised,xie2018tempoGAN,deng2019super}, and diffusion models~\cite{shu2023physics,shan2024pird,du2024conditional}, to further improve reconstruction quality. 
Despite these advances, many approaches primarily optimize point-wise fidelity and treat the mapping as a black box, raising concerns about whether the reconstructed flows remain consistent with the underlying governing physics. 
In light of this limitation, a growing trend is to incorporate physical priors into scientific machine learning~\cite{raissi2019physics,karniadakis2021physics, meng2025when,xu2026ai}. 
Kochkov \textit{et al.}~\cite{kochkov2021machine} introduce a hybrid solver--learning pipeline to enforce physical consistency. 
Subsequent studies adopt more flexible strategies by incorporating physical constraints into the training objective.
For example, Chung \textit{et al.}~\cite{chung2023turbulence} explicitly integrate gradient-based constraints, while Ren \textit{et al.}~\cite{ren2025superbench} propose a soft-constraint approach by adding a divergence-free loss as a regularization term. 

Nevertheless, three challenges remain in time-resolved volumetric turbulence super-resolution.
First, although previous studies have demonstrated the feasibility of reconstructing three-dimensional turbulent fields~\cite{fukami2021machine,chung2023turbulence}, the task requires more than predicting values on a volumetric grid: it demands recovering the multiscale spatial structure and coherent temporal evolution of the underlying flow.
In three dimensions, vortical structures form connected geometries throughout the volume, and their local strain--rotation interplay is characterized by invariants of the velocity-gradient tensor~\cite{pope2000turbulent}.
Because these connected structures span a broad range of scales and evolve continuously, they cannot be fully represented by isolated two-dimensional slices, which motivates super-resolution directly on time-resolved three-dimensional fields.
Second, although physical priors are increasingly adopted, they are typically introduced through frame-wise spatial constraints (\eg, gradient penalties~\cite{chung2023turbulence} or divergence-free losses~\cite{ren2025superbench}).
These formulations target kinematic properties at individual time steps, without accounting for the dynamical evolution of turbulent flows. As a result, reconstructions can appear spatially plausible while deviating from consistent dynamical trajectories. 
Third, many learning-based methods are trained under a fixed, predefined coarse-graining (\eg, average/max pooling~\cite{fukami2019super} or specific LES--DNS pairings~\cite{kim2021unsupervised}), which limits robustness when the coarse-graining process differs at inference time.

Addressing these limitations goes beyond improving point-wise accuracy, and requires super-resolution methods that maintain multiscale structure and coherent temporal evolution in dynamical systems.
To this end, we introduce a unified framework for time-resolved three-dimensional turbulence reconstruction.
First, the framework operates directly on time-resolved three-dimensional fields and jointly models their multiscale spatial structure and temporal evolution.
Temporal evolution is modeled on the Eulerian grid using a state space model, which promotes coherent evolution of reconstructed volumetric structures over time.
To recover multiscale spatial structure intrinsic to turbulence, the framework further incorporates a multiscale neural operator that couples spectral modeling of large-scale interactions with convolutional refinement of local structures. 
Second, 
we incorporate Navier--Stokes residuals as soft constraints to promote dynamical consistency in the reconstructed flows.
Finally, a mixture-of-experts (MoE) mechanism enables adaptation to unknown coarse-graining by combining predictions from pretrained experts, with Navier--Stokes residuals guiding the combination without HR supervision.

The framework is trained and evaluated on a curated benchmark for time-resolved volumetric turbulence super-resolution (TurbSR3D) spanning diverse flow regimes, including forced isotropic turbulence~\cite{li2008public}, forced magnetohydrodynamic turbulence~\cite{eyink2013flux}, and homogeneous buoyancy-driven turbulence~\cite{livescu2014homogeneous}, with multiple upscaling factors and coarse-graining settings.
A simulated decaying isotropic turbulence dataset is further used to test the framework under non-stationary conditions and on independently evolved LES inputs paired with DNS references.
To evaluate both numerical accuracy and physical fidelity, we employ a set of criteria that includes common error metrics, diagnostics of multiscale spatial structure (\eg, energy spectra, distributional statistics, and velocity-gradient invariant topology), and measures of temporal consistency.
Beyond canonical turbulence systems, we extend our analysis to complex atmospheric flows using the ERA5 data~\cite{hersbach2020era5}, where the task is framed as the downscaling of multi-level fields.
Across these evaluations, the proposed framework improves reconstruction accuracy relative to strong learning-based and classical baselines, while better preserving multiscale structure and dynamically consistent evolution.

In summary, our contributions are as follows: (i) \textbf{Spatiotemporal modeling}: 
A turbulence-aware architecture is introduced, which couples a multiscale neural operator with a state space model to jointly recover multiscale volumetric structures and capture their coherent temporal evolution. (ii) \textbf{Physics-informed learning}: Navier--Stokes residuals are incorporated as soft dynamical constraints and further used to guide unsupervised MoE adaptation to unknown coarse-graining. (iii) \textbf{Benchmark and evaluation}: TurbSR3D is introduced as a benchmark dataset for time-resolved three-dimensional turbulence super-resolution, with standardized data splits and multiple coarse-graining settings and upscaling factors. We evaluate the framework on canonical turbulent flows and atmospheric reanalysis data, assessing reconstruction accuracy, multiscale spatial structure, temporal consistency, and turbulence statistics.

\section{Results}\label{sec:results}
\begin{figure}[t]
    \centering
    \begin{overpic}[width=\textwidth, percent]{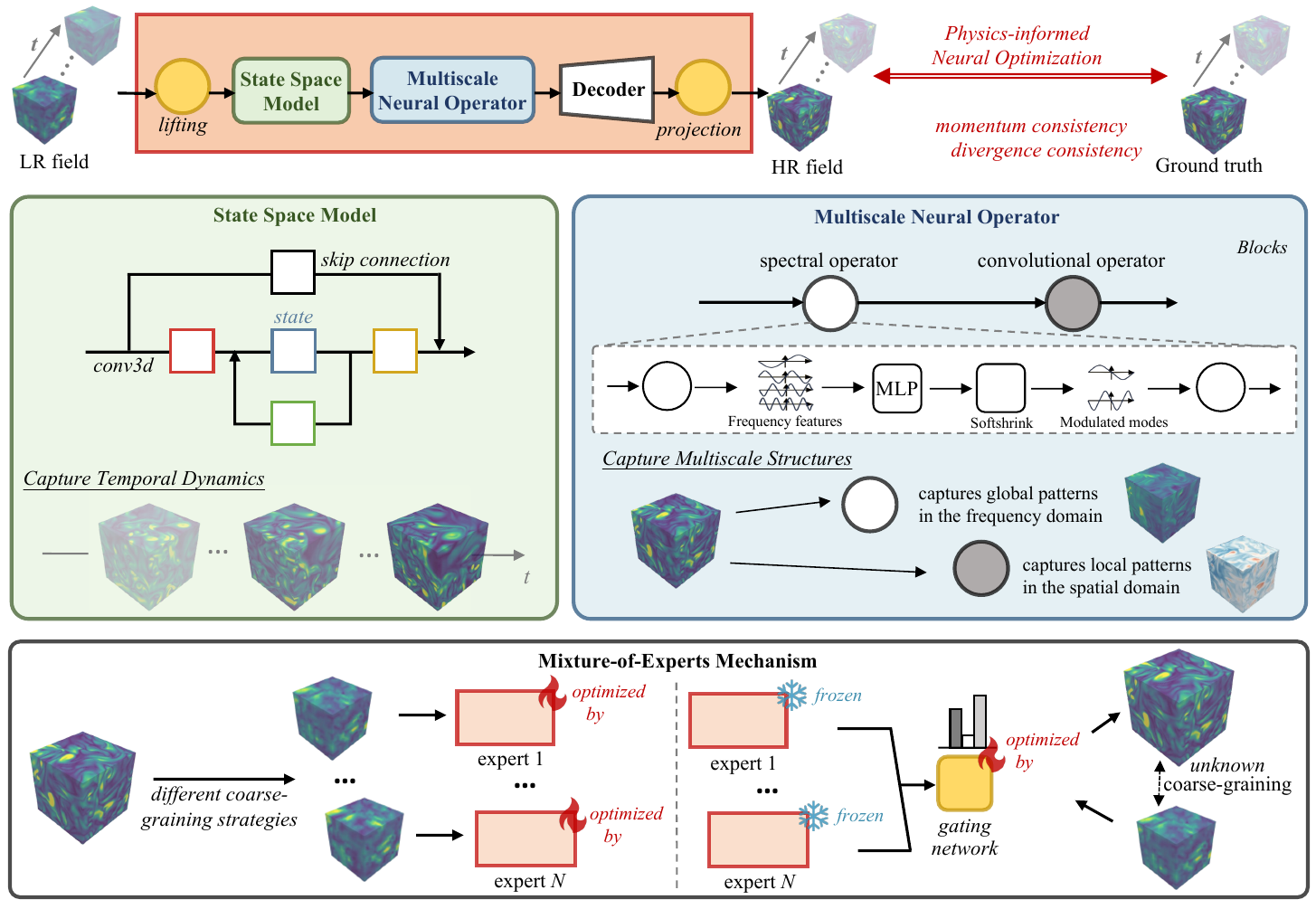}
    \put(1.39,67.50){\makebox(0,0){\large\bfseries a}}
    \put(2.44,52.56){\makebox(0,0){\large\bfseries b}}
    \put(45.39,52.56){\makebox(0,0){\large\bfseries c}}
    \put(2.39,18.67){\makebox(0,0){\large\bfseries d}}
    \put(13.83,62.44){\makebox(0,0){\footnotesize $P$}}
    \put(53.44,62.44){\makebox(0,0){\footnotesize $Q$}}
    \put(54.44,66.44){\makebox(0,0){\footnotesize $\mathcal{M}_{\bm{\theta}}$}}
    \put(5.5,58.3){\makebox(0,0){\tiny $\bm{X_{\scalebox{0.8}{$\mathrm{L}$}}}$}}
    \put(63.8,58.3){\makebox(0,0){\tiny $\bm{\hat{X}_{\scalebox{0.8}{$\mathrm{H}$}}}$}}
    \put(95,58.1){\makebox(0,0){\tiny $\bm{{X}_{\scalebox{0.8}{$\mathrm{H}$}}}$}}
    \put(78.33,61.56){\makebox(0,0){\tiny  $\mathcal{L} = \mathcal{L}_{\scalebox{0.8}{\text{mae}}} + \lambda_{\scalebox{0.8}{\text{phy}}}\,\mathcal{L}_{\scalebox{0.8}{\text{phy}}}$}}
    \put(66.94,59.44){\makebox(0,0){\tiny  $\mathcal{L}_{\scalebox{0.8}{\text{phy}}}$}}
    \put(69.44,59.44){\makebox(0,0){\tiny  $=$}}
    \put(70.28,57.67){\makebox(0,0){\tiny  $+$}}
    \put(5.7,42){\makebox(0,0){\footnotesize $\bm{z}_{\scalebox{0.6}{$t$}}$}}
    \put(37.72,42){\makebox(0,0){\footnotesize $\bm{s}_{\scalebox{0.6}{$t$}}$}}
    \put(14.67,42.39){\makebox(0,0){\footnotesize ${\bm{B}}_t$}}
    \put(22.33,42.39){\makebox(0,0){\footnotesize $\bm{h}_t$}}
    \put(22.28,48.39){\makebox(0,0){\footnotesize $\bm{D}$}}
    \put(22.28,36.83){\makebox(0,0){\footnotesize ${\bm{A}}$}}
    \put(30.06,42.39){\makebox(0,0){\footnotesize $\bm{C}_t$}}
    \put(51.00,46.06){\makebox(0,0){\footnotesize $\bm{s}_{\scalebox{0.6}{$t$}}$}}
    \put(91.00,46.06){\makebox(0,0){\footnotesize $\hat{\bm{s}}_{\scalebox{0.6}{$t$}}$}}
    \put(63.17,46.06){\makebox(0,0){\footnotesize $\mathcal{O}_s$}}
    \put(81.56,46.06){\makebox(0,0){\footnotesize $\mathcal{O}_c$}}
    \put(50.72,39.67){\makebox(0,0){\footnotesize $\mathcal{F}$}}
    \put(76.06,39.50){\makebox(0,0){\footnotesize $S_\lambda$}}
    \put(93.11,39.50){\makebox(0,0){\footnotesize $\mathcal{F}^{\scalebox{0.45}{$-1$}}$}}
    \put(66.11,30.72){\makebox(0,0){\footnotesize $\mathcal{O}_s$}}
    \put(74.61,25.83){\makebox(0,0){\footnotesize $\mathcal{O}_c$}}
    \put(92.50,50.28){\makebox(0,0){\tiny $\times K$}}
    \put(6.56,4.11){\makebox(0,0){\tiny $X_{\scalebox{0.8}{$\mathrm{H}$}}$}}
    \put(27.5,11.2){\makebox(0,0){\tiny $X_{\scalebox{0.8}{$\mathrm{L_{\scalebox{0.8}{$1$}}}$}}$}}
    \put(30,2){\makebox(0,0){\tiny $X_{\scalebox{0.8}{$\mathrm{L_{\scalebox{0.8}{$N$}}}$}}$}}
    \put(38.39,14.44){\makebox(0,0){\footnotesize $\mathcal{M}_{{\bm{\theta}}_{\scalebox{0.5}{$1$}}}$}}
    \put(39.94,5.28){\makebox(0,0){\footnotesize $\mathcal{M}_{{\bm{\theta}}_{\scalebox{0.5}{$N$}}}$}}
    \put(46.7,14.6){\makebox(0,0){\tiny $\mathcal{L}$}}
    \put(48,5.4){\makebox(0,0){\tiny $\mathcal{L}$}}
    \put(56.11,14.22){\makebox(0,0){\footnotesize $\mathcal{M}_{{\bm{\theta}}_{\scalebox{0.5}{$1$}}}$}}
    \put(57.61,5.06){\makebox(0,0){\footnotesize $\mathcal{M}_{{\bm{\theta}}_{\scalebox{0.5}{$N$}}}$}}
    \put(73.44,9.44){\makebox(0,0){\footnotesize ${G}_{\psi}$}}
    \put(80.5,11){\makebox(0,0){\tiny $\mathcal{L}_{\scalebox{0.8}{\text{phy}}}$}}
    \put(93.50,3.50){\makebox(0,0){\tiny $X_{\scalebox{0.8}{$\mathrm{L}$}}$}}
    \put(94.50,13.00){\makebox(0,0){\tiny $\hat{X}_{\scalebox{0.8}{$\mathrm{H}$}}$}}
    \end{overpic}
    \caption{\textbf{Overview of the proposed framework for three-dimensional turbulence super-resolution.}
    \textbf{a}, Set-up and overall framework. Given an LR three-dimensional sequence $X_{\mathrm{L}}$, the model learns a mapping $\mathcal{M}_\theta$ that produces a representative HR estimate conditioned on the LR input.
    \textbf{b}, Illustration of the state space model at time step $t$, which captures temporal dynamics of turbulent flows.  
    \textbf{c}, Illustration of the multiscale neural operator for recovering multiscale spatial structure of turbulent flows.
    \textbf{d}, Illustration of an MoE mechanism with unsupervised adaptation, enabling generalization to unknown coarse-graining.
    }
    \label{fig:method}
\end{figure}

This section is organized into three parts.
It begins with a high-level overview of the proposed framework and the inference pipeline (Fig.~\ref{fig:method}) to facilitate interpretation of the subsequent results.
Reconstruction performance is then evaluated across diverse turbulent regimes, including forced isotropic turbulence (FIT), forced magnetohydrodynamic turbulence (MHDT), homogeneous buoyancy-driven turbulence (HBDT), and decaying isotropic turbulence (DIT), as well as real-world atmospheric reanalysis data from ERA5~\cite{hersbach2020era5}.
The section concludes with an assessment of the framework behavior under unseen coarse-graining, including independently evolved LES inputs.

\subsection{Problem set-up}\label{ssec:problem_setup}
To improve the computational efficiency of turbulent flow simulation, one often resorts to coarse-grid discretization and/or reduced-resolution measurements, which inevitably introduce approximation errors relative to fully resolved DNS.
The objective is to reduce such errors by reconstructing high-fidelity, time-resolved volumetric turbulent dynamics from LR inputs.

The reconstruction of high-fidelity flow fields is cast as an inverse problem, wherein the objective is to recover the HR flow field representation $\bm{X}_\text{H}\!\in\!\mathbb{R}^{T\!\times\!H_\text{H}\!\times\!W_\text{H}\!\times\!D_\text{H}\!\times\!C}$ from its coarse-grid counterpart $\bm{X}_\text{L}\!\in\!\mathbb{R}^{T\!\times\!H_\text{L}\!\times\!W_\text{L}\!\times\!D_\text{L}\!\times\!C}$.
Here, $T$ denotes the temporal length of the sequence, $(H, W, D)$ the spatial resolution of the three-dimensional grid, subscripts $_\text{H}$ and $_\text{L}$ indicate HR and LR fields, respectively, and $C$ the number of physical variables (\eg, velocity components, pressure, or density).
For the volumetric turbulence experiments, the spatial resolutions satisfy $H_\text{H}=sH_\text{L}$, $W_\text{H}=sW_\text{L}$, and $D_\text{H}=sD_\text{L}$, where $s$ denotes the upscaling factor.

The degradation from HR to LR is formulated as
\begin{equation} \label{eq:deg}
    \bm{X}_{\mathrm{L}} = \mathcal{D}(\bm{X}_{\mathrm{H}}) + \bm{\xi},
\end{equation}
where $\mathcal{D}$ is referred to as the coarse-graining operator \cite{rudd1998coarse} and $\bm{\xi}$ quantifies the uncertainty arising from measurement or simulation inaccuracies.
The coarse-graining operator $\mathcal{D}$ removes fine-scale information and is generally many-to-one, so multiple HR fields can be consistent with the same LR sequence. This non-uniqueness is described by the conditional distribution $p(\bm{X}_{\mathrm{H}}\mid\bm{X}_{\mathrm{L}})$. The present framework targets a deterministic representative HR estimate under paired LR--HR supervision.
This estimator is parameterized by a neural network
$\mathcal{M}_{\bm{\theta}}:\mathbb{R}^{T\times H_\text{L}\times W_\text{L}\times D_\text{L}\times C}\to\allowbreak
\mathbb{R}^{T\times H_\text{H}\times W_\text{H}\times D_\text{H}\times C}$,
which maps an LR sequence $\bm{X}_\text{L}$ to a representative HR estimate $\hat{\bm{X}}_\text{H}=\mathcal{M}_{\bm{\theta}}(\bm{X}_\text{L})$.
The optimal parameters $\bm{\theta}^*$ are obtained via:
\begin{equation}\label{eq:optimization}
\bm{\theta}^* = \underset{\bm{\theta}}{\arg\min}\
\mathbb{E}_{(\bm{X}_\text{L}, \bm{X}_\text{H}) \sim \mathcal{T}}
\left[\mathcal{L}\!\left(\mathcal{M}_{\bm{\theta}}(\bm{X}_\text{L}),\, \bm{X}_\text{H}\right)\right],
\end{equation}
where $\mathcal{T}$ denotes the training data distribution,
$\mathcal{L}$ represents the loss function,
and $\mathbb{E}$ denotes the expectation over data pairs $(\bm{X}_\text{L}, \bm{X}_\text{H})$ drawn from $\mathcal{T}$.
The statistical role of the $L_1$ data term and the physical regularization is described in Section~\ref{ssec:optimization}.

As illustrated in Fig.~\ref{fig:method}a, given a temporal sequence of low-fidelity turbulent flow fields $\bm{X}_\text{L}$, the proposed model first lifts the input to a latent representation via a learnable lifting mapping $P$, yielding
$\bm{z} = P(\bm{X}_\text{L}) \in \mathbb{R}^{T \times H_\text{L} \times W_\text{L} \times D_\text{L} \times V}$, where $V$ denotes the latent channel dimension.
To preserve multiscale structures and coherent temporal evolution, the model explicitly performs spatiotemporal latent modeling of dynamics and cross-scale interactions.
Specifically, the latent representation evolves over time through a state space model, which captures temporal dynamics on the Eulerian grid and retains temporal memory, thereby promoting temporal consistency across the reconstructed sequence (Fig.~\ref{fig:method}b).
To characterize multiscale spatial structures in turbulent motions, the temporally evolved latent features are further processed by a multiscale neural operator that couples long-range interactions in the frequency domain with local refinement via convolution (Fig.~\ref{fig:method}c).
Finally, a decoder followed by a projection mapping $Q$ maps the latent representation back to the HR physical space, producing the reconstructed flow field $\hat{\bm{X}}_\text{H}$. The decoder architecture and output projection are specified for the turbulence and ERA5 tasks in Supplementary Notes~\ref{ssec:turb_decoder_config} and~\ref{ssec:era5_decoder_config}, respectively.
In addition to the architectural design, we impose spatiotemporal constraints derived from the governing equations by incorporating Navier--Stokes residuals into the training objective, encouraging physically consistent dynamics.

Most supervised learning-based super-resolution methods~\cite{fukami2021machine,shu2023physics} are trained under a fixed coarse-graining, which limits their ability to generalize when the degradation strategy differs at inference time.
To account for this, an unsupervised MoE mechanism is employed (Fig.~\ref{fig:method}d). Multiple expert models correspond to different coarse-graining hypotheses, and a gating network adaptively combines their predictions to accommodate unseen coarse-graining.

\begin{table*}[htbp]
\centering
\caption{\textbf{Numerical comparison of reconstruction metrics nRMSE (Eq.~\eqref{eq:nrmse}) between the proposed method and baseline methods on the TurbSR3D benchmark}.}
\label{tab:benchmark_results}
\sisetup{table-format=1.4}
\setlength{\tabcolsep}{6pt}
\renewcommand{\arraystretch}{1.20}
\resizebox{\textwidth}{!}{
	\begin{threeparttable}
    \begin{tabular}{lc S S S S S S}
    \toprule
    \multirow{2}{*}{\textbf{Methods}} & \multirow{2}{*}{\makecell{\textbf{Coarse-}\\\textbf{graining}}} & \multicolumn{2}{c}{\textbf{FIT}} & \multicolumn{2}{c}{\textbf{MHDT}} & \multicolumn{2}{c}{\textbf{HBDT}} \\
    \cmidrule(lr){3-4}\cmidrule(lr){5-6}\cmidrule(lr){7-8}
    & & {\textbf{4$\times$}} & {\textbf{8$\times$}} & {\textbf{4$\times$}} & {\textbf{8$\times$}} & {\textbf{8$\times$}} & {\textbf{16$\times$}} \\
    \midrule
    Tricubic & \multirow{9}{*}{\begin{tabular}{@{}c@{}}Box/ \\ Favre\end{tabular}} & 0.1204 & 0.2242 & 0.1838 & 0.3372 & 0.0933 & 0.1666 \\
    RRDB     &                             & 0.0460 & 0.1712 &  {0.0678} & 0.2205 & 0.0307 &  {0.0940} \\
    RCAN     &                             &  {0.0415} & 0.1753 & 0.0688 & 0.2285 &  {0.0262} & 0.1396 \\
    EDSR     &                             & 0.0423 & 0.1693 & 0.0699 &  {0.2250} & 0.0286 & 0.1301 \\
    FNO      &                             & 0.0730 & 0.1657 & 0.1166 & 0.2525 & 0.0544 & 0.1224 \\
    PhySR    &                             & 0.0476 &  {0.1486} & 0.0840 & 0.2443 & 0.0316 & 0.1159 \\
    DiffSR & & \multicolumn{1}{c}{0.0423} & \multicolumn{1}{c}{0.1440} & \multicolumn{1}{c}{0.0731} & \multicolumn{1}{c}{0.2153} & \multicolumn{1}{c}{0.0264} & \multicolumn{1}{c}{0.0860} \\
    SwinIR & & \multicolumn{1}{c}{0.0421} & \multicolumn{1}{c}{0.1499} & \multicolumn{1}{c}{0.0792} & \multicolumn{1}{c}{0.2447} & \multicolumn{1}{c}{0.0353} & \multicolumn{1}{c}{0.1228} \\
    \textbf{Ours} &                        & \textbf{0.0291}$^{\downarrow\text{29.9\%}}$ & \textbf{0.1195}$^{{\downarrow\text{17.0\%}}}$ & \textbf{0.0603}$^{\downarrow\text{11.1\%}}$ & \textbf{0.2094}$^{{\downarrow\text{2.7\%}}}$ & \textbf{0.0195}$^{\downarrow\text{25.6\%}}$ & \textbf{0.0739}$^{{\downarrow\text{14.1\%}}}$ \\
    \midrule
    Tricubic & \multirow{9}{*}{Gaussian}   & 0.1325 & 0.2278 & 0.2000 & 0.3327 & 0.0959 & 0.2074 \\
    RRDB     &                             & 0.0494 & 0.1749 &  {0.0692} & 0.2246 & 0.0316 &  {0.0985} \\
    RCAN     &                             &  {0.0421} & 0.1802 & 0.0712 & 0.2330 &  {0.0275} & 0.1055 \\
    EDSR     &                             & 0.0445 & 0.1752 & 0.0718 &  {0.2290} & 0.0299 & 0.1452 \\
    FNO      &                             & 0.0745 & 0.1679 & 0.1188 & 0.2559 & 0.0535 & 0.1224 \\
    PhySR    &                             & 0.0494 &  {0.1517} & 0.0852 & 0.2479 & 0.0389 & 0.1165 \\
    DiffSR & & \multicolumn{1}{c}{0.0438} & \multicolumn{1}{c}{0.1456} & \multicolumn{1}{c}{0.0756} & \multicolumn{1}{c}{0.2175} & \multicolumn{1}{c}{0.0300} & \multicolumn{1}{c}{0.0984} \\
    SwinIR & & \multicolumn{1}{c}{0.0429} & \multicolumn{1}{c}{0.1535} & \multicolumn{1}{c}{0.0804} & \multicolumn{1}{c}{0.2496} & \multicolumn{1}{c}{0.0354} & \multicolumn{1}{c}{0.1242} \\
    \textbf{Ours} &                        & \textbf{0.0300}$^{\downarrow\text{28.7\%}}$ & \textbf{0.1173}$^{{\downarrow\text{19.4\%}}}$ & \textbf{0.0620}$^{\downarrow\text{10.4\%}}$ & \textbf{0.2099}$^{{\downarrow\text{3.5\%}}}$ & \textbf{0.0200}$^{\downarrow\text{27.3\%}}$ & \textbf{0.0754}$^{\downarrow\text{23.4\%}}$ \\
    \midrule
    Tricubic & \multirow{9}{*}{Spectral}   & 0.1501 & 0.2505 & 0.2204 & 0.3640 & 0.1196 & 0.1671 \\
    RRDB     &                             & 0.0770 & 0.2156 &  {0.1130} &  {0.3013} & 0.0432 &  {0.1354} \\
    RCAN     &                             & 0.0755 & 0.2209 & 0.1135 & 0.3105 &  {0.0410} & 0.1438 \\
    EDSR     &                             & 0.0759 & 0.2140 & 0.1163 & 0.3036 & 0.0428 & 0.1476 \\
    FNO      &                             & 0.1036 & 0.2098 & 0.1612 & 0.3141 & 0.0812 & 0.1767 \\
    PhySR    &                             &  {0.0722} &  {0.2002} & 0.1188 & 0.3117 & 0.0452 & 0.1511 \\
    DiffSR & & \multicolumn{1}{c}{0.0810} & \multicolumn{1}{c}{0.2246} & \multicolumn{1}{c}{0.1339} & \multicolumn{1}{c}{0.3226} & \multicolumn{1}{c}{0.0496} & \multicolumn{1}{c}{0.1366} \\
    SwinIR & & \multicolumn{1}{c}{0.0704} & \multicolumn{1}{c}{0.2067} & \multicolumn{1}{c}{0.1190} & \multicolumn{1}{c}{0.3628} & \multicolumn{1}{c}{0.0521} & \multicolumn{1}{c}{0.1602} \\
    \textbf{Ours} &                        & \textbf{0.0615}$^{{\downarrow\text{12.6\%}}}$ & \textbf{0.1884}$^{\downarrow\text{5.9\%}}$ & \textbf{0.1071}$^{\downarrow\text{5.2\%}}$ & \textbf{0.2957}$^{\downarrow\text{1.9\%}}$ & \textbf{0.0348}$^{\downarrow\text{15.1\%}}$ & \textbf{0.1127}$^{\downarrow\text{16.8\%}}$ \\
    \bottomrule
    \end{tabular}
Three turbulent regimes are considered: incompressible forced isotropic turbulence (FIT), incompressible magnetohydrodynamic turbulence (MHDT), and homogeneous buoyancy-driven turbulence (HBDT).
Coarse-graining is simulated using three filtering strategies: box/Favre, Gaussian, and spectral cutoff.
We evaluate multiple upscaling factors (4$\times$, 8$\times$, and 16$\times$), depending on the regime.
	\end{threeparttable}
} %
\end{table*}

\subsection{Data preparation}\label{sec:data_prep}
The model is trained and evaluated on three classes of time-resolved volumetric turbulence from the TurbSR3D dataset (FIT, MHDT, and HBDT), covering distinct physical mechanisms and modeling challenges.
We also use DIT data generated by DNS, alongside real-world atmospheric data from ERA5~\cite{hersbach2020era5}.
The turbulence data and construction protocols are detailed below.

\noindent\textbf{FIT}~\cite{li2008public}.
As a canonical setting in turbulence research, forced isotropic turbulence is widely used to study multiscale spatiotemporal dynamics in incompressible flows~\cite{ten2004fully,carroll2013proposed}. 
We consider the three-component velocity and pressure fields, denoted as $\{\bm{u},\,p\}$.

\noindent\textbf{MHDT}~\cite{eyink2013flux}.
To assess performance in a multi-physics regime, we consider incompressible MHD turbulence, which describes the coupled dynamics of an electrically conducting fluid and a magnetic field. 
The velocity, magnetic field, and pressure fields $\{\bm{u},\,\bm{b},\,p\}$ are used to jointly evaluate kinetic and magnetic structures.

\noindent\textbf{HBDT}~\cite{livescu2014homogeneous}.
HBDT describes buoyancy-driven mixing between miscible fluids of different densities.
Density, velocity, and pressure fields $\{\rho,\,\bm{u},\,p\}$ are used to capture variable-density dynamics.

\noindent\textbf{DIT}.
It shares the variables of FIT but is unforced, so its statistics evolve as the kinetic energy decays.
We perform DNS of this flow and use it, separately from the TurbSR3D benchmark, to examine the physical constraints and generalization to LES inputs under statistically non-stationary conditions.
We consider the three-component velocity and pressure fields $\{\bm{u},\,p\}$.
The simulation setup is described in Supplementary Note~\ref{ap:dit_dns}.

\noindent\textbf{Coarse-graining simulation protocols.}
In practice, coarse-graining can arise from coarse-grid discretization (\eg, LES or RANS), limited-resolution measurements, or alternative numerical settings~\cite{bauer2015efficiency}.
Since aligned low- and high-fidelity flow fields generated by such processes are generally unavailable, we construct paired training data by applying coarse-graining to high-fidelity data.
Specifically, we consider three coarse-graining strategies for supervised training:
(i) box/Favre filtering (box for FIT, MHDT and Favre for HBDT)~\cite{fukami2019super,chung2023turbulence}, which mimics volumetric or density-weighted averaging,
(ii) Gaussian smoothing~\cite{bai2021predicting}, which approximates smooth low-pass effects, and
(iii) spectral cutoff~\cite{du2024conditional}, which models sharp wavenumber truncation.
To evaluate robustness to unseen coarse-graining, we further test (without retraining) on three additional strategies: tricubic downsampling, Lanczos filtering, and uniform subsampling  with $5\%$ additive noise.

\noindent\textbf{Experimental configuration.}
The evaluation covers FIT, MHDT, and HBDT under two upscaling factors per regime: 4$\times$/8$\times$ for FIT and MHDT, and 8$\times$/16$\times$ for HBDT.
For each supervised evaluation setting, data are split into training, validation, and test sets with a 6:2:2 ratio, and models are trained from scratch.
Additional details are provided in Section~\ref{sec:methods} and Supplementary Note~\ref{ap:data}.

\subsection{Accurate reconstruction of turbulent flows} \label{ssec:numerical_approximation}
Numerical reconstruction accuracy is assessed by benchmarking the proposed method against representative deep learning baselines, including RRDB~\cite{wang2018esrgan}, EDSR~\cite{bee2017enhanced}, RCAN~\cite{zhang2018image}, FNO~\cite{li2021fourier}, PhySR~\cite{ren2023physr}, DiffSR~\cite{zhu2026diffusion} and SwinIR~\cite{liang2021swinir} (see Section~\ref{ssec:baselines}).
We utilize established metrics in super-resolution, including the normalized root mean square error (nRMSE) presented in Table~\ref{tab:benchmark_results}, as well as the peak signal-to-noise ratio (PSNR), structural similarity index measure (SSIM), and Fourier RMSE (fRMSE) reported in Supplementary Tables~\ref{tab:fit_detailed}--\ref{tab:hbdt_detailed}.
As shown in Table~\ref{tab:benchmark_results}, the proposed method achieves the lowest nRMSE across all evaluated settings. 
On FIT, our method outperforms the second-best results in terms of nRMSE by 29.9\% (4$\times$, box), 28.7\% (4$\times$, Gaussian), and 12.6\% (4$\times$, spectral), with improvements persisting at 8$\times$ (17.0\%, 19.4\%, and 5.9\%).
MHDT presents the greatest challenge for joint restoration of its kinetic-magnetic structures, which are characterized by multi-physics interactions and non-local coupling. 
Nevertheless, our method remains superior, albeit with a smaller margin than in FIT. 
For HBDT, which features variable-density mixing and buoyancy-driven turbulence, the proposed method achieves significant error reductions in nRMSE (14.1\%--27.3\%).
The proposed method also achieves the highest SSIM and PSNR and the lowest fRMSE across all 18 settings (Supplementary Tables~\ref{tab:fit_detailed}--\ref{tab:hbdt_detailed}).
To quantify training uncertainty, we conduct multi-seed experiments. The proposed method achieves the best mean performance across all five metrics with low variability across seeds (Supplementary Table~\ref{tab:seed_fit}).
These results corroborate the importance of injecting spatiotemporal inductive biases for turbulence, where coherent evolution and cross-scale interactions are central to reconstruction fidelity.
Moreover, our method has a parameter count comparable to or smaller than most competing baselines (see Supplementary Fig.~\ref{fig:params}), indicating that the improvements do not arise from increased model capacity, but from the proposed spatiotemporal design.

\begin{figure}[htbp]
    \centering
    \begin{overpic}[width=0.87\textwidth]{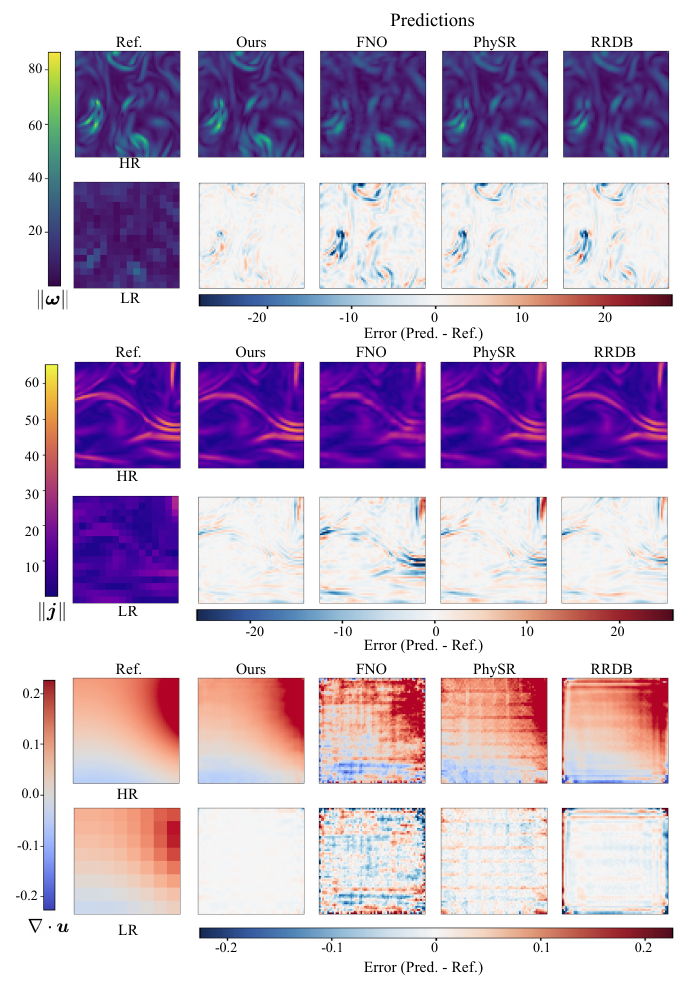}
    \put(2,95){\large\bfseries a}
    \put(2,64){\large\bfseries b}
    \put(2,31.5){\large\bfseries c}
    \end{overpic}
\caption{\textbf{Point-wise fidelity across three turbulence regimes and representative field quantities.}
\textbf{a,} Predicted vorticity magnitude $\|\bm{\omega}\|=\|\nabla\times\bm{u}\|$ and corresponding residual maps on FIT with $4\times$ upscaling.
\textbf{b,} Predicted current magnitude $\|\bm{j}\|=\|\nabla\times\bm{b}\|$ and residual maps on MHDT with $4\times$ upscaling.
\textbf{c,} Predicted velocity divergence $\nabla\!\cdot\!\bm{u}$ and residual maps on HBDT with $8\times$ upscaling.}
    \label{fig:spatial_fidelity_1}
\end{figure}

\subsection{Recovering multiscale spatial structure from coarse-grained data} \label{ssec:spatial_structure}
Beyond numerical accuracy, we further examine whether the proposed method produces high-fidelity reconstructions that preserve {multiscale} spatial structure and maintain coherent structures inherent to turbulent flows.

\noindent\textbf{Point-wise fidelity.}
Fig.~\ref{fig:spatial_fidelity_1} provides qualitative, point-wise comparisons using representative field quantities across three regimes, specifically targeting the recovery of challenging small-scale features.  
First, we examine the vorticity magnitude $\|\bm{\omega}\|$ on FIT to probe recovery of small-scale rotational filaments and intermittent structures.
As shown in Fig.~\ref{fig:spatial_fidelity_1}a, the proposed method yields more continuous filaments with reduced residual artifacts.
In contrast, FNO~\cite{li2021fourier}, RRDB~\cite{wang2018esrgan}, and PhySR~\cite{ren2023physr} exhibit red--blue halo patterns around vortex cores, suggesting incomplete recovery of fine-scale structures.
Second, we examine the current magnitude $\|\bm{j}\|$ on MHDT to probe thin current sheets and magnetic shear layers (Fig.~\ref{fig:spatial_fidelity_1}b).  
The proposed method recovers sharper and more continuous sheet structures, whereas competing baselines tend to produce thicker, fragmented sheets with orientation deviations.
Third, we use the velocity divergence field $\nabla\!\cdot\!\bm{u}$ on HBDT to characterize mixing-induced dilatation in variable-density flow under challenging upscaling (Fig.~\ref{fig:spatial_fidelity_1}c).
The proposed method better preserves the spatial organization of divergence features, while baselines introduce structural distortions and unphysical patterns.
For example, FNO~\cite{li2021fourier} and PhySR~\cite{ren2023physr} show pronounced grid-like artifacts and discontinuities, whereas RRDB~\cite{wang2018esrgan} exhibits boundary-related ringing.
These results align with the local refinement mechanism of the multiscale neural operator, which facilitates the recovery of small-scale details from coarse inputs. 

\begin{figure}[htbp]
    \centering
    \begin{overpic}[width=0.91\textwidth,percent]{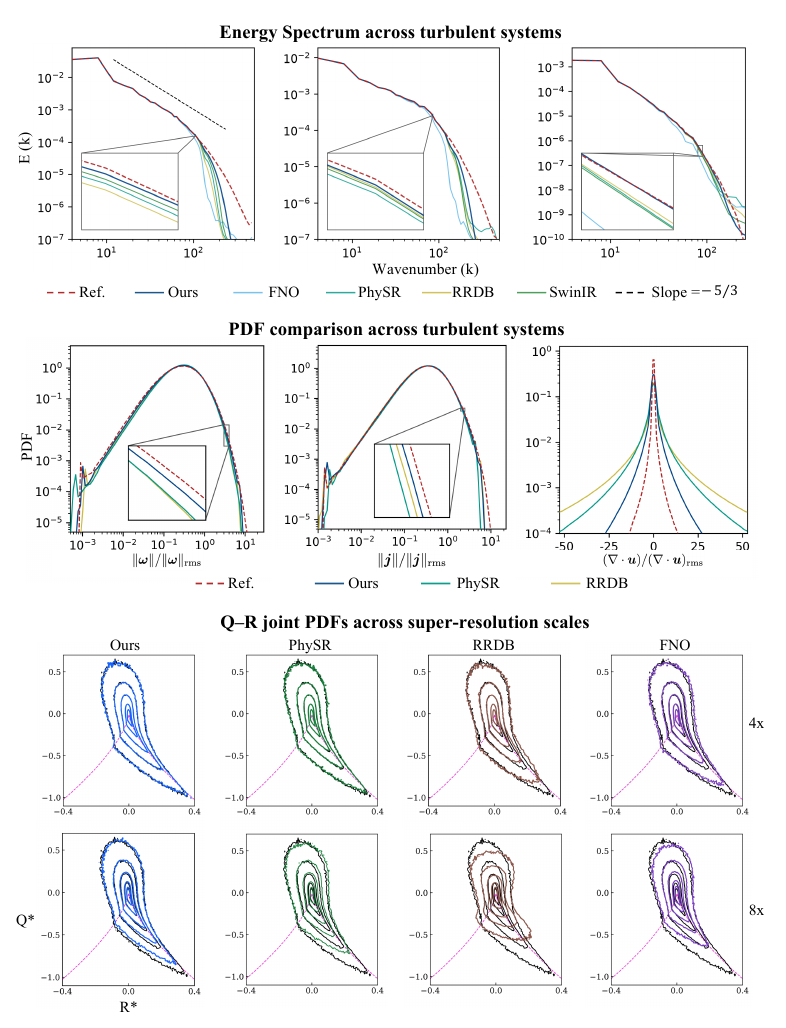}
    \put(2,96.3){\large\bfseries a}
    \put(2,67.3){\large\bfseries b}
    \put(2,38.6){\large\bfseries c}
    \end{overpic}
    \caption{\textbf{Statistical fidelity across three turbulence regimes.}
    \textbf{a,} Energy spectra for FIT ($4\times$), MHDT ($4\times$), and HBDT ($8\times$) under spectral-cutoff coarse-graining.
    \textbf{b,} Probability density functions (PDFs) of representative field quantities: vorticity magnitude $\|\bm{\omega}\|$ (FIT, $4\times$), current magnitude $\|\bm{j}\|$ (MHDT, $4\times$), and velocity divergence $\nabla\!\cdot\!\bm{u}$ (HBDT, $8\times$) under box/Favre coarse-graining. $(\cdot)_{\mathrm{rms}}$ denotes the root-mean-square magnitude.
    \textbf{c,} Joint PDFs of normalized velocity-gradient invariants $(Q^*,R^*)$ for FIT at $4\times$ and $8\times$ with box-filtered inputs. Black and colored contours denote DNS and reconstructions, respectively; the magenta dashed curve marks the zero-discriminant boundary. Evaluation details are given in Supplementary Note~\ref{ssec:spatial_structure_appendix}.}
    \label{fig:spatial_fidelity_2}
\end{figure}

\noindent\textbf{Energy spectra.}
The energy spectrum describes how energy is distributed over spatial scales, with higher wavenumbers corresponding to finer structures.
Fig.~\ref{fig:spatial_fidelity_2}a compares energy spectra across the three turbulence regimes, with insets highlighting the inertial--dissipation transition. 
We focus on spectral-cutoff coarse-graining, which explicitly removes high-wavenumber content and therefore poses a stringent test for spectrum recovery.
Across FIT, MHDT, and HBDT, the proposed method improves the recovery of high-wavenumber energy, reducing the underestimation exhibited by FNO, RRDB, PhySR, and SwinIR in the magnified ranges. In HBDT, the proposed method also better reproduces the decay of the reference spectrum towards the upper end of the displayed wavenumber range, where the baselines exhibit excess energy. Together, these results indicate that the proposed method more accurately reproduces the distribution of energy across scales, reducing both the energy deficit and the excess observed in the baseline spectra. This spectral behavior is consistent with the spatial comparisons in Fig.~\ref{fig:spatial_fidelity_1}c, where the proposed reconstruction avoids the grid-like and boundary artifacts visible in baseline models.
Under box/Favre coarse-graining, similar behavior is observed, with improved spectral consistency relative to competing baselines (Supplementary Fig.~\ref{fig:energy_spectrum}).
These differences can be interpreted through the architectures' spatial inductive biases. RRDB uses residual dense convolutions~\cite{wang2018esrgan}, PhySR combines convolutional recurrence with convolutional reconstruction~\cite{ren2023physr}, and SwinIR uses windowed and shifted-window attention~\cite{liang2021swinir}. Their spatial context grows through successive local or windowed operations. FNO provides global coupling through a finite set of retained Fourier modes~\cite{li2021fourier}. Our multiscale neural operator combines global Fourier coupling~\cite{guibas2022efficient} with local three-dimensional convolutional refinement, providing both large-scale context and local corrections for recovering missing fine-scale fluctuations. In the ablation study of the multiscale neural operator, removing the spectral operator produces the larger increase in low-wavenumber velocity reconstruction error, while removing the convolutional operator produces the larger increases at intermediate and high wavenumbers (Supplementary Fig.~\ref{fig:mno_ablation} and Supplementary Table~\ref{tab:mno_ablation}). Both removals affect all three ranges, supporting complementary and overlapping contributions to reconstruction across scales. On FIT at $4\times$, replacing the multiscale neural-operator blocks with MS-ESRGAN-derived convolutional blocks~\cite{yousif2021highfidelity} at closely matched capacity increases nRMSE from $0.0615$ to $0.0739$ ($20.1\%$) and produces greater high-wavenumber energy underestimation (Supplementary Table~\ref{tab:mno_msesrgan_ablation} and Supplementary Fig.~\ref{fig:mno_msesrgan_ablation}).

\noindent\textbf{Probability density functions.}
Probability density functions (PDFs) provide a complementary statistical perspective on whether reconstructed fields recover not only typical values but also intermittency.
Fig.~\ref{fig:spatial_fidelity_2}b shows PDFs of representative quantities: vorticity magnitude $\|\bm{\omega}\|$ (FIT, $4\times$), current magnitude $\|\bm{j}\|$ (MHDT, $4\times$), and velocity divergence $\nabla\!\cdot\!\bm{u}$ (HBDT, $8\times$).
Each quantity is normalized by its root-mean-square (rms) value to emphasize the distributional shape.
Across all three regimes, the proposed method closely follows the HR reference, including the tails that reflect intermittency and rare extremes.
For FIT, it reproduces the vorticity distribution over a wide dynamic range.
For MHDT, it recovers the tail decay of $\|\bm{j}\|$, consistent with localized current-sheet statistics.
For HBDT, the proposed method better reproduces the central peak and reduces the excess probability in the tails.
In contrast, baselines align with the reference mainly near the mode but exhibit increasing tail deviations, indicating reduced fidelity in intermittent small-scale statistics.
Distributional similarity is further quantified by the Jensen--Shannon (JS) divergence (Supplementary Fig.~\ref{fig:jsd}), where the proposed method achieves the lowest values in all cases.
These results indicate {improved recovery of tail behavior and intermittent statistics}.

\noindent\textbf{Topology of velocity-gradient invariants.}
To assess local deformation and rotation, we decompose the velocity gradient $\bm{A}=\nabla\bm{u}=\bm{S}+\bm{\Omega}$ into its symmetric strain tensor $\bm{S}$ and antisymmetric rotation tensor $\bm{\Omega}$. For incompressible flow, the invariants are
\(
Q=\tfrac12\left(\|\bm{\Omega}\|_F^2-\|\bm{S}\|_F^2\right),\quad R=-\det(\bm{A}).
\)
Thus, $Q>0$ identifies rotation-dominated regions and $Q<0$ strain-dominated regions, while $R$ combines the strain configuration with strain--vorticity coupling~\cite{chong1990classification,meneveau2011lagrangian}. Their joint PDF describes how frequently these local configurations occur. Its characteristic teardrop shape reflects preferred combinations of strain and rotation in turbulence~\cite{chevillard2006lagrangian}: a broad rotation-dominated region and an elongated strain-dominated lower-right branch. The latter includes motions that extend a fluid element in two directions and compress it in the third, locally flattening it. These statistics probe small-scale physics because strain governs viscous dissipation and vortex stretching amplifies vorticity (Supplementary Note~\ref{ssec:spatial_structure_appendix}).
Fig.~\ref{fig:spatial_fidelity_2}c compares joint-PDF contours of the normalized invariants $(Q^*,R^*)$ for FIT with box-filtered inputs. A common DNS strain scale and common density thresholds preserve differences in gradient intensity and occurrence frequency across reconstructions (Supplementary Note~\ref{ssec:spatial_structure_appendix}). At $4\times$, all four methods retain the basic teardrop shape, with our method and PhySR closely following the reference. 
At $8\times$, FNO and PhySR show pronounced contour contraction along the strain-dominated lower-right branch, while RRDB additionally shifts the rotation-dominated contours towards larger $R^*$ and distorts the shortened lower-right branch. These deviations indicate altered strain--rotation statistics and deficient recovery of strongly strain-dominated states. Stronger coarse-graining removes more fine-scale information, and the quadratic and cubic dependence of $Q$ and $R$ on velocity gradients amplifies reconstruction errors. Our method better preserves both features, consistent with global spatial context supporting local velocity refinement and physical constraints regularizing derivative relationships. The ablation reveals distinct regional changes (Supplementary Fig.~\ref{fig:ap_qr_ablation}): removing the local operator or physical loss shifts the right flank of the rotation-dominated contours towards larger $R^*$, whereas removing the spectral operator contracts the negative-$R^*$ side across both rotation- and strain-dominated regions. All three removals also impair the lower-right strain-dominated branch, with several intermediate contours showing insufficient extension and larger departures from the reference than in the full model. Extended baseline comparisons appear in Supplementary Fig.~\ref{fig:ap_qr_jpdf}.

Taken together, these diagnostics demonstrate that the proposed method preserves multiscale spatial structure across turbulent regimes. In particular, the multiscale neural operator plays a central role in this capability by combining global spectral interactions with local convolutional refinement, enabling complementary recovery of large-scale organization and fine-scale fluctuations.
The resulting improvements are consistently reflected across complementary measures of spatial fidelity, including point-wise reconstruction accuracy, energy spectra, distributional statistics, and velocity-gradient invariant topology.

\subsection{Capturing the temporal evolution of turbulent flows}\label{ssec:temporal_evolution}
Turbulent dynamics are time dependent. 
This motivates the design of the state space model, which maintains a latent state to encode temporal memory and updates it recurrently on the Eulerian grid.
This temporal context helps reconstruct the evolving turbulent flow.

\begin{figure}[t]
    \centering
    \begin{overpic}[width=1\textwidth]{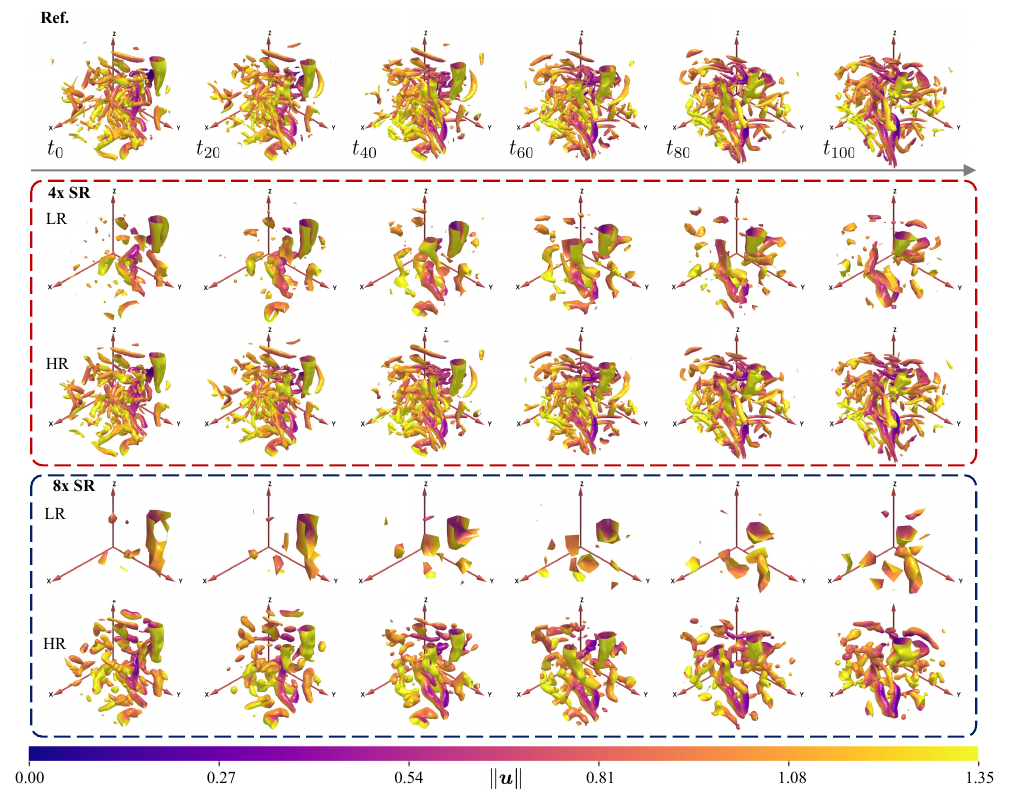}
    \put(1,76.3){\large\bfseries a}
    \put(1,59.2){\large\bfseries b}
    \put(1,30.5){\large\bfseries c}
    \end{overpic}
    \caption{\textbf{Q-criterion structures over time.} 
    \textbf{a}, Reference HR flow at six instants \(t_0,\dots,t_{100}\) (left to right) on FIT, with a uniform cadence \(\Delta t \! = \! 2 \!\! \times \!\! 10^{-3}\). 
    For each time, an isosurface of the $Q$-criterion highlights rotation-dominated regions; surfaces are color-mapped by speed $\|\bm{u}\|$.
    \textbf{b} and \textbf{c}, Super-resolution experiments at $4\times$ and $8\times$: for each SR factor, the top row shows the LR field, and the bottom row shows the reconstruction results by our method, evaluated at the same times as \textbf{a}. 
    The LR $Q$ fields are computed from the coarse-grained velocity components, with velocity gradients evaluated using the LR grid spacing.
    } 
    \label{fig:q-iso}
\end{figure}

\noindent\textbf{Qualitative evolution of vortex structures.}
Fig.~\ref{fig:q-iso} visualizes Q-criterion isosurfaces for FIT at six instants ($t_0$--$t_{100}$). 
The HR reference (Fig.~\ref{fig:q-iso}a) shows smoothly advecting and deforming vortex tubes. 
At $4\times$ super-resolution (Fig.~\ref{fig:q-iso}b), the reconstructed vortex tubes closely follow the changes in geometry and connectivity seen in the HR reference. 
At the more demanding $8\times$ setting (Fig.~\ref{fig:q-iso}c), small-scale filaments are partially attenuated by coarse-graining, yet the large-scale vortical motion remains temporally aligned with the HR sequence.

\begin{figure}[t]
    \centering
    \begin{overpic}[width=0.85\textwidth]{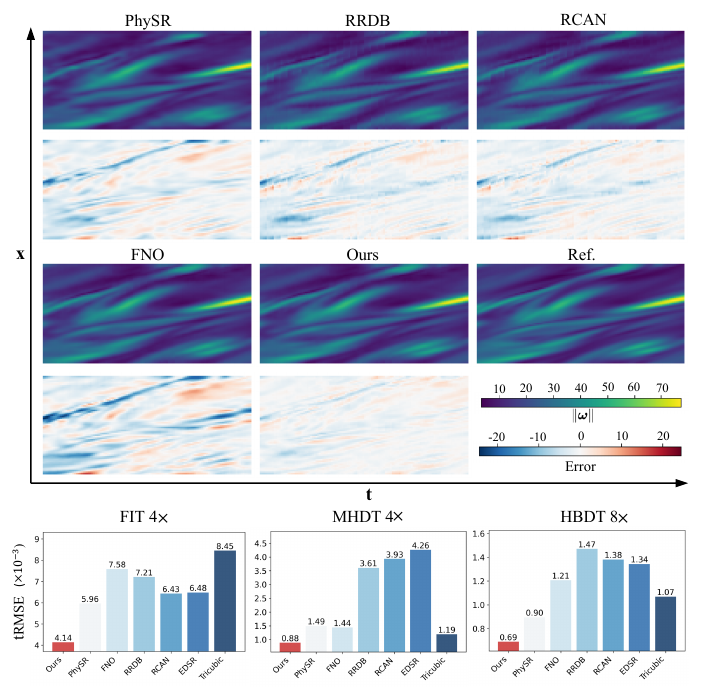}
    \put(2,95){\large\bfseries a}
    \put(2,22.5){\large\bfseries b}
    \end{overpic}
    \caption{\textbf{Temporal coherence of reconstructions.}
    \textbf{a,} Time--space ($t$--$x$) slices of vorticity magnitude $\|\bm{\omega}\|$ for a representative FIT sequence. The horizontal axis denotes time $t$ and the vertical axis denotes the spatial coordinate $x$. The top row shows the reference (HR) field and reconstructions from different methods, and the bottom row shows error maps.
    \textbf{b}, Quantitative temporal error measured by tRMSE (lower is better; values reported in units of $\times 10^{-3}$), computed using Eq.~\eqref{eq:trmse}, for FIT ($4\times$, left), MHDT ($4\times$, middle), and HBDT ($8\times$, right). 
    } 
    \label{fig:s-t}
\end{figure}

\noindent\textbf{Quantitative evaluation of temporal coherence.}
To complement the visual evidence, Fig.~\ref{fig:s-t}a shows $t$--$x$ slices of vorticity magnitude $\|\bm{\omega}\|$. 
Our reconstructions form continuous diagonal ridges with slope and phase consistent with the reference, whereas RRDB, RCAN and PhySR exhibit fragmented or discontinuous streaks, indicative of temporal inconsistency. 
We quantify errors in frame-to-frame changes using temporal RMSE (tRMSE; Eq.~\eqref{eq:trmse}) in Fig.~\ref{fig:s-t}b. 
Across FIT, MHDT, and HBDT, our method achieves the lowest tRMSE, with substantial reductions (23\%--31\%) relative to the best-performing baseline shown in Fig.~\ref{fig:s-t}b. 
Together, the structural comparisons and lower temporal-increment errors indicate closer agreement with the reference flow evolution.  

A state space model ablation on FIT supports the contribution of temporal modeling: relative to the variant trained without the state space model, the full model reduces nRMSE and tRMSE by $14.36\%$ and $16.05\%$ at $4\times$, and by $12.87\%$ and $6.52\%$ at $8\times$, respectively. The time--space slices show that the full model closely matches the reference structures and preserves their smooth evolution, while the model without the state space model and the other baselines exhibit varying degrees of local blurring, deformation, and reduced temporal smoothness (Supplementary Fig.~\ref{fig:ssm_removal_tx}). The state space model also outperforms other temporal baselines at closely matched parameter counts in the $4\times$ comparison (Supplementary Note~\ref{ssec:temporal_ablation}).

\clearpage
\begin{figure}[!ht]
    \centering
    \begin{overpic}[width=\textwidth]{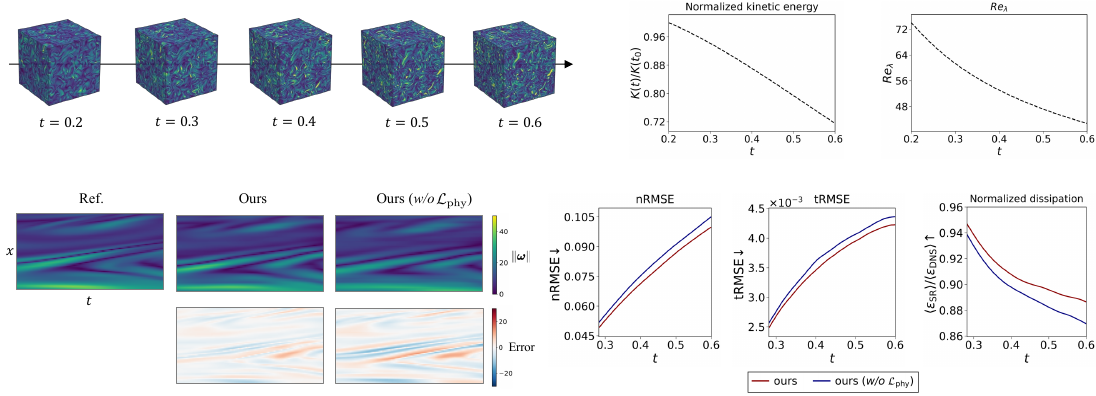}
    \put(0.4,34.5){\color{black}\normalfont\bfseries a}
    \put(56.5,34.5){\color{black}\normalfont\bfseries b}
    \put(79.0,34.5){\color{black}\normalfont\bfseries c}
    \put(0.4,17.5){\color{black}\normalfont\bfseries d}
    \put(50.0,17.5){\color{black}\normalfont\bfseries e}
    \put(66.4,17.5){\color{black}\normalfont\bfseries f}
    \put(83.5,17.5){\color{black}\normalfont\bfseries g}
    \end{overpic}
    \caption{\textbf{Dynamical consistency in DIT.} Reconstructions are evaluated at $4\times$ upscaling with spectral-cutoff inputs on DIT dataset (Supplementary Note~\ref{ap:dit_dns}).
    \textbf{a}, Reference flow fields at five times from $t=0.2$ to $t=0.6$.
    \textbf{b,c}, Evolution of normalized kinetic energy $K(t)/K(t_0)$ and Taylor-scale Reynolds number $Re_\lambda$ in the DNS.
    \textbf{d}, Space--time slices of vorticity magnitude $\|\bm{\omega}\|$ for the DNS reference (Ref.), the full model (Ours), and the model trained without $\mathcal{L}_{\mathrm{phy}}$, with reconstruction error maps beneath the predictions.
    \textbf{e,f}, Reconstruction error (nRMSE) and temporal-increment error (tRMSE); lower values indicate greater fidelity.
    \textbf{g}, Spatially averaged viscous dissipation normalized by the DNS value at the corresponding time. All displayed values are below unity, so higher ratios indicate closer agreement with DNS. Red and blue curves in \textbf{e--g} denote the full model and the variant without $\mathcal{L}_{\mathrm{phy}}$, respectively.}
    \label{fig:dynamical_consistency}
\end{figure}

\noindent\textbf{Dynamical consistency in decaying turbulence.}
To test whether physical soft constraints promote dynamical consistency under non-stationary conditions, we evaluate reconstructions of DIT (Supplementary Note~\ref{ap:dit_dns}) at $4\times$ upscaling with spectral-cutoff inputs. The sustained decline in kinetic energy and Taylor-scale Reynolds number (Fig.~\ref{fig:dynamical_consistency}a--c) requires reconstructions to track evolving flow statistics, providing a stringent test of fidelity throughout turbulent decay. We compare the full model with a variant trained without the physical loss $\mathcal{L}_{\mathrm{phy}}$ (Eq.~\eqref{eq:loss_p}).
The space--time slices show that the full model better preserves the position and continuity of the inclined vorticity streaks, with reduced local distortions and weaker error bands along the evolving structures (Fig.~\ref{fig:dynamical_consistency}d). Consistently lower tRMSE indicates closer agreement with reference frame-to-frame increments (Fig.~\ref{fig:dynamical_consistency}f), while lower nRMSE confirms improved reconstruction accuracy throughout the plotted interval (Fig.~\ref{fig:dynamical_consistency}e).
The viscous dissipation rate, $\varepsilon=2\nu S_{ij}S_{ij}$, where $\nu$ is the kinematic viscosity and $S_{ij}=\tfrac12(\partial u_i/\partial x_j+\partial u_j/\partial x_i)$ is the strain-rate tensor, measures the rate at which kinetic energy is converted into heat. Its dependence on velocity gradients makes dissipation a diagnostic of small-scale recovery. Although dissipation recovery becomes less complete as decay proceeds, the physical constraints maintain the mean dissipation ratio, $\langle\varepsilon_{\mathrm{SR}}\rangle/\langle\varepsilon_{\mathrm{DNS}}\rangle$ (angle brackets denote spatial averaging), closer to unity throughout the plotted interval, with the advantage over the variant without $\mathcal{L}_{\mathrm{phy}}$ widening towards later times (Fig.~\ref{fig:dynamical_consistency}g). Together with the improved temporal-increment fidelity, this sustained benefit supports the role of physical soft constraints in promoting dynamical consistency as turbulence decays.

\begin{figure}[htbp]
    \centering
    \begin{overpic}[width=1\textwidth]{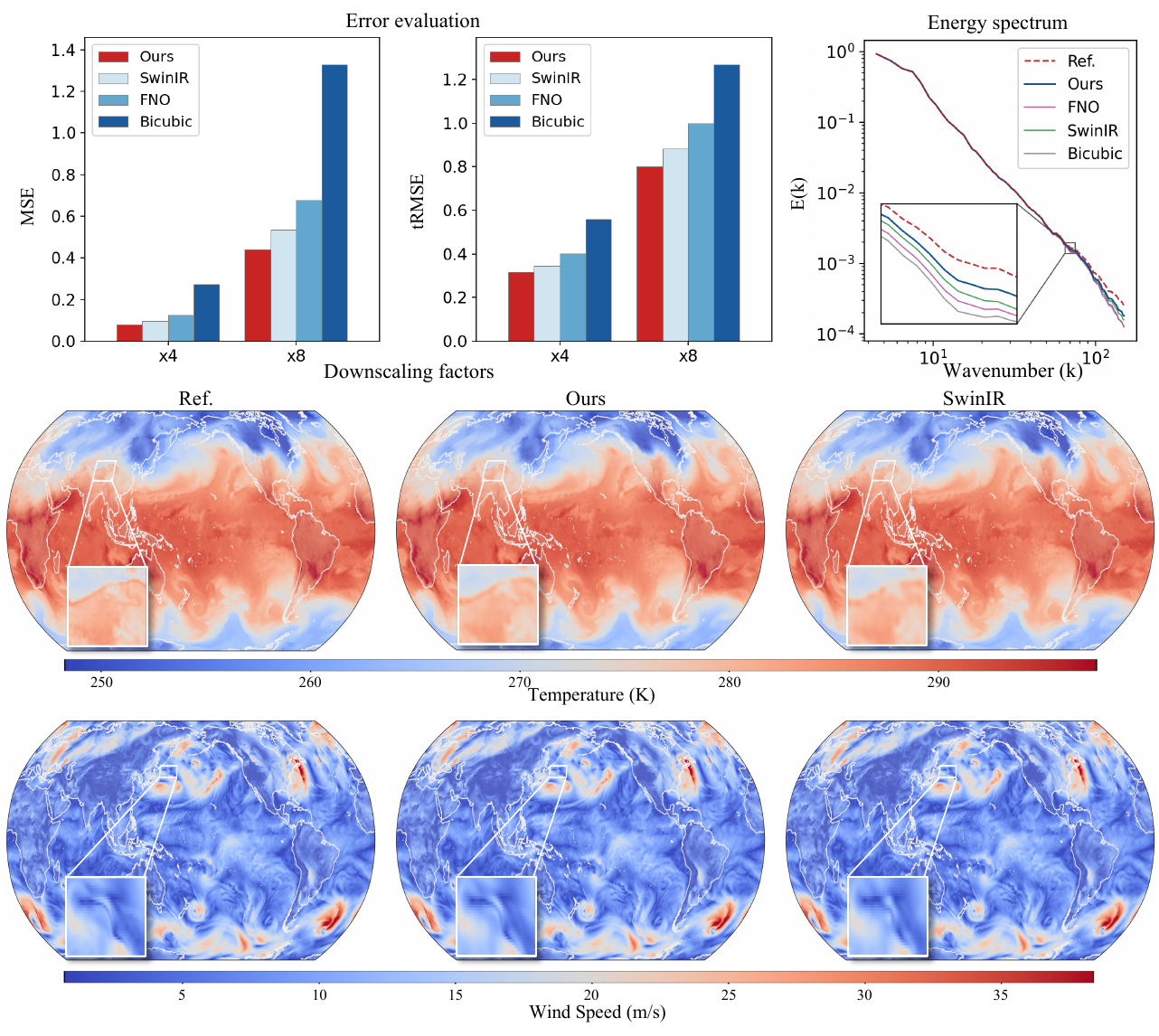}
    \put(0,87){\large\bfseries a}
    \put(69.5,87){\large\bfseries b}
    \put(0,55){\large\bfseries c}
    \end{overpic}
    \caption{\textbf{Atmospheric downscaling on ERA5 reanalysis data.}
    \textbf{a}, Mean Squared Error (MSE) and temporal error (tRMSE, Eq.~\eqref{eq:trmse}) computed at $4\times$ and $8\times$ downscaling factors. 
    \textbf{b}, Kinetic energy spectra $E(k)$ at 850~hPa plotted against wavenumber $k$.
    \textbf{c}, Visualizations of temperature (top) and wind speed (bottom) fields at 850~hPa. The panels display the HR reference (Ref.) alongside reconstructions from the proposed method and SwinIR. Insets display magnified views of local regions.
    }
    \label{fig:era5}
\end{figure}

\subsection{Atmospheric downscaling on ERA5}
This section evaluates atmospheric downscaling on ERA5 reanalysis fields~\cite{hersbach2020era5}, serving as an application-oriented validation in a real-world geophysical setting exhibiting multiscale structure and temporal coherence.
In this experiment, the LR input is a spatiotemporal multi-level atmospheric field 
$\bm{X}_{\mathrm{L}} \in \mathbb{R}^{T \times H_{\mathrm{L}} \times W_{\mathrm{L}} \times P \times C}$, where $H$ and $W$ correspond to latitude--longitude grids and $P$ indexes the pressure levels spanning 1000--300~hPa. The channel dimension $C$ includes temperature and horizontal wind components ($u, v$). Given the input $\bm{X}_{\mathrm{L}}$, the model reconstructs the corresponding HR sequence $\bm{X}_{\mathrm{H}} \in \mathbb{R}^{T \times H_{\mathrm{H}} \times W_{\mathrm{H}} \times P \times C}$ at higher spatial resolution.
As in the turbulence experiments, the third coordinate axis participates directly in the spatial representation. The three-dimensional backbone applies its Fourier and convolutional operators jointly over latitude, longitude, and pressure level, while $C$ indexes the physical variables. The ERA5 decoder increases the horizontal resolution through planar upsampling while retaining the prescribed pressure levels. Thus, the ERA5 task uses a three-dimensional cross-level representation with a horizontally refined output grid.
We compare against commonly used baselines for atmospheric downscaling~\cite{ren2025superbench,zhong2024investigating}, including bicubic interpolation, SwinIR~\cite{liang2021swinir}, and FNO~\cite{li2021fourier}. More details about experiment settings are provided in Section~\ref{ssec:datasets}.

Fig.~\ref{fig:era5}a reports MSE and tRMSE for 4$\times$ and 8$\times$ downscaling, measuring grid-point error and temporal fidelity, respectively. For both factors, the proposed method attains the lowest MSE among the compared methods. The same ordering holds for tRMSE, indicating smaller errors in frame-to-frame changes. This is consistent with the design goal of modeling temporal dynamics in latent space.
We also quantify training uncertainty through multi-seed experiments. The proposed method achieves the best mean performance across all five metrics at both downscaling factors, with low variability across seeds (Supplementary Table~\ref{tab:seed_era5}).

To further examine spatial fidelity, we compare spatial fields and spectra at 850~hPa. This level is widely adopted in synoptic and jet-related analyses~\cite{jones2023south} as a representative proxy for diagnosing lower-tropospheric wind and temperature structures. Fig.~\ref{fig:era5}c shows that the proposed method better preserves localized gradients and spatial organization in both temperature and wind-speed fields, with reduced artifacts in the zoomed regions. The kinetic energy spectra in Fig.~\ref{fig:era5}b are closer to the reference over a range of wavenumbers, indicating improved recovery of the multiscale energy distribution. These results align with the multiscale design of the proposed model, which couples global interactions with local refinement to reconstruct scale-dependent structures.

Vertical wind shear is a critical diagnostic in atmospheric analysis~\cite{Barton2025soil,jacob2024tropical}. To evaluate cross-level consistency in the reconstructed fields, we compute the 850--500~hPa bulk wind-vector difference and evaluate it using vCoh and vRMSE. Our method achieves the highest vCoh and lowest vRMSE at both downscaling factors, indicating more accurate preservation of the amplitude and spatial structure of this cross-level diagnostic than baselines (Supplementary Table~\ref{tab:vertical}).

\subsection{Generalizing to unknown coarse-graining}
\label{ssec:generalization}
This section evaluates generalization to unknown coarse-graining in two complementary settings. First, we study FIT under prescribed coarse-graining operators to assess adaptation to changes in filtering, resampling, and measurement noise. We then construct paired LES--DNS sequences of DIT to assess performance under a more realistic, simulation-generated degradation. In this second setting, coarse inputs arise from an independently evolved LES with a subgrid-scale (SGS) closure, introducing modeling and numerical errors that evolve over time.

\begin{figure}[hbtp]
    \centering
    \begin{overpic}[width=0.98\textwidth,percent]{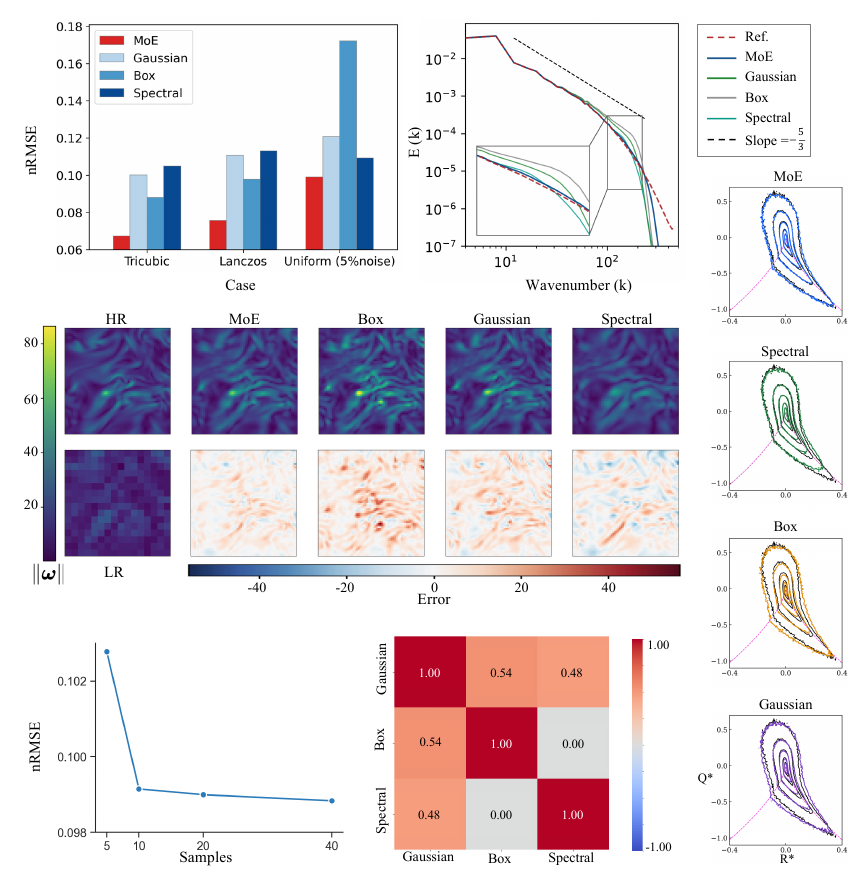}
    \put(2,97){\large\bfseries a}
    \put(48.5,97){\large\bfseries b}
    \put(2,64){\large\bfseries c}
    \put(78.5,79.5){\large\bfseries d}
    \put(2,29){\large\bfseries e}
    \put(42.5,29){\large\bfseries f}
    \end{overpic}
    \caption{\textbf{Generalization to unknown coarse-graining.}
        \textbf{a}, nRMSE of the Box, Gaussian, and Spectral experts and the MoE under three unseen coarse-graining settings: tricubic downsampling, Lanczos filtering, and uniform subsampling with $5\%$ additive noise.
        \textbf{b}, Energy spectrum $E(k)$ for the tricubic case.
        \textbf{c}, Reconstruction of vorticity magnitude $\|\bm{\omega}\|$ under tricubic downsampling, showing the HR reference, LR input, and outputs from the MoE and individual experts. 
        \textbf{d}, Joint PDFs of normalized velocity-gradient invariants $(Q^*,R^*)$ for the tricubic case (Supplementary Note~\ref{ssec:spatial_structure_appendix}).
        \textbf{e}, nRMSE of the MoE under uniform subsampling with $5\%$ noise as a function of the number of LR sequences used for adaptation.
        \textbf{f}, Pearson correlation coefficients of reconstruction error fields between pairs of experts for the tricubic case (Supplementary Note~\ref{ap:generalzation}).
    }
    \label{fig:generalization}
\end{figure}

We train three models under distinct coarse-graining strategies (Table~\ref{tab:benchmark_results}).
Box averaging and Gaussian smoothing attenuate a range of wavenumbers, whereas spectral cutoff removes modes above a prescribed threshold. Each expert therefore learns a reconstruction mapping conditioned on the spectral content retained by its training coarse-graining.
Although each model performs well under its training coarse-graining, performance degrades when the coarse-graining differs from the training setting, a challenge observed in turbulence super-resolution~\cite{nista2024influence}.
This limits practical applicability, where the specific coarse-graining is often unknown and can vary across data sources.
To address this issue, we introduce an MoE mechanism. 
Each pretrained model serves as an expert corresponding to a coarse-graining hypothesis, and its parameters are frozen.
At inference time, a lightweight gating network is optimized to combine the frozen expert predictions for given LR inputs (see Section~\ref{ssec:moe}). 

Evaluation is conducted on FIT under three representative unseen coarse-graining settings:
(i) tricubic downsampling, commonly used in rescaling-based super-resolution benchmarks~\cite{ren2025superbench};
(ii) Lanczos-filtered downsampling, a standard low-pass resampling strategy in signal processing~\cite{duchon1979lanczos,mottola2021reproducibility}; and 
(iii) uniform subsampling with $5\%$ additive Gaussian noise, mimicking noisy measurements~\cite{ren2025superbench}.
As shown in Fig.~\ref{fig:generalization}a, the MoE achieves lower numerical error than any single expert model across these settings.

Tricubic downsampling provides a representative case for examining multiscale spatial fidelity. 
The three experts share the same architecture, but each learns a different reconstruction mapping associated with its training coarse-graining. Their spectral differences on the same tricubic input therefore reveal the sensitivity of fine-scale reconstruction to the assumed coarse-graining. In Fig.~\ref{fig:generalization}b, all three experts overestimate energy over part of the high-wavenumber range, with the largest excess in the box-trained model. At still higher wavenumbers, their spectra decay prematurely. This scale-dependent bias is consistent with a mismatch between the spectral content retained in the test inputs and that encountered during training~\cite{nista2024influence}. By adapting the combination of expert predictions to the unseen inputs, the MoE reduces the energy excess and better reproduces the reference spectral decay across the magnified range.
Consistent trends are also observed in structure-sensitive diagnostics: the MoE maintains the characteristic shape and extent of the DNS $Q$--$R$ topology (Fig.~\ref{fig:generalization}d) and avoids excessive smoothing in vortical structures, resulting in improved structural continuity (Fig.~\ref{fig:generalization}c).

To assess the sample efficiency of adaptation, we vary the number of LR sequences used to optimize the gating network (Fig.~\ref{fig:generalization}e). 
The error decreases sharply as the number of sequences increases from 5 to 10, with only small improvements thereafter. 
This trend indicates that adaptation can be achieved with limited unpaired LR data under unseen coarse-graining.
The low correlation between the box and spectral reconstruction errors further supports the use of these experts as complementary sources of information (Fig.~\ref{fig:generalization}f; Supplementary Note~\ref{ap:generalzation}).

\begin{figure}[hbtp]
    \centering
    \begin{overpic}[width=0.98\textwidth,percent]{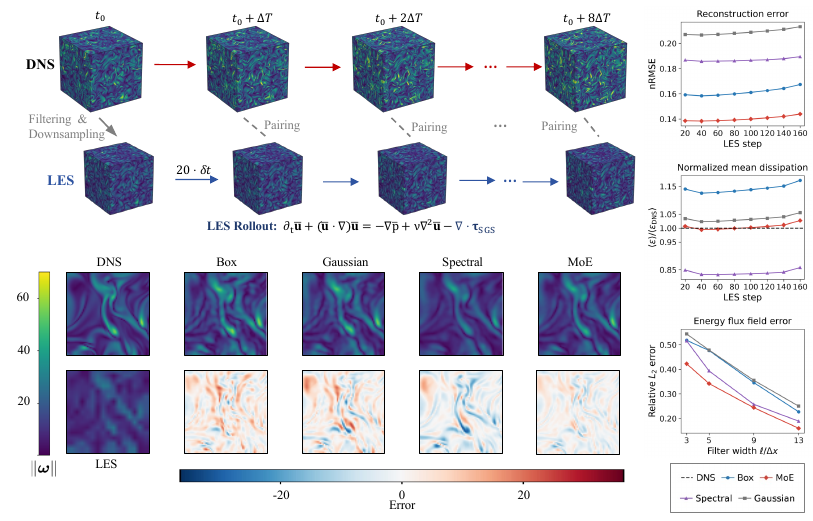}
    \put(0.5,62.2){\color{black}\normalfont\large\bfseries a}
    \put(0.5,31.4){\color{black}\normalfont\large\bfseries b}
    \put(79.2,62.2){\color{black}\normalfont\large\bfseries c}
    \put(79.2,43.3){\color{black}\normalfont\large\bfseries d}
    \put(79.2,24.5){\color{black}\normalfont\large\bfseries e}
    \end{overpic}
    \caption{\textbf{Generalization to LES inputs in DIT.}
        \textbf{a}, A filtered and downsampled DNS state initializes an independently evolved LES with an SGS closure. LES and DNS fields are paired at matching physical times, separated by $\Delta T=20\delta t$, where $\delta t$ is the LES integration time step.
        \textbf{b}, Vorticity magnitude $\|\bm{\omega}\|$ at LES step 160 for DNS, LES, the individual experts and the MoE, with reconstruction error maps beneath the corresponding predictions.
        \textbf{c}, Reconstruction nRMSE against the paired DNS references over 160 LES steps.
        \textbf{d}, Spatially averaged viscous dissipation normalized by the DNS value at each corresponding time; the dashed line denotes unity.
        \textbf{e}, Relative $L_2$ error of the local energy flux field, $\|\Pi_\ell-\Pi_\ell^{\mathrm{DNS}}\|_2/\|\Pi_\ell^{\mathrm{DNS}}\|_2$, aggregated over all evaluated times and spatial locations. Filter width $\ell$ is normalized by the DNS grid spacing $\Delta x$.
    }
    \label{fig:les_dns}
\end{figure}

For the paired LES--DNS test in DIT, a DNS velocity field is filtered and downsampled to initialize the LES, which subsequently advances independently with an SGS closure (Fig.~\ref{fig:les_dns}a). LES inputs and DNS references are paired at matching physical times. This construction captures the combined effects of limited spatial resolution, SGS modeling, and numerical evolution encountered in coarse-grid simulations. The DNS simulation and LES--DNS pairing are described in Supplementary Notes~\ref{ap:dit_dns} and~\ref{ap:les_dns_pairing}.

A central challenge is the chaotic sensitivity of turbulent trajectories to perturbations. SGS-model and discretization errors can be amplified by nonlinear evolution, progressively separating the LES trajectory from the corresponding DNS trajectory~\cite{pope2000turbulent,boffetta2017chaos}. Reconstruction must therefore accommodate an evolving dynamical mismatch in addition to recovering unresolved spatial content. With all experts frozen, we adapt the gating network without HR supervision using only 10 LES patches. The MoE consistently achieves the lowest reconstruction nRMSE throughout the LES evolution (Fig.~\ref{fig:les_dns}c). The MoE also better recovers the fine-scale vortical structures of the DNS field (Fig.~\ref{fig:les_dns}b).
These improvements extend to the physical processes of energy dissipation and interscale transfer. Using the dissipation diagnostic introduced in Section~\ref{ssec:temporal_evolution}, the MoE maintains a normalized mean dissipation, $\langle\varepsilon_{\mathrm{SR}}\rangle/\langle\varepsilon_{\mathrm{DNS}}\rangle$, close to unity, while the Box and Gaussian experts overestimate dissipation and the Spectral expert underestimates it (Fig.~\ref{fig:les_dns}d). This agreement is consistent with the MoE more faithfully recovering the small-scale velocity gradients responsible for viscous dissipation.

We further examine the local energy flux, $\Pi_\ell=-\tau_{ij}^{(\ell)}\bar S_{ij}^{(\ell)}$, which characterizes kinetic-energy transfer across scale $\ell$. Here, $\tau_{ij}^{(\ell)}$ and $\bar S_{ij}^{(\ell)}$ denote the subfilter stress and filtered strain-rate tensor, respectively. The MoE yields the lowest relative flux-field error at every examined filter scale (Fig.~\ref{fig:les_dns}e), indicating improved recovery of the magnitude and spatial distribution of multiscale energy transfer.
Together, these results demonstrate that the MoE enables data-efficient generalization to unseen coarse-graining and SGS-modeled LES inputs without HR supervision.

\section{Discussion}\label{sec:discussion}

This work presents a turbulence-aware deep learning framework for super-resolution reconstruction of time-resolved three-dimensional flows from coarse-grained inputs. The framework is designed to improve numerical fidelity while better preserving turbulence-relevant spatiotemporal structure.
The framework combines a multiscale neural operator with a state space model to reconstruct spatial structures and their temporal evolution, while Navier--Stokes residuals provide physical regularization during training. An MoE mechanism further enables adaptation to unknown coarse-graining using LR data alone.

The framework achieves lower reconstruction errors than the compared learning-based baselines across FIT, MHDT, and HBDT under box/Favre filtering, Gaussian filtering, and spectral cutoff, with nRMSE reductions reaching approximately 30\% relative to the best-performing baseline. Energy spectra, PDFs, and joint distributions of velocity-gradient invariants further show that the reconstructed fields recover multiscale energy distributions and key features of turbulence statistics. The evolution of vortical structures and space--time patterns demonstrates improved temporal coherence relative to the baselines. In DIT, physical regularization improves agreement with the reference temporal evolution and mean viscous dissipation, promoting dynamical consistency under non-stationary conditions. Results on ERA5 atmospheric reanalysis data show that these improvements carry over to geophysical flows. Under the tested unseen coarse-graining settings and for independently evolved LES inputs, the MoE achieves lower reconstruction errors than the individual experts, with adaptation using only unpaired LR data.

Despite these promising results, several limitations warrant future work. 
First, the framework provides a single deterministic reconstruction for each LR sequence and does not quantify the variability among compatible HR realizations. Agreement with reference fields and selected turbulence statistics therefore does not establish recovery of the full conditional distribution. Conditional generative approaches~\cite{zhu2026diffusion,du2024conditional,raonic2026generative} offer a possible way to represent this variability, but may introduce additional training and sampling costs, particularly when iterative sampling and multiple realizations are required. Such overhead could compromise the computational efficiency that motivates the present framework. Future work could therefore explore computationally efficient approaches to uncertainty quantification that complement the present deterministic framework.
Second, generalization across physically distinct turbulent systems remains a fundamental challenge.
Turbulent flows exhibit substantial regime-to-regime variability in dynamics, scale interactions, and statistics, making universal generalization difficult without exposure to diverse physical regimes during training.
One possible path forward is large-scale pretraining on substantially broader and more heterogeneous turbulence datasets~\cite{taira2025machine}. In practice, however, such datasets remain expensive and difficult to obtain, particularly for high-fidelity three-dimensional turbulence~\cite{mcconkey2021curated,duraisamy2021perspectives}.
An alternative direction is the development of hybrid approaches that tightly integrate data-driven models with physical structure or constraints, thereby reducing reliance on exhaustive multi-regime data coverage~\cite{Sharma2023review,taira2025machine}.  
Third, extending the framework to unstructured and curvilinear meshes in complex engineering flows presents additional challenges associated with mesh geometry and spatial discretization~\cite{barwey2025mesh}. Addressing these challenges warrants dedicated investigation for specific flow configurations and forms an important direction for future work.
Fourth, the framework leverages knowledge of governing equations to define physical constraints, which may be less effective when the governing dynamics are partially unknown.
A promising alternative is to derive effective governing equations from data, and incorporate the inferred dynamics back into the learning process~\cite{gurevich2024learning}.
Fifth, training multiple experts increases the one-time training cost; however, experts are trained once and then frozen, while unsupervised adaptation requires optimizing only a lightweight gating network without HR supervision. 
Finally, the proposed framework serves as a post-processing component that complements high-fidelity simulations by reconstructing HR fields from coarse-grained data (\eg, LES or coarse-grid simulations; Section~\ref{ssec:generalization}).
By enhancing the effective resolution of existing numerical data, this approach reduces the reliance on prohibitively expensive high-fidelity simulations or measurements when detailed flow information is required.
In this way, the framework complements conventional numerical pipelines and provides a practical pathway toward accessing HR turbulence statistics at reduced computational cost.

\section{Methods}\label{sec:methods}
This section presents the methodology for time-resolved volumetric turbulence super-resolution, addressing the challenge of recovering high-fidelity dynamics from coarse-grained inputs.
The proposed framework integrates four complementary components:
(i) a state space model to capture temporal evolution on Eulerian grids,  
(ii) a multiscale neural operator to model cross-scale spatial structure interactions,  
(iii) Navier--Stokes-based soft constraints to encourage dynamical consistency,
and (iv) an unsupervised MoE mechanism to maintain robustness against unknown coarse-graining.
Together, these components form a unified framework for high-fidelity reconstruction of turbulent flow dynamics.
The following subsections describe these components, the evaluation metrics, the datasets, and the baseline methods.

\subsection{State space model}\label{ssec:ssm}
Turbulent flows exhibit temporal dependencies, where the evolution of flow structures depends on historical states.
To capture this coherent temporal evolution for turbulence, we adopt a state space model~\cite{gu2021combining} as the core mechanism for temporal modeling. 
A state space model provides a principled strategy for capturing temporal dependencies and has been adopted in a variety of sequence modeling tasks~\cite{rangapuram2018deep,fu2022hungry,li2024videomamba}.
In its standard form, it maps an input sequence $\{\bm{x}_t\}_{t}$ to outputs $\{\bm{y}_t\}_{t}$ via a latent state $\bm{h}_t$:
\begin{equation}
\begin{split}
        \bm{h}_t &= \bm{A}\bm{h}_{t-1} + \bm{B}\bm{x}_t, \\
        \bm{y}_t &= \bm{C}\bm{h}_t + \bm{D}\bm{x}_t,
\end{split}
\end{equation} 
where $\bm{A}, \bm{B}, \bm{C}, \bm{D}$ are learnable parameters.

To account for the intermittent and spatially varying dynamics of three-dimensional turbulence~\cite{sreenivasan1999fluid},
we extend the state space model with an input-conditioned parameterization~\cite{gu2023mamba}. The input features dynamically modulate the system matrices $\bm{B}$ and $\bm{C}$, allowing the model to adapt its state updates and readout to evolving flow patterns.
Specifically, we apply the state space model on the 3D Eulerian grid by maintaining a local latent state $\bm{h}_t(\bm{r}) \in \mathbb{R}^M$ at each spatial coordinate $\bm{r} = (x, y, z)$. 
Given a latent representation $\bm{z}_t$ from the lifting mapping $P$,
we first apply a 3D convolution (conv3d) to $\bm{z}_t$ for local spatial aggregation. The latent state then evolves independently over time at each spatial location $\bm{r}$:
\begin{equation}
\begin{split}
    \bm{h}_t(\bm{r})
    &= {\bm{A}} \, \bm{h}_{t-1}(\bm{r})
     + {\bm{B}}_t(\bm{r}) \, \bm{z}_t(\bm{r}), \\
    \bm{y}_t(\bm{r})
    &= \bm{C}_t(\bm{r}) \, \bm{h}_t(\bm{r})
     + \bm{D} \, \bm{z}_t(\bm{r}),
\end{split}
\end{equation}
where $\bm{D}$ serves as an input-independent skip connection that preserves immediate signal information.
The parameter ${\bm{A}}$ governs the temporal memory of the system, while the coefficients ${\bm{B}}_t$ and $\bm{C}_t$ are dynamically modulated by $\bm{z}_t$ to adapt to local flow variations.
To enhance representational capability, we apply a spatially varying selection gate to $\bm{y}_t$, yielding the feature $\bm{s}_t$ that serves as the input for the  multiscale neural operator.
Further details are provided in Supplementary Note~\ref{ap:ssm}.

\subsection{Multiscale neural operator}
Modeling the multiscale physics of turbulence requires capturing both global correlations (\eg, large-scale coherent structures) and local variations (\eg, small dissipative eddies)~\cite{duraisamy2019turbulence}. Standard convolutional networks, while efficient for local patterns, often struggle to model long-range dependencies due to their limited effective receptive fields~\cite{lecun1998gradient}. Conversely, spectral operators offer global receptive fields but may lack the inductive bias needed for precise local feature refinement~\cite{li2021fourier}. To synergize these complementary strengths, we employ a stack of $K$ hybrid blocks that couple a spectral operator for large-scale structure modeling with a convolutional operator for local refinement.
Each hybrid block is defined as:
\begin{equation}
\hat{\bm{s}}_t = \bm{s}_t + \mathcal{O}_{c}(\mathcal{O}_{s}(\bm{s}_t)),
\end{equation}
where $\bm{s}_t$ denotes the temporally evolved feature yielded by the state space model, and $\mathcal{O}_{s}$ and $\mathcal{O}_{c}$ provide global spectral coupling and local convolutional refinement, respectively.  
For the global coupling $\mathcal{O}_{s}$, we adopt the Adaptive Fourier Neural Operator~\cite{guibas2022efficient}. By performing global spatial coupling in the frequency domain, this operator enjoys a global receptive field, allowing it to efficiently capture long-range interactions across the entire domain:
\begin{equation}
\mathcal{O}_{s}(\bm{s}_t):=\mathcal{F}^{-1}\!\left(\mathcal{S}_{\lambda}\big(\text{MLP}(\mathcal{F}(\bm{s}_t))\big)\right) + \bm{s}_t,
\end{equation}
where $\mathcal{F}$ denotes the 3D discrete Fourier transform and $\mathcal{F}^{-1}$ its inverse operation. The soft-thresholding operation $\mathcal{S}_{\lambda}$ applies element-wise shrinkage in the spectral domain:
\begin{equation} \label{eq:soft-shrinkage}
\mathcal{S}_{\lambda}(\bm{x}):=\frac{\bm{x}}{|\bm{x}|}\odot \max\!\big(|\bm{x}|-\lambda,\,0\big),
\end{equation}
with threshold $\lambda$ controlling the amount of shrinkage. 
This shrinkage regularizes small-magnitude spectral coefficients and improves stability in practice.
However, relying solely on spectral operations, especially with sparsity constraints, may risk smoothing out fine-scale turbulent fluctuations. To refine these local details, we parameterize the convolution operator $\mathcal{O}_c$ as:
\begin{equation}
\mathcal{O}_{c}(\bm{s}_t)
:= \bm{W}_2 \!\left( \text{Conv}_{\text{dep}}\!\big(\phi(\bm{W}_1 \bm{s}_t)\big) \right),
\end{equation}
where $\bm{W}_1$ and $\bm{W}_2$ are point-wise linear projections, $\phi(\cdot)$ is a nonlinear activation, and $\text{Conv}_{\text{dep}}$ denotes a 3D depthwise convolution.
This design supplies a local inductive bias for spatial feature refinement.
The contributions of the two operators across spatial scales
are evaluated through component-removal experiments in Supplementary
Note~\ref{ssec:mno_ablation}.

\subsection{Physics-informed neural optimization}
\label{ssec:optimization}
We use a voxel-wise $L_1$ loss to supervise reconstruction from paired LR--HR data. Minimizing the expected absolute error yields componentwise conditional medians~\cite{gneiting2011point}, providing a statistical basis for this deterministic point-estimation objective. The linear growth of the $L_1$ penalty limits the influence of large reconstruction errors compared with a squared-error penalty. We combine this data term with differentiable physical penalties that act as soft constraints and encourage consistency with the governing equations.
The total training objective is defined as
\begin{equation}
    \mathcal{L} = \mathcal{L}_{\mathrm{mae}} + \lambda_{\mathrm{phy}}\,\mathcal{L}_{\mathrm{phy}},
    \label{eq:loss}
\end{equation}
where $\mathcal{L}_{\mathrm{mae}}$ measures voxel-wise reconstruction error between the
prediction $\mathcal{M}_{\bm{\theta}}(\bm{X}_{\mathrm{L}})$ and the HR target
$\bm{X}_{\mathrm{H}}$, and $\mathcal{L}_{\mathrm{phy}}$ penalizes violations of the governing
equations with a weighting factor $\lambda_{\mathrm{phy}}$.
The physical loss combines a momentum term with a divergence-related constraint:
\begin{equation}
    \mathcal{L}_{\mathrm{phy}} =
    \beta \mathcal{L}_{\mathrm{mom}}
    + \|\mathcal{R}_{\mathrm{div}}\|_{1},
    \label{eq:loss_p}
\end{equation}
where $\beta$ controls the relative strength of the momentum regularizer. The loss weights and their sensitivity are reported in Supplementary Notes~\ref{ap:implementation} and~\ref{ssec:physical_loss_sensitivity}. We use $L_1$ penalties for the physical residuals because their linear growth limits the influence of large local residuals compared with squared $L_2$ penalties. The residuals and corresponding momentum penalties are defined below for each flow system.

\noindent\textbf{FIT.}
We define the source-free momentum defect and divergence residual as
\begin{equation}
\begin{split}
    \mathcal{R}_{0}^{\mathrm{FIT}} &=
        \frac{\partial \bm{u}}{\partial t}
        + (\bm{u}\cdot\nabla)\bm{u}
        + \frac{1}{\rho}\nabla p
        - \nu \nabla^{2}\bm{u}, \\
    \mathcal{R}_{\mathrm{div}} &= \nabla\cdot\bm{u}.
\end{split}
\end{equation}
Here, \(\bm{u}\) denotes the velocity field, \(p\) the pressure, and \(\nu\) the kinematic viscosity. \(\nabla\) denotes the spatial gradient operator.

\noindent\textbf{MHDT.}
The source-free momentum defect additionally includes the Lorentz-force contribution, and both velocity and magnetic-field divergence enter the solenoidal constraint:
\begin{equation}
\begin{split}
    \mathcal{R}_{0}^{\mathrm{MHDT}} &=
        \frac{\partial \bm{u}}{\partial t}
        + (\bm{u}\cdot\nabla)\bm{u}
        + \nabla p
        - \nu \nabla^{2}\bm{u}
        - (\nabla\times\bm{b})\times\bm{b}, \\
    \mathcal{R}_{\mathrm{div}} &= 
        (\nabla\cdot\bm{u},\, \nabla\cdot\bm{b}) .
\end{split}
\end{equation}
Here, \(p=p_{\mathrm{th}}/\rho_0\) denotes the kinematic pressure, with \(p_{\mathrm{th}}\) the thermodynamic pressure and \(\rho_0\) the constant mass density. The magnetic field \(\bm{b}\) is expressed in Alfv\'en velocity units.

For FIT and MHDT, we apply a spatial high-pass filter $\mathcal{H}_{\sigma}$ to the momentum residual before evaluating the loss:
\begin{equation}
    \mathcal{L}_{\mathrm{mom}}
    = \left\|\mathcal{H}_{\sigma}
      [\mathcal{R}_{0}]\right\|_1.
    \label{eq:scale_selective_momentum}
\end{equation}
The complete momentum residual is $\mathcal{R}_0-\bm{f}$, so omitting external forcing changes the filtered residual by $\mathcal{H}_{\sigma}[\bm{f}]$. In both datasets, forcing is confined to low wavenumbers~\cite{li2008public,eyink2013flux}, and its direct contribution is strongly attenuated by the high-pass filter. This yields a scale-selective momentum penalty that emphasizes finer spatial scales. The nonlinear terms are evaluated on the full reconstructed fields before filtering, so interactions across scales remain included. A comparison on FIT supports this formulation: the high-pass $L_1$ momentum loss yields lower high-wavenumber spectral error than both unfiltered $L_1$ and high-pass squared $L_2$ penalties (Supplementary Fig.~\ref{fig:fit_momentum_norm}). Supplementary Note~\ref{ssec:scale_selective_momentum} provides the filter implementation, quantifies the remaining forcing contribution, and reports the loss comparison.

\noindent\textbf{HBDT.}
The residuals follow the variable-density, low-Mach-number Navier--Stokes formulation. The momentum residual includes the density-dependent inertial and viscous terms as well as the buoyancy contribution:
\begin{equation}
\begin{split}
    \mathcal{R}_{\mathrm{mom}} &=
        \frac{\partial (\rho\bm{u})}{\partial t}
        + \nabla\cdot(\rho\,\bm{u}\otimes\bm{u})
        + \nabla p
        - \nabla\cdot\bm{\tau}
        - \rho\bm{g}, \\
    \mathcal{R}_{\mathrm{div}} &=
        \frac{\partial \rho}{\partial t}
        + \nabla\cdot(\rho\bm{u}).
\end{split}
\end{equation}
Here, \(\rho\) is the variable density, \(\bm{g}\) the gravitational acceleration, and \(\bm{\tau}\) the viscous stress tensor.
The momentum penalty is evaluated without filtering, $\mathcal{L}_{\mathrm{mom}}=\|\mathcal{R}_{\mathrm{mom}}\|_1$.

\noindent\textbf{DIT.}
DIT is unforced, so its momentum residual has the same form as $\mathcal{R}_0^{\mathrm{FIT}}$, with $\mathcal{R}_{\mathrm{div}}=\nabla\cdot\bm{u}$. The momentum penalty is evaluated without filtering, $\mathcal{L}_{\mathrm{mom}}=\|\mathcal{R}_{\mathrm{mom}}\|_1$.

The residual penalties above encourage consistency with the governing equations during training. Incompressibility can also be enforced directly by projecting the reconstructed velocity onto the divergence-free subspace. Theory-guided hard constraint projection (HCP)~\cite{chen2021hcp} provides a general formulation for this approach: it maps a prediction onto a constraint set defined by discretized physical equations while minimizing the change to that prediction.

For this dataset, we also evaluate HCP as an optional hard incompressibility constraint. The projection acts on the reconstructed physical velocity $\bm{v}$, after the output mapping $Q$ and inverse normalization, and returns the closest velocity satisfying the discrete continuity equation:
\begin{equation}
    \bm{u}^{\mathrm{H}}
    =\bm{v}-\mathsf{D}_h^{\mathsf T}
      (\mathsf{D}_h\mathsf{D}_h^{\mathsf T})^{\dagger}\mathsf{D}_h\bm{v},
    \label{eq:hcp_methods}
\end{equation}
where $\mathsf{D}_h$ is the discrete divergence operator and $\dagger$ denotes the Moore--Penrose pseudoinverse. The projection removes only the component of $\bm{v}$ orthogonal to the divergence-free subspace and introduces no learnable parameters. When HCP is used, the reconstruction loss is evaluated on the projected velocity $\bm{u}^{\mathrm{H}}$. Supplementary Note~\ref{ssec:hard_incompressibility} provides the derivation and a comparison of models trained with and without HCP.

\subsection{Unsupervised generalization with mixture-of-experts} \label{ssec:moe}
In practical scenarios, the coarse-graining process that generates low-fidelity turbulent flows can vary across simulations, sensors, and numerical settings, and is often unknown at inference time. 
To improve robustness under such coarse-graining shifts, we propose an unsupervised MoE mechanism~\cite{shazeer2017outrageously} that enables adaptation to unseen coarse-graining using only LR inputs.

Given $N$ representative coarse-graining strategies, $N$ paired training sets are constructed and one expert model is trained per strategy, yielding pretrained experts $\{\mathcal{M}_{\bm{\theta}_n}\}_{n=1}^{N}$.
After training, all expert parameters are frozen.
Adaptation to unknown coarse-graining is then achieved by learning a gating network $\mathcal{G}_{\psi}$ that combines these frozen experts, as freezing the experts and adapting only the gating weights provides a stable and data-efficient way to handle unknown coarse-graining when HR supervision is unavailable.
Specifically, $\mathcal{G}_{\psi}$ maps an LR input sequence $\bm{X}_{\mathrm{L}}$ to a set of spatially varying mixture weights
\begin{equation}
    \bm{\alpha}(\bm{X}_{\mathrm{L}}) = \mathrm{softmax}(\mathcal{G}_{\psi}(\bm{X}_{\mathrm{L}})),
    \label{eq:moe_alpha}
\end{equation}
where $\bm{\alpha}(\bm{X}_{\mathrm{L}})\in\mathbb{R}^{H_{\mathrm{H}}\times W_{\mathrm{H}}\times D_{\mathrm{H}}\times N}$
denotes a voxel-wise weight field satisfying $\sum_{n=1}^{N}\alpha_n(\bm{r})=1$
and $\alpha_n(\bm{r})\ge 0$ at each spatial location $\bm{r}$. 
The final HR reconstruction is obtained as a combination of expert predictions at each voxel:
\begin{equation}
    \hat{\bm{X}}_{\mathrm{H}}(\bm{r})
    = \sum\nolimits_{n=1}^{N}
    \alpha_n(\bm{r};\bm{X}_{\mathrm{L}})
    \, \mathcal{M}_{\bm{\theta}_n}(\bm{X}_{\mathrm{L}})(\bm{r}).
    \label{eq:moe_out}
\end{equation}
The gating network $\mathcal{G}_{\psi}$  is optimized at inference time without access to HR supervision by minimizing the physics-residual objective defined in Eq.~\eqref{eq:loss_p}, encouraging the combined reconstruction to satisfy the governing equations while remaining compatible with the observed LR input. In practice, the unsupervised adaptation requires only a small number of unlabeled LR sequences (see Fig.~\ref{fig:generalization}e).

\subsection{Evaluation metrics} \label{ssec:metrics}

We assess reconstruction quality using complementary measures of pointwise accuracy,
structural similarity, temporal consistency, and multiscale turbulence statistics.
Pointwise metrics quantify agreement with the paired HR reference. Energy spectra, PDFs, and velocity-gradient diagnostics characterize the scale-dependent and statistical properties of the reconstructed fields.

We report nRMSE, PSNR, and 3D SSIM, following standard definitions in the super-resolution literature~\cite{bee2017enhanced,chung2023turbulence}. The definitions of PSNR and SSIM are provided in Supplementary Note~\ref{ap:metrics}.
The nRMSE measures the relative error between the prediction
$\hat{\bm{X}}$ and the ground truth $\bm{X}$:
\begin{equation} \label{eq:nrmse}
    \mathrm{nRMSE}(\hat{\bm{X}}, \bm{X}) =
    \frac{\|\hat{\bm{X}} - \bm{X}\|_2}{\|\bm{X}\|_2}.
\end{equation}
To probe temporal consistency, we compute temporal RMSE (tRMSE) on frame-to-frame
increments:
\begin{equation} \label{eq:trmse}
\mathrm{tRMSE} =
\sqrt{
\frac{1}{T-1}
\sum\nolimits_{t=2}^{T}
\|\Delta \hat{\bm{X}}_t - \Delta \bm{X}_t\|_2^2
},
\end{equation}
where $\Delta \bm{X}_t=\bm{X}_t-\bm{X}_{t-1}$.
This metric measures agreement in frame-to-frame changes.

We also compute Fourier-space RMSE (fRMSE) between the three-dimensional Fourier coefficients of $\hat{\bm{X}}$ and $\bm{X}$ (Supplementary Note~\ref{ap:metrics}). To assess the distribution of energy across scales, we compare shell-integrated energy spectra $E(k)$. For FIT, MHDT, and HBDT, we use kinetic energy, total kinetic and magnetic energy (with the magnetic field in Alfv\'en units), and density-weighted kinetic energy, respectively (Supplementary Note~\ref{ap:metrics}).

\subsection{Datasets} \label{ssec:datasets}
\noindent\textbf{TurbSR3D.}
We construct a time-resolved volumetric turbulence super-resolution dataset, termed {TurbSR3D}, comprising over 3,500 samples (310\,GB).  
High-fidelity data are drawn from canonical simulations in the Johns Hopkins Turbulence Databases (JHTDB)~\cite{perlman2007data,li2008public,eyink2013flux,livescu2014homogeneous}.  
LR inputs are obtained by applying the coarse-graining operator $\mathcal{D}$ (Eq.~\eqref{eq:deg}) at factors of $4\times$, $8\times$, or $16\times$.
The dataset covers three representative turbulence regimes: forced isotropic turbulence, forced magnetohydrodynamic turbulence, and homogeneous buoyancy-driven turbulence.

\noindent\textbf{DIT.}
We provide an additional dataset of decaying isotropic turbulence to evaluate physical constraints and generalization to LES inputs under non-stationary conditions. For the LES evaluation, independently evolved LES fields are paired with DNS references at matching physical times. Details of the DNS and LES simulations are provided in Supplementary Notes~\ref{ap:dit_dns} and~\ref{ap:les_dns_pairing}, respectively.

\noindent\textbf{ERA5.} 
An atmospheric downscaling benchmark is constructed from the ERA5 reanalysis~\cite{hersbach2020era5}. 
ERA5 data span 10 consecutive days in January 2022 (January 1--10), sampled every 6 hours.
The benchmark covers all longitudes between $75^\circ$S and $75^\circ$N and includes temperature and horizontal wind components ($u, v$) at eight pressure levels from 1000 to 300~hPa.
LR inputs are generated by bicubic downsampling of the HR fields on the horizontal grid at factors of $4\times$ and $8\times$. The first six days are used for training, the next two for validation, and the final two for testing.

\subsection{{Baselines}}
\label{ssec:baselines}
We evaluate our framework against well-established baselines in two categories: 
i) classical numerical interpolation methods, \ie, tricubic and bicubic interpolation, and 
ii) state-of-the-art deep learning models, \ie, RRDB~\cite{wang2018esrgan}, EDSR~\cite{bee2017enhanced}, RCAN~\cite{zhang2018image}, FNO~\cite{li2021fourier}, PhySR~\cite{ren2023physr}, SwinIR~\cite{liang2021swinir} and DiffSR~\cite{zhu2026diffusion}.
Tricubic interpolation is one of the traditional super-resolution methodologies based on the filter operation~\cite{keys2003cubic}, typically serving as a baseline in turbulence reconstruction studies \cite{fukami2019super,shu2023physics}. 
Bicubic interpolation, a widely used 2D interpolation method, is also commonly applied in atmospheric downscaling tasks~\cite{ren2025superbench}.
EDSR~\cite{bee2017enhanced}, RRDB \cite{wang2018esrgan}, RCAN \cite{zhang2018image} are three representative neural network architectures in the image super-resolution domain.
FNO~\cite{li2021fourier} enables learning PDE solution operators via directly parameterizing the integral kernel in frequency space, rendering it an ideal choice for modeling physical systems.
PhySR~\cite{ren2023physr} is a physics-informed super-resolution framework designed for modeling HR response of spatiotemporal data. 
SwinIR~\cite{liang2021swinir} is a transformer-based model that has demonstrated strong performance in image super-resolution tasks and serves as a robust baseline for comparison. DiffSR~\cite{zhu2026diffusion} employs diffusion-based generative modeling for three-dimensional turbulence super-resolution and serves as a representative generative baseline.
More details are provided in Supplementary Note~\ref{ap:baselines}.

\clearpage

\bibliographystyle{sn-mathphys-num}
\bibliography{sn-bibliography}
\clearpage

\begin{supplementaryinformation}

\section{Implementation and methodology details}\label{ap:implementation}
\subsection{Benchmark experimental settings of TurbSR3D}
We adopt a unified training protocol across all datasets and baselines.
For lower upscaling factors (\ie, \(4\times\)), each learning-based method is trained for 1000 epochs; for higher factors (\ie, \(8\times/16\times\)), we train for 500 epochs due to earlier convergence and overfitting risks. All implementations use PyTorch and are run on an Ubuntu 20.04 server with 8$\times$ RTX 4090 GPUs. Unless otherwise specified, each run uses 2 GPUs with a per-GPU batch size of 4 (effective batch size 8). Mixed-precision (AMP) training is enabled for all deep-learning models, reducing memory footprint and wall-clock time,  except FNO due to code-level constraints. Training-seed uncertainty is quantified on FIT at $4\times$ with box-filtered inputs using independent seeds 42, 43, and 44 for all learned methods. The data partitions and model-specific training and evaluation protocols are fixed across seeds; Supplementary Table~\ref{tab:seed_fit} reports the mean and sample standard deviation over the three runs.
We optimize the model using AdamW with a weight decay of \(10^{-3}\). The learning rate follows an exponential decay schedule, starting from an initial value of \(10^{-4}\) and decaying by a factor of $0.997$ at each epoch.
Training minimizes the composite loss in Eq.~\eqref{eq:loss} with the physics-informed neural optimization described in Section~\ref{ssec:optimization}. Both the data term \(\mathcal{L}_{\text{mae}}\) and the physical terms in \(\mathcal{L}_{\text{phy}}\) use $L_1$ penalties. For TurbSR3D experiments, we set the balancing coefficient in Eq.~\eqref{eq:loss} to $\lambda_{\mathrm{phy}}=10^{-3}$ and the momentum weight in Eq.~\eqref{eq:loss_p} to $\beta=0.3$. For decaying isotropic turbulence (DIT), we use $\lambda_{\mathrm{phy}}=10^{-2}$ and $\beta=0.3$. The sensitivity to these weights is examined in Supplementary Note~\ref{ssec:physical_loss_sensitivity}.

\subsection{Sensitivity to physical-loss weights}
\label{ssec:physical_loss_sensitivity}

We examine the sensitivity to $\lambda_{\mathrm{phy}}$ and $\beta$ in Eqs.~\eqref{eq:loss} and~\eqref{eq:loss_p} on DIT and FIT (Supplementary Tables~\ref{tab:physical_loss_sensitivity_dit} and~\ref{tab:physical_loss_sensitivity_fit}). With $\beta=0.3$, $\lambda_{\mathrm{phy}}=0.01$ for DIT and $0.001$ for FIT give the best nRMSE, SSIM, and PSNR among the tested settings. Increasing $\lambda_{\mathrm{phy}}$ to $0.1$ worsens all three metrics in both tasks. With $\lambda_{\mathrm{phy}}$ fixed at these respective values, $\beta=0.3$ performs best across all three metrics compared with $0.1$ and $1$.

\begin{table}[htbp]
\centering
\caption{\textbf{Sensitivity to physical-loss weights on DIT at $4\times$ with spectral-cutoff coarse-graining.} Best values for each metric are in \textbf{bold}. nRMSE and SSIM are reported to four decimal places and PSNR to two decimal places.}
\label{tab:physical_loss_sensitivity_dit}
{\small
\setlength{\tabcolsep}{12pt}
\renewcommand{\arraystretch}{1.12}
\begin{tabular}{rrrrr}
\toprule
$\lambda_{\mathrm{phy}}$ & $\beta$ & nRMSE $\downarrow$ & SSIM $\uparrow$ & PSNR (dB) $\uparrow$ \\
\midrule
0     & 0.3 & 0.0817 & 0.9402 & 43.32 \\
0.001 & 0.3 & 0.0807 & 0.9415 & 43.44 \\
0.01  & 0.3 & \textbf{0.0773} & \textbf{0.9463} & \textbf{43.82} \\
0.1   & 0.3 & 0.1130 & 0.9022 & 40.45 \\
0.01  & 0.1 & 0.0789 & 0.9438 & 43.64 \\
0.01  & 1   & 0.0841 & 0.9372 & 43.17 \\
\bottomrule
\end{tabular}
}
\end{table}

\begin{table}[htbp]
\centering
\caption{\textbf{Sensitivity to physical-loss weights on FIT at $8\times$ with Gaussian coarse-graining.} Best values for each metric are in \textbf{bold}. nRMSE and SSIM are reported to four decimal places and PSNR to two decimal places.}
\label{tab:physical_loss_sensitivity_fit}
{\small
\setlength{\tabcolsep}{12pt}
\renewcommand{\arraystretch}{1.12}
\begin{tabular}{rrrrr}
\toprule
$\lambda_{\mathrm{phy}}$ & $\beta$ & nRMSE $\downarrow$ & SSIM $\uparrow$ & PSNR (dB) $\uparrow$ \\
\midrule
0      & 0.3 & 0.1233 & 0.7957 & 42.20 \\
0.0001 & 0.3 & 0.1232 & 0.7960 & 42.20 \\
0.001  & 0.3 & \textbf{0.1173} & \textbf{0.8093} & \textbf{42.61} \\
0.01   & 0.3 & 0.1216 & 0.7985 & 42.32 \\
0.1    & 0.3 & 0.1810 & 0.6722 & 38.66 \\
0.001  & 0.1 & 0.1174 & 0.8090 & 42.59 \\
0.001  & 1   & 0.1195 & 0.8044 & 42.46 \\
\bottomrule
\end{tabular}
}
\end{table}
\FloatBarrier

\subsection{Benchmark experimental settings of ERA5}
For the ERA5 downscaling task, the training protocol is adjusted to accommodate the specific scale and characteristics of the data. Each model is trained for \(T_{\max}=100\) epochs on a single GPU with a batch size of 4. Consistent with the TurbSR3D experiments, AMP is enabled to improve computational efficiency. Training-seed uncertainty is quantified for Ours, FNO, and SwinIR at both $4\times$ and $8\times$ using independent seeds 42, 43, and 44. The chronological data split and all other model-specific protocols are fixed across seeds; Supplementary Table~\ref{tab:seed_era5} reports the resulting mean and sample standard deviation.
We optimize the models using AdamW with a weight decay of \(10^{-3}\). 
Regarding the training objective, considering the coarse spatial resolution of the ERA5 reanalysis data (\(\sim\)30 km), we do not incorporate explicit governing equations into the loss. Consequently, the physics-informed term is set to \(\lambda_{\text{phy}}=0\), and the optimization minimizes solely the data fidelity term \(\mathcal{L}_{\text{mae}}\), formulated as the L1 loss between the predicted and ground-truth fields.   

\subsection{Details of state space model}\label{ap:ssm}
\noindent\textbf{Preliminaries.}
In the channelwise implementation, each internal feature channel maintains an $M$-dimensional recurrent state at each Eulerian grid location. The channel index is omitted in the main-text equations for clarity. Here, $M$ corresponds to the state dimension $d_s$ used in Supplementary Note~\ref{ssec:ssm_sensitivity}.

We consider a continuous-time state space model
\begin{align}
  \dot{\bm{h}}(t) &= \bm{A}\bm{h}(t) + \bm{B}\bm{x}(t), \\
  \bm{y}(t) &= \bm{C}\bm{h}(t) + \bm{D}\bm{x}(t),
\end{align}
where $\bm{A}, \bm{B}, \bm{C}, \bm{D}$ are learnable parameters, and $\dot{\bm{h}}(t)$ denotes the time derivative of the latent state.
Following standard practice~\cite{gu2021combining}, the continuous-time system is discretized using a zero-order hold with step size $\Delta$, yielding
\begin{align}
  \bar{\bm{A}} &= \exp(\Delta \bm{A}), \\
  \bar{\bm{B}} &= (\Delta \bm{A})^{-1}\bigl(\exp(\Delta \bm{A}) - I\bigr)\,(\Delta \bm{B}).
\end{align}
The resulting discrete-time formulation is
\begin{align}
  \bm{h}_t &= \bar{\bm{A}}\bm{h}_{t-1} + \bar{\bm{B}}\bm{x}_t, \\
  \bm{y}_t &= \bm{C}\bm{h}_t + \bm{D}\bm{x}_t.
\end{align}
For notational simplicity, we omit the bar notation in the main text.

\noindent\textbf{Gating mechanism.}
To enhance representational capability, we introduce a spatially varying selection gate
$\bm{g}_t(\bm{r}) = \phi\left(\bm{v}_t(\bm{r})\right)$,
where $\bm{v}_t$ is another locally computed transformation of $\bm{z}_t$ by a 3D convolution and $\phi$ denotes a smooth gating nonlinearity (implemented as the SiLU activation). The output is thus given by:
\begin{align}
{\bm{s}}_t(\bm{r})
= \bm{g}_t(\bm{r}) \odot \bm{y}_t(\bm{r}),  
\end{align}
where the gating mechanism modulates local temporal updates according to the instantaneous flow state.
A subsequent point-wise linear projection is applied to $\bm{s}_t$ to align channel dimensions.

\subsection{Methods and training configurations of TurbSR3D}\label{ap:baselines}\label{ssec:turb_decoder_config}
The proposed framework is benchmarked against classical interpolation and representative learning-based super-resolution models.
Unless otherwise noted, all learnable baselines are re-trained under a unified protocol to ensure comparability.
For {RRDB}, {RCAN}, {EDSR}, the implementations and hyperparameter guidelines are referenced from the public BlastNet~2.0 repository~\cite{chung2023turbulence} while being adapted to our setting.

\noindent\textbf{Tricubic.}
Implemented using \texttt{scipy.ndimage}, the tricubic operator is applied to each LR frame independently to reconstruct HR fields.

\noindent\textbf{RRDB}~\cite{wang2018esrgan}.
The architecture is configured with two residual dense blocks (channel width $56$) followed by a trilinear upsampling head.

\noindent\textbf{RCAN}~\cite{zhang2018image}.
The model features a channel width of $64$, a reduction ratio of $16$, and $10$ residual channel--attention blocks. Spatial upsampling is performed via a staged convolutional head with pixel-shuffle.

\noindent\textbf{EDSR}~\cite{bee2017enhanced}.
It uses a channel width of $44$, residual scaling of $0.1$, and $32$ residual blocks, with a pixel-shuffle upsampling head.

\noindent\textbf{FNO}~\cite{li2021fourier}.
This baseline is instantiated as a low-resolution spatiotemporal feature trunk paired with a pixel-shuffle upsampling head. It consists of two hidden layers with a channel width of $128$, retaining $3$ spatial spectral modes (reduced to $2$ for HBDT at $16\times$) and $2$ temporal modes.

\noindent\textbf{PhySR}~\cite{ren2023physr}.
This model comprises one ConvLSTM~\cite{shi2015convolutional} layer, eight CNN residual blocks (channel width $40$), and a pixel-shuffle upsampling head.

\noindent\textbf{SwinIR}~\cite{liang2021swinir}.
This model is extended to 3D for volumetric turbulence super-resolution by adopting 3D convolutions, window-based self-attention with alternating regular and shifted windows, and 3D relative position biases. The architecture consists of six Residual Swin Transformer Blocks (RSTBs), each containing six Swin Transformer Layers (STLs), with a channel width of 72 and four attention heads. The attention window size is set to $4\times4\times4$ for $4\times$ super-resolution and $2\times2\times2$ for $8\times$ and $16\times$ super-resolution.

\noindent\textbf{DiffSR}~\cite{zhu2026diffusion}.
DiffSR is adapted from the original work, with the auxiliary ESRGAN prediction omitted. It uses a 3D U-Net and follows the physical-loss settings of the original study. Training uses a diffusion schedule of 400 time steps. Inference uses DDIM~\cite{song2021denoising} with a time-step stride of 10, yielding 40 denoising refinement steps.

\noindent\textbf{Ours.}
The proposed model comprises one layer of the state space model and six multiscale neural-operator blocks.
The state space model uses a feature width of $64$ and a state dimension of $16$, while the neural-operator layers use a feature width of $256$.
In the spectral neural operator, the soft-shrinkage threshold in Eq.~(\ref{eq:soft-shrinkage}) is set to $\lambda=0.01$.
Block-diagonalisation (four blocks) is further employed to reduce the parameter count and improve computational efficiency.

\noindent\textbf{Decoder and output projection.}
The decoder operates frame by frame with parameters shared across time. A \(3\times3\times3\) convolution first transforms the multiscale neural-operator features, followed by successive upsampling stages.
Each stage uses a \(3\times3\times3\) convolution to expand the feature channels by a factor of eight, followed by 3D pixel shuffle, which rearranges these channels to double the resolution along all three Cartesian coordinates.
An upscaling factor \(s=2^m\) therefore requires \(m\) stages: two, three, and four for \(4\times\), \(8\times\), and \(16\times\) reconstruction, respectively.
The learned projection \(Q\) is a final \(3\times3\times3\) convolution that maps the decoded features to the physical variables.

It is worth noting that maintaining identical hyperparameter settings across experimental setups leads to an inherent increase in model parameters for higher upscaling factors (Supplementary Tables~\ref{tab:fit_detailed}--\ref{tab:hbdt_detailed}), although the magnitude of this growth depends on the architectural design. As illustrated in Supplementary Fig.~\ref{fig:params}, the proposed framework achieves superior performance while maintaining the fewest average parameters across all test cases.

\begin{figure}[htbp]
    \centering
    \includegraphics[width=0.72\textwidth]{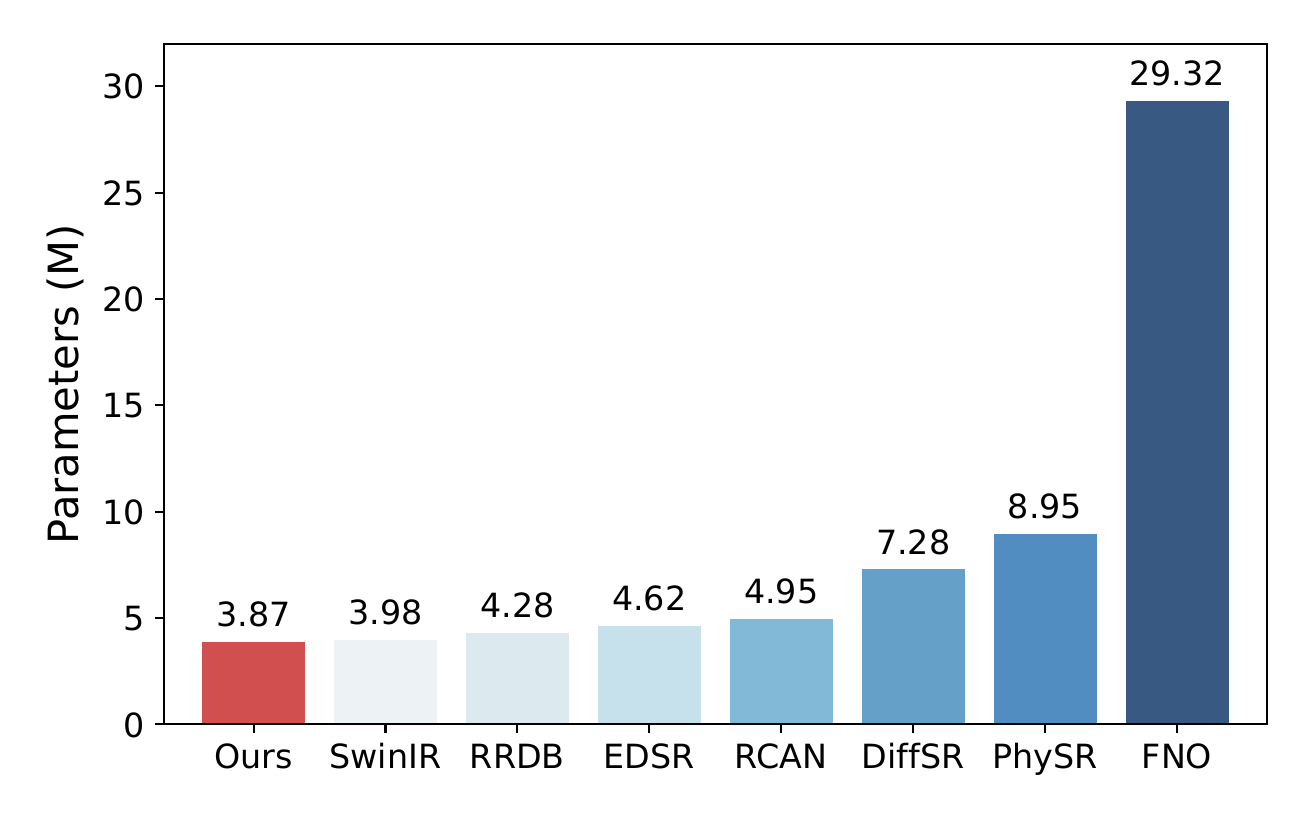}
    \caption{Average parameter counts (M) for eight super-resolution reconstruction methods: Ours, SwinIR, RRDB, EDSR, RCAN, DiffSR, PhySR, and FNO. Each bar shows the equally weighted mean across six configurations: FIT at $4\times$ and $8\times$, MHDT at $4\times$ and $8\times$, and HBDT at $8\times$ and $16\times$, using the parameter counts reported in Supplementary Tables~\ref{tab:fit_detailed}--\ref{tab:hbdt_detailed}. Values above the bars are rounded to two decimal places. The proposed method has the lowest average parameter count (3.87~M).}
    \label{fig:params}
\end{figure}

\subsection{Scale-selective momentum regularization}
\label{ssec:scale_selective_momentum}
The main FIT and MHDT benchmarks use the high-pass momentum penalty in Eq.~\eqref{eq:scale_selective_momentum}. The high-pass operator is implemented as
\begin{equation}
    \mathcal{H}_{\sigma}=I-G_{\sigma},
    \label{eq:high_pass_implementation}
\end{equation}
where $I$ is the identity operator and $G_{\sigma}$ is a local three-dimensional Gaussian low-pass filter. Thus, the filtered residual is obtained by subtracting its Gaussian-smoothed component from the original residual. The Gaussian low-pass component is implemented with a discrete separable kernel truncated at $3\sigma$ and valid convolution along all three spatial directions. At $4\times$ and $8\times$ reconstruction, $\sigma/\Delta x_{\mathrm{HR}}$ is $1.4991$ and $2.9983$, respectively, with kernel radii of $5$ and $9$ HR grid points.
The FIT forcing is supported on $0<|\bm{k}|^2\leq6$~\cite{li2008public}, and the MHDT Taylor--Green forcing is supported at $|\bm{k}|=\sqrt{12}$~\cite{eyink2013flux}. Applying the training filter directly to the forcing fields gives the RMS values in Supplementary Table~\ref{tab:forcing_rms}. For each dataset, the pre-filter RMS from the $4\times$ evaluation serves as the common reference at both reconstruction factors. The filtered forcing RMS is at most $0.1989\%$ of this reference, demonstrating strong suppression of its direct contribution to the scale-selective constraint.

\begin{table}[!htbp]
\centering
\caption{\textbf{Forcing RMS before and after high-pass filtering.} The same Gaussian filter is used in training and in this diagnostic. Retained percentages are computed from unrounded values relative to the dataset-specific $4\times$ pre-filter reference.}
\label{tab:forcing_rms}
\small
\begin{tabular}{lcrrr}
\toprule
Dataset & Scale & \makecell[r]{Reference forcing\\$\mathrm{RMS}_{\mathrm{ref}}(\bm{f})$} & \makecell[r]{After filtering\\$\mathrm{RMS}(\mathcal{H}_{\sigma}\bm{f})$} & RMS retained \\
\midrule
FIT  & $4\times$ & $0.0847$ & $1.11\times10^{-5}$ & $0.0131\%$ \\
FIT  & $8\times$ & $0.0847$ & $4.38\times10^{-5}$ & $0.0517\%$ \\
MHDT & $4\times$ & $0.0545$ & $2.76\times10^{-5}$ & $0.0506\%$ \\
MHDT & $8\times$ & $0.0545$ & $1.08\times10^{-4}$ & $0.1989\%$ \\
\bottomrule
\end{tabular}
\end{table}

The implemented $L_1$ momentum term provides robust physical residual regularization. Motivated by the intermittency of turbulent velocity gradients and small-scale velocity increments~\cite{meneveau2011lagrangian,pope2000turbulent}, we use a linear-growth penalty that limits the influence of large local residuals compared with a squared $L_2$ penalty. The momentum residual is evaluated using numerical derivatives of reconstructed fields, so large local residuals may also contain contributions from numerical errors.

Gaussian residual statistics naturally motivate a quadratic penalty. For additive Gaussian forcing that is white in time, the Onsager--Machlup action contains a quadratic residual term weighted by the forcing covariance. The present benchmarks use state-dependent low-mode energy maintenance in FIT and deterministic Taylor--Green forcing in MHDT~\cite{li2008public,eyink2013flux}; the implemented $L_1$ term serves as a soft dynamical constraint in these systems.

We compare three momentum-loss formulations on FIT at $4\times$ super-resolution with box-filtered inputs: unfiltered $L_1$, high-pass $L_1$, and high-pass squared $L_2$ (mean squared residual). As shown in Supplementary Fig.~\ref{fig:fit_momentum_norm}, all three closely follow the reference energy spectrum at low and intermediate wavenumbers. The high-pass $L_1$ variant shows closer agreement in the enlarged high-wavenumber region and yields the lowest relative $L_1$ error in the high-wavenumber energy spectrum: 0.1736, compared with 0.2125 for unfiltered $L_1$ and 0.1920 for high-pass squared $L_2$. These comparisons support both high-pass filtering of the momentum residual and the use of $L_1$ for high-wavenumber spectral reconstruction.

\begin{figure}[!htbp]
  \centering
  \includegraphics[width=0.85\linewidth]{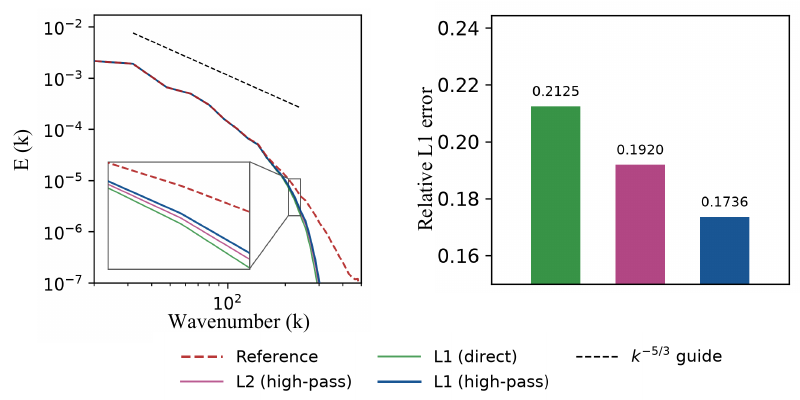}
  \caption{\textbf{Sensitivity to the momentum-loss formulation on FIT at $4\times$ super-resolution with box-filtered inputs.} \textbf{Left}, Kinetic-energy spectra of the reference and reconstructions using an unfiltered $L_1$ penalty (``L1 (direct)''), a high-pass squared $L_2$ penalty (``L2 (high-pass)''), and the implemented high-pass $L_1$ penalty (``L1 (high-pass)''). The inset highlights differences at high wavenumbers; the black dashed line is a $k^{-5/3}$ guide. \textbf{Right}, Relative $L_1$ errors of the reconstructed energy spectra over $k>128$. Lower values indicate closer agreement with the reference in this high-wavenumber range.}
  \label{fig:fit_momentum_norm}
\end{figure}
\FloatBarrier
\color{black}

\subsection{Hard incompressibility constraint on the physical output}
\label{ssec:hard_incompressibility}

\noindent\textbf{HCP formulation.}
Theory-guided hard constraint projection (HCP)~\cite{chen2021hcp} embeds physical knowledge by discretizing a governing equation, separating its coefficients from the predicted physical quantities, and projecting those predictions onto the resulting constraint set. For a prediction $\bm{y}$ and a linear constraint $\mathsf{A}\bm{y}=\bm{b}$ with full-row-rank $\mathsf{A}$, the closest admissible prediction is
\begin{equation}
 \bm{y}^{\mathrm{H}}
 =\bm{y}-\mathsf{A}^{\mathsf T}(\mathsf{A}\mathsf{A}^{\mathsf T})^{-1}
    (\mathsf{A}\bm{y}-\bm{b}).
 \label{eq:hcp_linear_constraint}
\end{equation}
The original HCP construction couples hydraulic-head predictions through a discretized subsurface-flow equation. Here, the constraint couples the three components of the reconstructed turbulent velocity through instantaneous continuity. The network provides the reconstruction, and continuity determines the correction.

\noindent\textbf{Projection onto the divergence-free subspace.}
At each reconstructed time, let $\bm{v}$ denote the HR velocity after the learned output mapping $Q$ and inverse normalization using the training statistics in Eq.~\eqref{eq:norm}. The latent channels encode learned features, so incompressibility is imposed on this final physical velocity. For a uniform grid with $N$ voxels, write $\bm{v}=(\bm{v}_x^{\mathsf T},\bm{v}_y^{\mathsf T},\bm{v}_z^{\mathsf T})^{\mathsf T}\in\mathbb{R}^{3N}$. The discrete divergence operator $\mathsf{D}_h=[\mathsf{D}_x\;\mathsf{D}_y\;\mathsf{D}_z]$ combines the spatial derivatives as $\mathsf{D}_h\bm{v}=\mathsf{D}_x\bm{v}_x+\mathsf{D}_y\bm{v}_y+\mathsf{D}_z\bm{v}_z$, so continuity constrains the three components jointly. Each $\mathsf{D}_\alpha$ is a fourth-order central difference evaluated at interior grid points. Imposing $\mathsf{D}_h\bm{w}=0$ gives
\begin{equation}
 \begin{aligned}
  \bm{u}^{\mathrm{H}}
  &=\underset{\bm{w}:\,\mathsf{D}_h\bm{w}=0}{\arg\min}
    \frac{1}{2}\|\bm{w}-\bm{v}\|_2^2
  =\mathsf{P}_h\bm{v}, \\
  \mathsf{P}_h&=\mathsf{I}-\mathsf{D}_h^{\mathsf T}
       (\mathsf{D}_h\mathsf{D}_h^{\mathsf T})^{\dagger}\mathsf{D}_h.
 \end{aligned}
  \label{eq:hcp_projection}
\end{equation}
Here $\dagger$ denotes the Moore--Penrose pseudoinverse. For the interior operator used here, $\mathsf{D}_h$ has full row rank and $\dagger$ reduces to the ordinary inverse; the pseudoinverse also covers operators with linearly dependent rows, such as periodic differences. In addition, $\mathsf{D}_h^{\mathsf T}$ is the discrete adjoint under the uniform-grid Euclidean inner product. The stationarity condition for the constrained minimum gives $\bm{u}^{\mathrm{H}}-\bm{v}=-\mathsf{D}_h^{\mathsf T}\bm{\lambda}$, with $\mathsf{D}_h\mathsf{D}_h^{\mathsf T}\bm{\lambda}=\mathsf{D}_h\bm{v}$; eliminating the multiplier yields Eq.~\eqref{eq:hcp_projection}.

The operator $\mathsf{P}_h$ is an orthogonal projection onto the discrete divergence-free subspace $\ker\mathsf{D}_h$, whose dimension is $3N-\operatorname{rank}(\mathsf{D}_h)$. It leaves admissible velocities unchanged and makes the smallest correction in the discrete $L_2$ norm needed to satisfy the continuity constraint. The effects on reconstruction accuracy and energy spectra are evaluated below.

\noindent\textbf{Use in the reconstruction model.}
HCP is an optional extension to the reconstruction model, evaluated only in the dedicated DIT comparison below. The projection is determined by the physical constraint and introduces no learnable parameters or penalty weight. It enforces discrete continuity to numerical precision, while the soft momentum term continues to regularize the dynamics during training. In the DIT comparison, the model trained with HCP evaluates the reconstruction loss on the projected velocity.

\noindent\textbf{Evaluation on DIT.}
We compare models trained with and without HCP on DIT at $4\times$ super-resolution with spectral-cutoff inputs. The divergence RMS measures compliance with incompressibility:
\begin{equation}
  \delta_{\mathrm{div}}(\bm{w})
  =\left[\frac{1}{N_{\mathrm{c}}}\|\mathsf{D}_h\bm{w}\|_2^2\right]^{1/2},
  \label{eq:hcp_div_metric}
\end{equation}
where $N_{\mathrm{c}}$ is the number of evaluated divergence values. Reconstruction accuracy is measured by nRMSE, defined in Eq.~\eqref{eq:nrmse}, and the energy spectra assess the recovered energy distribution across scales.

As shown in Supplementary Table~\ref{tab:dit_hcp}, incorporating HCP reduces the divergence RMS from $1.41\times10^{-2}$ to $1.52\times10^{-6}$, with a small increase in nRMSE from 0.0773 to 0.0779. HCP therefore substantially improves compliance with incompressibility at a small cost in reconstruction accuracy. The energy spectra in Supplementary Fig.~\ref{fig:dit_hcp} show that both models closely follow the reference at low and intermediate wavenumbers. The model trained with HCP also shows closer agreement with the reference over part of the high-wavenumber range, while retaining the main spectral characteristics of the reconstructed flow.

\begin{figure}[!htbp]
 \centering
 \includegraphics[width=0.65\linewidth]{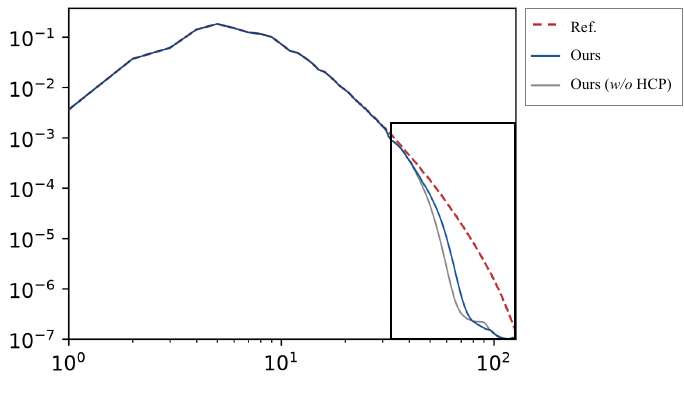}
 \caption{\textbf{Effect of HCP on the reconstructed energy spectrum.} Reference spectrum and predictions from models trained with and without HCP on DIT at $4\times$ super-resolution with spectral-cutoff inputs.}
 \label{fig:dit_hcp}
\end{figure}
\begin{table}[!htbp]
 \centering
 \caption{\textbf{Effect of HCP on divergence RMS and nRMSE.} Models trained with and without HCP on DIT at $4\times$ super-resolution with spectral-cutoff inputs.}
 \label{tab:dit_hcp}
 \small\setlength{\tabcolsep}{12pt}
 \begin{tabular}{lcc}
 \toprule
 Model & Divergence RMS & nRMSE \\
 \midrule
 HCP & $1.52\times10^{-6}$ & 0.0779 \\
 w/o HCP & $1.41\times10^{-2}$ & 0.0773 \\
 \bottomrule
 \end{tabular}
\end{table}
\FloatBarrier
\color{black}

\subsection{Methods and training configurations of ERA5}\label{ssec:era5_decoder_config}
For the ERA5 dataset, the model configurations are adjusted to handle the specific data characteristics.

\noindent\textbf{Bicubic.}
Implemented using \texttt{scipy.ndimage}, the bicubic operator is applied to each 2D slice independently to reconstruct HR fields.

\noindent\textbf{SwinIR}~\cite{liang2021swinir}.
Designed as a 2D baseline, this model performs reconstruction in a slice-wise manner.
The numbers of RSTBs and STLs, window size, channel width, and number of attention heads are set to $2$, $2$, $4$, $64$, and $4$, respectively. The reconstruction is handled by a pixel-shuffle head.

\noindent\textbf{FNO}~\cite{li2021fourier}.
The model is configured with two hidden layers and a channel width of $64$. In the spectral domain, $4$ horizontal spatial modes, $2$ vertical spatial modes, and $2$ temporal modes are retained.

\noindent\textbf{Ours.}
For this dataset, the proposed framework is instantiated with four multiscale neural-operator blocks. The channel width of the neural-operator layers is set to $224$.

\noindent\textbf{Decoder and output projection.}
The ERA5 decoder operates frame by frame with parameters shared across time. A \(3\times3\times3\) convolution first transforms the multiscale neural-operator features.
Each upsampling stage then applies a shared \(3\times3\) convolution at every pressure level to expand the feature channels by a factor of four, followed by 2D pixel shuffle, which doubles the resolution along both horizontal coordinates.
The \(4\times\) and \(8\times\) configurations use two and three stages, respectively, while retaining all eight pressure levels.
The learned projection \(Q\) is a final \(3\times3\times3\) convolution that maps the decoded multi-level features to the physical variables.

\subsection{Supplementary definitions of evaluation metrics} \label{ap:metrics}
Let $\hat{\bm{X}}$ denote the prediction field and $\bm{X}$ the ground truth.

\noindent\textbf{Peak signal-to-noise ratio (PSNR).}
PSNR evaluates reconstruction quality on a logarithmic scale based on mean squared
error (MSE):
\begin{equation}
    \mathrm{PSNR}(\hat{\bm{X}}, \bm{X}) =
    10 \cdot \log_{10} \Bigl( \frac{L^2}{\mathrm{MSE}(\hat{\bm{X}}, \bm{X})} \Bigr),
\end{equation}
where $L$ denotes the data range of $\bm{X}$.

\noindent\textbf{Structural similarity (SSIM).}
We adopt the 3D extension of SSIM~\cite{chung2023turbulence} to assess local structural
fidelity:
\begin{equation}
\mathrm{SSIM}(\hat{\bm{X}},\bm{X})
=\frac{(2\mu_{\hat{\bm{X}}}\mu_{\bm{X}}+c_1)(2\sigma_{\hat{\bm{X}}\bm{X}}+c_2)}
{(\mu_{\hat{\bm{X}}}^2+\mu_{\bm{X}}^2+c_1)(\sigma_{\hat{\bm{X}}}^{2}+\sigma_{\bm{X}}^{2}+c_2)},
\end{equation}
where $\mu$ and $\sigma^2$ denote local means and variances, $\sigma_{\hat{\bm{X}}\bm{X}}$
is the local covariance, and $c_1$, $c_2$ are constants for numerical stability.

\noindent\textbf{Fourier-space RMSE (fRMSE).}
To quantify scale-dependent errors, we compute Fourier-space RMSE between the 3D
spectra of $\hat{\bm{X}}$ and $\bm{X}$:
\begin{equation}
\mathrm{fRMSE} =
\sqrt{
\frac{1}{|\mathcal{K}|}
\sum\nolimits_{\bm{\kappa}\in\mathcal{K}}
\left\|
\mathcal{F}_{3d}(\hat{\bm{X}})(\bm{\kappa})
-
\mathcal{F}_{3d}(\bm{X})(\bm{\kappa})
\right\|_2^2
},
\end{equation}
where $\mathcal{F}_{3d}$ denotes the 3D discrete Fourier transform over spatial
dimensions and $\mathcal{K}$ is the set of frequency indices.

\noindent\textbf{Energy spectra.}
We compute energy spectra by summing Fourier-mode energies within spherical wavenumber shells and dividing by the shell width $\Delta k$. The shell $\mathcal{S}_k$ contains wavevectors $\bm{\kappa}$ satisfying $k-\Delta k/2\leq|\bm{\kappa}|<k+\Delta k/2$. Fourier coefficients are normalized by the total number of spatial grid points.

For FIT, we use the kinetic energy spectrum
\begin{equation}
    E(k)=\frac{1}{2\Delta k}\sum_{\bm{\kappa}\in\mathcal{S}_k}
    |\widehat{\bm{u}}(\bm{\kappa})|^2,
\end{equation}
where $\bm{u}$ is the velocity field and $\widehat{(\cdot)}$ denotes the normalized Fourier transform.

For MHDT, we use the total kinetic plus magnetic spectrum
\begin{equation}
    E(k)=\frac{1}{2\Delta k}\sum_{\bm{\kappa}\in\mathcal{S}_k}
    \left(|\widehat{\bm{u}}(\bm{\kappa})|^2+|\widehat{\bm{b}}(\bm{\kappa})|^2\right),
\end{equation}
where $\bm{b}$ is the magnetic field expressed in Alfv\'en units.

For HBDT, we adopt the density-weighted kinetic spectrum
\begin{equation}
    E(k)=\frac{1}{2\Delta k}\sum_{\bm{\kappa}\in\mathcal{S}_k}
    \left|\widehat{\sqrt{\rho}\,\bm{u}}(\bm{\kappa})\right|^2,
\end{equation}
where $\rho$ denotes the mass density. With this normalization, summing over all Fourier shells gives $\sum_k E(k)\Delta k=\tfrac{1}{2}\langle\rho|\bm{u}|^2\rangle_{\bm{x}}$, where $\langle\cdot\rangle_{\bm{x}}$ denotes spatial averaging.

\clearpage
\section{Datasets and preprocessing}\label{ap:data}
\subsection{JHTDB data sources and target variables}
We construct a volumetric turbulence super-resolution corpus from the JHTDB. 
Three canonical systems are used: forced isotropic turbulence (FIT), forced magnetohydrodynamic turbulence (MHDT), and homogeneous buoyancy-driven turbulence (HBDT). 
For completeness, we summarize the governing equations associated with each turbulence
regime in JHTDB.

\noindent\textbf{FIT.}
It is a canonical incompressible turbulent flow driven by large-scale forcing and widely used for studying spatiotemporal turbulence dynamics.
The governing equations are the incompressible Navier--Stokes equations:
\begin{equation}
\begin{split}
    \frac{\partial \bm{u}}{\partial t} + (\bm{u} \cdot \nabla) \bm{u}
    &= -\frac{1}{\rho_0}\nabla p + \nu \nabla^2 \bm{u} + \bm{f}, \\
    \nabla \cdot \bm{u} &= 0,
\end{split}
\end{equation}
where $\bm{u}$ and $p$ denote velocity and pressure, and $\bm{f}$ is an external
forcing term used to maintain statistical stationarity. Its Fourier support is restricted to the lowest modes, $0<|\bm{k}|^2\leq6$~\cite{li2008public}; its treatment in the scale-selective training objective is given in Section~\ref{ssec:optimization}.

\noindent\textbf{MHDT.}
It describes incompressible conducting fluids with coupled velocity and magnetic dynamics.
The equations couple the velocity field $\bm{u}$ and magnetic field $\bm{b}$:
\begin{equation}
\begin{split}
    \frac{\partial \bm{u}}{\partial t} + (\bm{u} \cdot \nabla) \bm{u}
    &= -\nabla p + (\nabla\times\bm{b})\times\bm{b}
    + \nu \nabla^2 \bm{u} + \bm{f}, \\
    \frac{\partial \bm{b}}{\partial t} + (\bm{u} \cdot \nabla) \bm{b}
    &= (\bm{b} \cdot \nabla) \bm{u} + \eta \nabla^2 \bm{b}, \\
    \nabla \cdot \bm{u} &= 0, \quad \nabla \cdot \bm{b} = 0,
\end{split}
\end{equation}
Here, $p=p_{\mathrm{th}}/\rho_0$ denotes the kinematic pressure, with $p_{\mathrm{th}}$ the thermodynamic pressure and $\rho_0$ the constant mass density. The magnetic field $\bm{b}$ is expressed in Alfv\'en velocity units, and $\eta$ denotes the magnetic diffusivity. The external stirring is the Taylor--Green forcing with amplitude $1/4$ at wavenumber $2$~\cite{eyink2013flux}.

\noindent\textbf{HBDT.}
HBDT describes buoyancy-driven mixing between two miscible incompressible fluids of different densities. The mixture obeys the variable-density, low-Mach-number Navier--Stokes equations, with nonzero velocity divergence arising from mixing:
\begin{equation}
\begin{split}
    \frac{\partial \rho}{\partial t} + \nabla \cdot (\rho \bm{u}) &= 0, \\
    \frac{\partial (\rho \bm{u})}{\partial t}
    + \nabla \cdot (\rho \bm{u} \otimes \bm{u})
    &= -\nabla p + \nabla \cdot \bm{\tau} + \rho \bm{g}, \\
    \bm{\tau}
    &= \frac{\rho}{Re_0}
    \left[\nabla \bm{u} + (\nabla \bm{u})^T
    - \frac{2}{3}(\nabla \cdot \bm{u})\bm{I}\right], \\
    \nabla \cdot \bm{u} + \frac{1}{Re_0 Sc}\nabla^2(\ln \rho) &= 0,
\end{split}
\end{equation}
where $\rho$ denotes density, $\bm{\tau}$ the viscous stress tensor,
$\bm{g}$ the gravitational acceleration,
$\bm{I}$ the identity tensor,
$Re_0$ the reference Reynolds number,
and $Sc$ the Schmidt number.
For each turbulence regime, the set of target (super-resolved) variables is listed in Supplementary Table~\ref{tab:variables_per_system}.

\begin{table}[htbp]
\centering
\caption{Reconstructed variables per turbulent system.}
\label{tab:variables_per_system}
\begin{tabular}{lccc}
\toprule
System & Velocity $\bm{u}$ & Pressure $p$ & Extra variable \\
\midrule
FIT   & \checkmark & \checkmark & -- \\
MHDT & \checkmark & \checkmark & Magnetic field $\bm{b}$ \\
HBDT  & \checkmark & \checkmark & Density $\rho$ \\
\bottomrule
\end{tabular}
\end{table}

\subsection{System-specific configurations and parameters}
Unless otherwise specified, flows are periodic in all directions; HR spatial spacing is $\Delta x=\Delta y=\Delta z=\pi/512$. 
Supplementary Tables~\ref{tab:fit_params}--\ref{tab:hbdt_params} summarise key variables and parameters used in our experiments.

\begin{table}[htbp]
\centering
\caption{FIT dataset: reconstructed variables, governing equations, and numerical parameters.}
\label{tab:fit_params}
\begin{tabular}{ll}
\toprule
Category & Specification \\
\midrule
\textbf{Reconstructed variables} 
    & Velocity $\bm{u}$ (3 components), pressure $p$ \\[4pt]

\textbf{Governing model} 
    & Incompressible Navier--Stokes with large-scale forcing \\[4pt]

\textbf{Transport coefficient} 
    & Kinematic viscosity $\nu = 1.85\times10^{-4}$ \\[4pt]

\textbf{Spatial grid spacing} 
    & $\Delta x = \Delta y = \Delta z = \pi / 512$ \\[4pt]

\textbf{Temporal resolution} 
    & $\Delta t = 2\times10^{-3}$ \\[4pt]

\bottomrule
\end{tabular}
\end{table}

\begin{table}[htbp]
\centering
\caption{MHDT dataset: reconstructed variables, governing equations, and numerical parameters.}
\label{tab:mhd_params}
\begin{tabular}{ll}
\toprule
Category & Specification \\
\midrule
\textbf{Reconstructed variables} 
    & Velocity $\bm{u}$, magnetic field $\bm{b}$, pressure $p$ \\[4pt]

\textbf{Governing model} 
    & Incompressible MHD equations \\[4pt]

\textbf{Transport coefficients} 
    & Kinematic viscosity $\nu = 1.1\times10^{-4}$ \\
    & Magnetic diffusivity $\eta = 1.1\times10^{-4}$ \\[4pt]

\textbf{Spatial grid spacing} 
    & $\Delta x = \Delta y = \Delta z = \pi / 512$ \\[4pt]

\textbf{Temporal resolution} 
    & $\Delta t = 2.5\times10^{-3}$ \\[4pt]

\bottomrule
\end{tabular}
\end{table}

\begin{table}[htbp]
\centering
\caption{HBDT dataset: reconstructed variables, governing equations, and numerical parameters.}
\label{tab:hbdt_params}
\begin{tabular}{ll}
\toprule
Category & Specification \\
\midrule
\textbf{Reconstructed variables} & Density $\rho$, velocity $\bm{u}$, pressure $p$ \\[2pt]

\textbf{Governing model} & Variable-density, low-Mach-number Navier--Stokes \\
                         & with buoyancy production term \\[2pt]

\textbf{Control parameters} 
    & Reynolds number $Re_0 = 12500$ \\
    & Froude number $Fr = 1.0$ \\
    & Schmidt number $Sc = 1.0$ \\[2pt]

\textbf{Gravity vector} 
    & $\bm{g} = (1.0,\, 0.0,\, 0.0)$\\[4pt]

\textbf{Spatial grid spacing} 
    & $\Delta x = \Delta y = \Delta z = \pi / 512$ \\[2pt]

\textbf{Temporal resolution} 
    & $\Delta t = 4\times10^{-2}$ \\[2pt]

\bottomrule
\end{tabular}
\end{table}

\subsection{TurbSR3D corpus statistics and splits}
We aggregate a total of $3600+$ high-quality spatiotemporal HR samples, organised into non-overlapping train/validation/test splits per system. 
Supplementary Table~\ref{tab:dataset_stats} reports the sample counts and storage footprints.

\begin{table}[htbp]
\centering
\caption{Dataset statistics and splits (samples: train/val/test/total; storage size).}
\label{tab:dataset_stats}
\begin{tabular}{lcc}
\toprule
System & Samples (train/val/test/total) & Storage \\
\midrule
FIT   & 653/218/218/1089 & $\sim$70 GB \\
MHDT & 707/236/236/1179 & $\sim$132 GB \\
HBDT  & 822/274/274/1370 & $\sim$108 GB \\
\bottomrule
\end{tabular}
\end{table}

\subsection{Coarse-graining and LR generation}

Given a high-resolution spatiotemporal field 
$\bm{X}_{\mathrm{H}} \in \mathbb{R}^{T \times H_{\!H} \times W_{\!H} \times D_{\!H} \times C}$, 
we obtain its low-resolution counterpart 
$\bm{X}_{\mathrm{L}} \in \mathbb{R}^{T \times H_{\!L} \times W_{\!L} \times D_{\!L} \times C}$ 
through a generic coarse-graining process:
\begin{equation}
    \bm{X}_{\mathrm{L}} = \mathcal{D}(\bm{X}_{\mathrm{H}}) + \bm{\xi},
\end{equation}
where $\mathcal{D}$ denotes a spatial filtering operator followed by uniform 
subsampling, and $\bm{\xi}$ represents additive measurement noise.

We employ standard coarse-graining commonly used in turbulence modeling:
\begin{itemize}
\item \textbf{Box/Favre filtering}: with width $\Delta$ (in voxels). 
      For variable-density flows, Favre filtering uses 
      $\tilde{\phi}=\overline{\rho \phi}/\,\overline{\rho}$.
\item \textbf{Gaussian filtering}: isotropic or anisotropic, with 
      standard deviations $\sigma_{x,y,z}$ and optional random in-plane rotation.
\item \textbf{Spectral cutoff}: removing Fourier modes with 
      $\|\bm{k}\| > k_c$ prior to inverse FFT.
\end{itemize}
In addition, we include three additional coarse-graining strategies for evaluation:
\begin{itemize}
\item \textbf{Tricubic interpolation}: a smooth volumetric interpolation operator commonly used as a numerical baseline.
\item \textbf{Lanczos filtering}: a windowed sinc-based low-pass filter offering sharper frequency truncation.
\item \textbf{Uniform downsampling with noise}: voxel decimation followed by additive perturbations to simulate sensor-level degradation.
\end{itemize}
Upscaling factors $s \in \{4, 8, 16\}$ are used unless otherwise stated. 
Noise is modeled as additive Gaussian noise $\bm{\xi} \sim \mathcal{N}(0, \sigma_\xi^2)$. 

\subsection{Normalization and windowing}\label{app:split_norm}

Each variable $v \in \{\rho, \bm{u}, \bm{b}, p\}$ is standardized using 
training-set statistics via a z-score transformation:
\begin{equation} \label{eq:norm}
\hat{v} = \frac{v - \mu_v}{\sigma_v},
\end{equation}
where $(\mu_v, \sigma_v)$ are computed from the training partition and retained 
for inverse normalization during inference.

For TurbSR3D, we extract spatiotemporal windows of size 
$\langle T_{\mathrm{win}}, H, W, D \rangle$, following deterministic train/validation/test 
splits (60\%/20\%/20\%) over non-overlapping volumetric regions to prevent data leakage. 
In these experiments, we set $T_{\mathrm{win}} = 8$ and $H = W = D = 64$.
Zero padding is used in all convolutional layers that preserve spatial dimensions, maintaining feature-map sizes throughout convolutional processing. This convention is consistent with related studies and implementations in flow reconstruction. Morimoto \textit{et al.}~\cite{morimoto2021cnn} reported comparable reconstruction errors with zero and periodic padding for filter widths \(H \leq 9\) in the biperiodic turbulence case they examined. Zero padding has also been used for compressible-flow reconstruction by Sofos \textit{et al.}~\cite{sofos2023compressible} and in the released BLASTNet~2.0 3D CNN implementation~\cite{chung2023turbulence}. These precedents motivate its use as a practical implementation choice in the present framework.

\subsection{JHTDB acquisition notes}
Data are retrieved via the JHTDB SOAP services (Python client), with high-order spatial interpolation (\ie, Lag8) and PCHIP temporal interpolation for point queries. 
To ensure reliable large-scale acquisition, we employ block-wise sampling without overlap, request retry with backoff, and execution on the SciServer platform~\cite{popp2020sciserver} for high-throughput access. 

\subsection{DNS of decaying homogeneous isotropic turbulence}\label{ap:dit_dns}

Decaying homogeneous isotropic turbulence was simulated by solving the unforced incompressible Navier--Stokes equations in the triply periodic domain $[0,2\pi)^3$, with nondimensional viscosity $\nu=0.0033$. The initial velocity was a solenoidal Gaussian random field with target shell spectrum
\begin{equation}
E(k)\propto k^4\exp[-2(k/k_0)^2],\qquad k_0=6,
\label{eq:dit_initial_spectrum}
\end{equation}
normalized to $u_{\mathrm{rms}}=(\langle u_i u_i\rangle/3)^{1/2}=1$. The initial Taylor-scale Reynolds number was $Re_\lambda=101.25$, defined by
\begin{equation}
Re_\lambda=\frac{u_{\mathrm{rms}}\lambda}{\nu},
\qquad
\lambda=\left(\frac{15\nu u_{\mathrm{rms}}^2}{\varepsilon}\right)^{1/2},
\label{eq:dit_taylor_reynolds}
\end{equation}
where $\varepsilon=\nu\langle\partial_j u_i\,\partial_j u_i\rangle$.

The equations were discretized on a uniform $256^3$ collocated grid using fourth-order centred differences and a direct fourth-order Laplacian for viscous diffusion. Classical fourth-order Runge--Kutta integration used $\Delta t=10^{-4}$. Incompressibility was enforced at each stage through an FFT-based projection using the compatible discrete divergence--gradient operator, with its null modes handled by a pseudoinverse. No subgrid closure, dealiasing or artificial filtering was applied.

The dataset comprised 201 velocity--pressure frames over $t=0.200$--$0.600$, sampled at intervals of $0.002$, with two additional frames at each endpoint supporting centred temporal derivatives. Across the dataset, $k_{\max}\eta$ ranged from $1.716$ to $1.917$, where $\eta=(\nu^3/\varepsilon)^{1/4}$, $\varepsilon$ was evaluated using Fourier derivatives, and $k_{\max}=\pi/h=128$ is the axial Nyquist wavenumber. This quantity characterizes nominal dissipation-scale resolution; the effective resolution depends on the finite-difference operators. Grid convergence was not assessed.

\subsection{LES simulation and LES--DNS pairing}\label{ap:les_dns_pairing}\label{ap:les_dns}

LES was initialized from the DNS velocity at $t_0=0.522$ on a $64^3$ periodic grid with spacing $H=4h$. Initialization used a normalized, separable periodic Lanczos-3 filter~\cite{duchon1979lanczos} followed by fourfold subsampling. The one-dimensional kernel was
\begin{equation}
\ell(q)=
\begin{cases}
\operatorname{sinc}(q)\operatorname{sinc}(q/3), & |q|<3,\\
0, & |q|\geq3,
\end{cases}
\qquad
\operatorname{sinc}(q)=\frac{\sin(\pi q)}{\pi q}.
\label{eq:les_dns_lanczos}
\end{equation}
Discrete weights were proportional to $\ell((m-1.5)/4)$ and normalized to sum to unity. Coarse-grid samples were centred at $x_j=(4j+1.5)h$ along each coordinate, and the restricted velocity was projected onto the coarse-grid discrete divergence-free subspace.

The resolved velocity satisfied
\begin{equation}
\partial_t U_i+U_j\partial_jU_i
=-\partial_i\tilde{p}+\nu\nabla^2U_i+\partial_j(2\nu_tS_{ij}),
\qquad
\partial_iU_i=0,
\label{eq:les_dns_evolution}
\end{equation}
with the Smagorinsky closure~\cite{smagorinsky1963general}
\begin{equation}
\nu_t=(C_sH)^2(2S_{ij}S_{ij})^{1/2},
\qquad
S_{ij}=\tfrac12(\partial_jU_i+\partial_iU_j),
\qquad C_s=0.16.
\label{eq:les_dns_smagorinsky}
\end{equation}
Here, $\tilde{p}$ is the modified kinematic pressure incorporating the isotropic SGS stress. The LES retained the DNS viscosity, fourth-order spatial discretization, compatible pressure projection and Runge--Kutta scheme, with the SGS stress recomputed at every integration stage.

Following initialization, LES evolved independently for 160 steps with $\Delta t=10^{-4}$. Eight outputs at $t=0.524,0.526,\ldots,0.538$ were paired with DNS fields at the same physical times.

The local energy flux $\Pi_\ell$ was computed using a top-hat filter of width $\ell$, with $\tau_{ij}^{(\ell)}=\overline{u_iu_j}^{(\ell)}-\bar u_i^{(\ell)}\bar u_j^{(\ell)}$.

\subsection{ERA5 data access and experimental protocol}
\label{ap:era5_data}

\noindent\textbf{Data access.}
ERA5 fields are retrieved via WeatherBench2~\cite{rasp2024weatherbench2} for efficient data access and preprocessing.

\noindent\textbf{Downscaling setting.}
In the downscaling experiment, the model maps a low-resolution spatiotemporal multi-level field
\(
\bm{X}_{\mathrm{L}} \in \mathbb{R}^{T \times H_{\mathrm{L}} \times W_{\mathrm{L}} \times P \times C}
\)
to the corresponding high-resolution field
\(
\bm{X}_{\mathrm{H}} \in \mathbb{R}^{T \times H_{\mathrm{H}} \times W_{\mathrm{H}} \times P \times C}.
\)
Unless stated otherwise, \(T=8\) and \(H_{\mathrm{H}}=W_{\mathrm{H}}=128\).
The vertical dimension uses \(P=8\) pressure levels:
1000, 925, 850, 700, 600, 500, 400, and 300~hPa.
The channel dimension \(C\) includes temperature and horizontal wind components \((u,v)\).

\noindent\textbf{Normalization.}
Normalization follows Supplementary Eq.~\eqref{eq:norm}.

\clearpage
\section{Additional details on experiments}\label{ap:exp}
\subsection{Accurate reconstruction of turbulent flows}

To complement the quantitative results in the main text (Table~\ref{tab:benchmark_results}), we report the full set of quantitative evaluations across all turbulent systems, coarse-graining strategies, and upscaling factors. 
Supplementary Tables~\ref{tab:fit_detailed}--\ref{tab:hbdt_detailed} present detailed comparisons between our framework and eight baselines, using four complementary metrics (PSNR, SSIM, nRMSE, and fRMSE; see Section~\ref{ssec:metrics}).
Higher PSNR and SSIM indicate better perceptual quality and spatial fidelity, while lower nRMSE and fRMSE reflect improved voxel-wise accuracy and spectral fidelity, respectively.
For every turbulent system (FIT, MHDT, HBDT) and each coarse-graining strategy, we independently train and evaluate all learning-based methods using the same LR--HR data pairs and training schedule. 
No cross-dataset fine-tuning or weight sharing is applied. 
This ensures that all methods are compared on equal footing and that the reported numbers reflect genuine model capability rather than transfer bias.
Supplementary Tables~\ref{tab:fit_detailed},~\ref{tab:mhdt_detailed}, and~\ref{tab:hbdt_detailed} provide the complete spatial and point-wise benchmark results. Our method achieves the highest PSNR and SSIM and the lowest nRMSE and fRMSE among the compared methods in all 18 evaluated settings, spanning three turbulent systems, three coarse-graining strategies, and two upscaling factors per system.

\begin{table*}[htbp]
    \centering
    \renewcommand{\arraystretch}{0.82}
    \caption{
    \textbf{Full quantitative comparison on FIT.}
    Detailed super-resolution results for $4\times$ and $8\times$ upscaling under three coarse-graining strategies (box, Gaussian, and spectral cutoff).
    Performance is evaluated using PSNR~($\uparrow$), SSIM~($\uparrow$), nRMSE~($\downarrow$), and fRMSE~($\downarrow$) (see Section~\ref{ssec:metrics}).
    For each upscaling factor and coarse-graining, the best result is shown in
    \textbf{bold}. Model sizes are reported as the number of parameters in millions (M).
}
    \label{tab:fit_detailed}
    \resizebox{0.92\textwidth}{!}{%
    \begin{tabular}{
        c %
        l %
        l %
        S[table-format=2.2] %
        S[table-format=1.4] %
        S[table-format=2.2] %
        S[table-format=1.4] %
        S[table-format=2.2] %
    }
    \toprule
    \textbf{Factor} & \textbf{Coarse-gr.} & \textbf{Method} & {\textbf{\#Params. (M)}} & {\textbf{SSIM↑}} & {\textbf{PSNR↑}} & {\textbf{nRMSE↓}} & {\textbf{fRMSE↓}} \\
    \midrule
    
    \multirow{27}{*}{\textbf{4$\times$}}
    & \multirow{9}{*}{Box} & Tricubic & \text{--} & 0.8321 & 42.28 & 0.1204 & 28.56 \\
    & & RRDB & 4.2 & 0.9614 & 50.90 & 0.0460 & 11.32 \\
    & & RCAN & 4.2 &  {0.9672} &  {51.92} &  {0.0415} &  {10.23} \\
    & & EDSR & 4.2 & 0.9661 & 51.77 & 0.0423 & 10.41 \\
    & & FNO & 34.0 & 0.9109 & 46.61 & 0.0730 & 17.27 \\
    & & PhySR & 3.3 & 0.9590 & 50.66 & 0.0476 & 11.63 \\
    & & DiffSR & \multicolumn{1}{c}{7.3} & \multicolumn{1}{c}{0.9651} & \multicolumn{1}{c}{51.62} & \multicolumn{1}{c}{0.0423} & \multicolumn{1}{c}{10.34} \\
    & & SwinIR & \multicolumn{1}{c}{3.8} & \multicolumn{1}{c}{0.9689} & \multicolumn{1}{c}{51.46} & \multicolumn{1}{c}{0.0421} & \multicolumn{1}{c}{10.23} \\
    & & {Ours} & 3.7 & \textbf{0.9834} & \textbf{55.09} & \textbf{0.0291} & \textbf{7.22} \\
    \cmidrule(l){2-8}
    & \multirow{9}{*}{Spectral} & Tricubic & \text{--} & 0.7630 & 40.26 & 0.1501 & 35.47 \\
    & & RRDB & 4.2 & 0.8919 & 46.41 & 0.0770 & 18.73 \\
    & & RCAN & 4.2 & 0.8932 & 46.60 & 0.0755 & 18.39 \\
    & & EDSR & 4.2 & 0.8930 & 46.54 & 0.0759 & 18.45 \\
    & & FNO & 34.0 & 0.8394 & 43.51 & 0.1036 & 24.45 \\
    & & PhySR & 3.3 &  {0.9044} &  {46.99} &  {0.0722} &  {17.56} \\
    & & DiffSR & \multicolumn{1}{c}{7.3} & \multicolumn{1}{c}{0.8889} & \multicolumn{1}{c}{45.84} & \multicolumn{1}{c}{0.0810} & \multicolumn{1}{c}{19.54} \\
    & & SwinIR & \multicolumn{1}{c}{3.8} & \multicolumn{1}{c}{0.9131} & \multicolumn{1}{c}{47.07} & \multicolumn{1}{c}{0.0704} & \multicolumn{1}{c}{17.16} \\
    & & {Ours} & 3.7 & \textbf{0.9251} & \textbf{48.41} & \textbf{0.0615} & \textbf{15.09} \\
    \cmidrule(l){2-8}
    & \multirow{9}{*}{Gaussian} & Tricubic & \text{--} & 0.8163 & 41.45 & 0.1325 & 31.39 \\
    & & RRDB & 4.2 & 0.9576 & 50.28 & 0.0494 & 12.13 \\
    & & RCAN & 4.2 &  {0.9682} & 51.78 &  {0.0421} &  {10.34} \\
    & & EDSR & 4.2 & 0.9645 &  {51.92} & 0.0445 & 10.93 \\
    & & FNO & 34.0 & 0.9100 & 46.43 & 0.0745 & 17.67 \\
    & & PhySR & 3.3 & 0.9585 & 50.33 & 0.0494 & 12.03 \\
    & & DiffSR & \multicolumn{1}{c}{7.3} & \multicolumn{1}{c}{0.9646} & \multicolumn{1}{c}{51.32} & \multicolumn{1}{c}{0.0438} & \multicolumn{1}{c}{10.69} \\
    & & SwinIR & \multicolumn{1}{c}{3.8} & \multicolumn{1}{c}{0.9698} & \multicolumn{1}{c}{51.29} & \multicolumn{1}{c}{0.0429} & \multicolumn{1}{c}{10.40} \\
    & & {Ours} & 3.7 & \textbf{0.9833} & \textbf{54.75} & \textbf{0.0300} & \textbf{7.43} \\
    \midrule
    
    \multirow{27}{*}{\textbf{8$\times$}}
    & \multirow{9}{*}{Box} & Tricubic & \text{--} & 0.5974 & 36.81 & 0.2242 & 52.83 \\
    & & RRDB & 4.3 & 0.6826 & 39.31 & 0.1712 & 40.77 \\
    & & RCAN & 5.1 & 0.6701 & 39.10 & 0.1753 & 41.81 \\
    & & EDSR & 4.7 & 0.6791 & 40.29 & 0.1693 & 39.36 \\
    & & FNO & 34.3 & 0.6936 & 39.49 & 0.1657 & 38.91 \\
    & & PhySR & 5.4 &  {0.7291} &  {40.54} &  {0.1486} &  {35.44} \\
    & & DiffSR & \multicolumn{1}{c}{7.3} & \multicolumn{1}{c}{0.7480} & \multicolumn{1}{c}{40.82} & \multicolumn{1}{c}{0.1440} & \multicolumn{1}{c}{34.46} \\
    & & SwinIR & \multicolumn{1}{c}{4.0} & \multicolumn{1}{c}{0.7300} & \multicolumn{1}{c}{40.39} & \multicolumn{1}{c}{0.1499} & \multicolumn{1}{c}{35.67} \\
    & & Ours & 3.9 & \textbf{0.7963} & \textbf{42.46} & \textbf{0.1195} & \textbf{28.63} \\
    \cmidrule(l){2-8}
    & \multirow{9}{*}{Spectral} & Tricubic & \text{--} & 0.5459 & 35.80 & 0.2505 & 58.87 \\
    & & RRDB & 4.3 & 0.5864 & 37.23 & 0.2156 & 51.06 \\
    & & RCAN & 5.1 & 0.5716 & 36.99 & 0.2209 & 52.51 \\
    & & EDSR & 4.7 & 0.5821 & 37.26 & 0.2140 & 50.64 \\
    & & FNO & 34.3 & 0.5963 & 37.42 & 0.2098 & 49.16 \\
    & & PhySR & 5.4 &  {0.6086} &  {37.90} &  {0.2002} &  {47.51} \\
    & & DiffSR & \multicolumn{1}{c}{7.3} & \multicolumn{1}{c}{0.5802} & \multicolumn{1}{c}{36.81} & \multicolumn{1}{c}{0.2246} & \multicolumn{1}{c}{52.41} \\
    & & SwinIR & \multicolumn{1}{c}{4.0} & \multicolumn{1}{c}{0.6085} & \multicolumn{1}{c}{37.49} & \multicolumn{1}{c}{0.2067} & \multicolumn{1}{c}{48.70} \\
    & & Ours & 3.9 & \textbf{0.6372} & \textbf{38.42} & \textbf{0.1884} & \textbf{44.69} \\
    \cmidrule(l){2-8}
    & \multirow{9}{*}{Gaussian} & Tricubic & \text{--} & 0.5991 & 36.68 & 0.2278 & 53.69 \\
    & & RRDB & 4.3 & 0.6815 & 39.12 & 0.1749 & 41.63 \\
    & & RCAN & 5.1 & 0.6682 & 38.86 & 0.1802 & 43.00 \\
    & & EDSR & 4.7 & 0.6743 & 39.05 & 0.1752 & 41.66 \\
    & & FNO & 34.3 & 0.6958 & 39.37 & 0.1679 & 39.37 \\
    & & PhySR & 5.4 &  {0.7290} &  {40.36} &  {0.1517} &  {36.17} \\
    & & DiffSR & \multicolumn{1}{c}{7.3} & \multicolumn{1}{c}{0.7499} & \multicolumn{1}{c}{40.72} & \multicolumn{1}{c}{0.1456} & \multicolumn{1}{c}{34.80} \\
    & & SwinIR & \multicolumn{1}{c}{4.0} & \multicolumn{1}{c}{0.7284} & \multicolumn{1}{c}{40.18} & \multicolumn{1}{c}{0.1535} & \multicolumn{1}{c}{36.52} \\
    & & {Ours} & 3.9 & \textbf{0.8093} & \textbf{42.61} & \textbf{0.1173} & \textbf{28.12} \\

    \bottomrule
    \end{tabular}
    }
\end{table*}

\begin{table*}[t]
    \centering
    \renewcommand{\arraystretch}{0.82}
\caption{
    \textbf{Full quantitative comparison on MHDT.}
    Detailed super-resolution results for $4\times$ and $8\times$ upscaling under three
    coarse-graining strategies (box, Gaussian, and spectral cutoff).
    Performance is evaluated using PSNR~($\uparrow$), SSIM~($\uparrow$), nRMSE~($\downarrow$),
    and fRMSE~($\downarrow$) (see Section~\ref{ssec:metrics}).
    For each upscaling factor and coarse-graining, the best result is shown in
    \textbf{bold}.
    Model sizes are reported as the number of parameters in millions (M). 
    }
    \label{tab:mhdt_detailed}
    
    \resizebox{0.92\textwidth}{!}{%
    \begin{tabular}{
        c %
        l %
        l %
        S[table-format=2.2] %
        S[table-format=1.4] %
        S[table-format=2.2] %
        S[table-format=1.4] %
        S[table-format=2.2] %
    }
    \toprule
    \textbf{Factor} & \textbf{Coarse-gr.} & \textbf{Method} & {\textbf{\#Params. (M)}} & {\textbf{SSIM↑}} & {\textbf{PSNR↑}} & {\textbf{nRMSE↓}} & {\textbf{fRMSE↓}} \\
    \midrule
    
    \multirow{27}{*}{\textbf{4$\times$}}
    & \multirow{9}{*}{Box} & Tricubic & \text{--} & 0.7643 & 38.39 & 0.1838 & 16.29 \\
    & & RRDB & 4.2 &  {0.9482} &  {47.34} &  {0.0678} &  {5.92} \\
    & & RCAN & 4.2 & 0.9464 & 47.28 & 0.0688 & 5.99 \\
    & & EDSR & 4.3 & 0.9448 & 47.15 & 0.0699 & 6.09 \\
    & & FNO & 34.0 & 0.8687 & 42.41 & 0.1166 & 10.28 \\
    & & PhySR & 3.5 & 0.9265 & 45.37 & 0.0840 & 7.39 \\
    & & DiffSR & \multicolumn{1}{c}{7.3} & \multicolumn{1}{c}{0.9395} & \multicolumn{1}{c}{46.58} & \multicolumn{1}{c}{0.0731} & \multicolumn{1}{c}{6.45} \\
    & & SwinIR & \multicolumn{1}{c}{3.8} & \multicolumn{1}{c}{0.9327} & \multicolumn{1}{c}{45.84} & \multicolumn{1}{c}{0.0792} & \multicolumn{1}{c}{6.95} \\
    & & {Ours} & 3.7 & \textbf{0.9579} & \textbf{48.28} & \textbf{0.0603} & \textbf{5.29} \\
    \cmidrule(l){2-8}
    & \multirow{9}{*}{Spectral} & Tricubic & \text{--} & 0.6816 & 36.74 & 0.2204 & 19.70 \\
    & & RRDB & 4.2 &  {0.8552} &  {42.71} &  {0.1130} &  {9.90} \\
    & & RCAN & 4.2 & 0.8536 & 42.69 & 0.1135 & 9.94 \\
    & & EDSR & 4.3 & 0.8484 & 42.47 & 0.1163 & 10.19 \\
    & & FNO & 34.0 & 0.7658 & 39.50 & 0.1612 & 14.27 \\
    & & PhySR & 3.5 & 0.8439 & 42.22 & 0.1188 & 10.46 \\
    & & DiffSR & \multicolumn{1}{c}{7.3} & \multicolumn{1}{c}{0.8168} & \multicolumn{1}{c}{41.16} & \multicolumn{1}{c}{0.1339} & \multicolumn{1}{c}{11.77} \\
    & & SwinIR & \multicolumn{1}{c}{3.8} & \multicolumn{1}{c}{0.8498} & \multicolumn{1}{c}{42.17} & \multicolumn{1}{c}{0.1190} & \multicolumn{1}{c}{10.48} \\
    & & {Ours} & 3.7 & \textbf{0.8671} & \textbf{43.12} & \textbf{0.1071} & \textbf{9.43} \\
    \cmidrule(l){2-8}
    & \multirow{9}{*}{Gaussian} & Tricubic & \text{--} & 0.7456 & 37.67 & 0.2000 & 17.78 \\
    & & RRDB & 4.2 &  {0.9489} &  {47.18} &  {0.0692} &  {6.03} \\
    & & RCAN & 4.2 & 0.9455 & 46.98 & 0.0712 & 6.20 \\
    & & EDSR & 4.3 & 0.9450 & 46.92 & 0.0718 & 6.25 \\
    & & FNO & 34.0 & 0.8689 & 42.26 & 0.1188 & 10.48 \\
    & & PhySR & 3.5 & 0.9288 & 45.25 & 0.0852 & 7.49 \\
    & & DiffSR & \multicolumn{1}{c}{7.3} & \multicolumn{1}{c}{0.9380} & \multicolumn{1}{c}{46.28} & \multicolumn{1}{c}{0.0756} & \multicolumn{1}{c}{6.66} \\
    & & SwinIR & \multicolumn{1}{c}{3.8} & \multicolumn{1}{c}{0.9335} & \multicolumn{1}{c}{45.71} & \multicolumn{1}{c}{0.0804} & \multicolumn{1}{c}{7.06} \\
    & & {Ours} & 3.7 & \textbf{0.9582} & \textbf{48.06} & \textbf{0.0620} & \textbf{5.43} \\
    \midrule
    
    \multirow{27}{*}{\textbf{8$\times$}}
    & \multirow{9}{*}{Box} & Tricubic & \text{--} & 0.4908 & 33.05 & 0.3372 & 30.22 \\
    & & RRDB & 4.3 &  {0.6560} &  {36.85} &  {0.2205} &  {19.52} \\
    & & RCAN & 5.1 & 0.6414 & 36.55 & 0.2285 & 20.23 \\
    & & EDSR & 4.7 & 0.6435 & 36.68 & 0.2250 & 19.88 \\
    & & FNO & 34.2 & 0.6009 & 35.62 & 0.2525 & 22.44 \\
    & & PhySR & 7.5 & 0.6036 & 35.91 & 0.2443 & 21.70 \\
    & & DiffSR & \multicolumn{1}{c}{7.3} & \multicolumn{1}{c}{0.6629} & \multicolumn{1}{c}{37.02} & \multicolumn{1}{c}{0.2153} & \multicolumn{1}{c}{18.96} \\
    & & SwinIR & \multicolumn{1}{c}{4.0} & \multicolumn{1}{c}{0.6047} & \multicolumn{1}{c}{35.87} & \multicolumn{1}{c}{0.2447} & \multicolumn{1}{c}{21.67} \\
    & & {Ours} & 3.9 & \textbf{0.6713} & \textbf{37.28} & \textbf{0.2094} & \textbf{18.50} \\
    \cmidrule(l){2-8}
    & \multirow{9}{*}{Spectral} & Tricubic & \text{--} & 0.4327 & 32.35 & 0.3640 & 32.69 \\
    & & RRDB & 4.3 &  {0.5003} &  {34.06} &  {0.3013} &  {26.83} \\
    & & RCAN & 5.1 & 0.4843 & 33.78 & 0.3105 & 27.67 \\
    & & EDSR & 4.7 & 0.4931 & 33.99 & 0.3036 & 26.98 \\
    & & FNO & 34.2 & 0.4822 & 33.68 & 0.3141 & 27.97 \\
    & & PhySR & 7.5 & 0.4731 & 33.75 & 0.3117 & 27.69 \\
    & & DiffSR & \multicolumn{1}{c}{7.3} & \multicolumn{1}{c}{0.4739} & \multicolumn{1}{c}{33.45} & \multicolumn{1}{c}{0.3226} & \multicolumn{1}{c}{28.45} \\
    & & SwinIR & \multicolumn{1}{c}{4.0} & \multicolumn{1}{c}{0.4212} & \multicolumn{1}{c}{32.40} & \multicolumn{1}{c}{0.3628} & \multicolumn{1}{c}{32.06} \\
    & & {Ours} & 3.9 & \textbf{0.5052} & \textbf{34.21} & \textbf{0.2957} & \textbf{26.20} \\
    \cmidrule(l){2-8}
    & \multirow{9}{*}{Gaussian} & Tricubic & \text{--} & 0.4881 & 33.16 & 0.3327 & 29.80 \\
    & & RRDB & 4.3 &  {0.6602} &  {36.71} &  {0.2246} &  {19.87} \\
    & & RCAN & 5.1 & 0.6465 & 36.38 & 0.2330 & 20.64 \\
    & & EDSR & 4.7 & 0.6479 & 36.53 & 0.2290 & 20.24 \\
    & & FNO & 34.2 & 0.6050 & 35.51 & 0.2559 & 22.73 \\
    & & PhySR & 7.5 & 0.6079 & 35.79 & 0.2479 & 22.00 \\
    & & DiffSR & \multicolumn{1}{c}{7.3} & \multicolumn{1}{c}{0.6699} & \multicolumn{1}{c}{36.94} & \multicolumn{1}{c}{0.2175} & \multicolumn{1}{c}{19.15} \\
    & & SwinIR & \multicolumn{1}{c}{4.0} & \multicolumn{1}{c}{0.6045} & \multicolumn{1}{c}{35.70} & \multicolumn{1}{c}{0.2496} & \multicolumn{1}{c}{22.07} \\
    & & {Ours} & 3.9 & \textbf{0.6825} & \textbf{37.27} & \textbf{0.2099} & \textbf{18.54} \\

    \bottomrule
    \end{tabular}
    } %
\end{table*}

\begin{table*}[t]
    \centering
    \renewcommand{\arraystretch}{0.82}
\caption{
            \textbf{Full quantitative comparison on HBDT.}
    Detailed super-resolution results for $8\times$ and $16\times$ upscaling under three
    coarse-graining strategies (Favre, Gaussian, and spectral cutoff).
    Performance is evaluated using PSNR~($\uparrow$), SSIM~($\uparrow$), nRMSE~($\downarrow$),
    and fRMSE~($\downarrow$) (see Section~\ref{ssec:metrics}).
    For each upscaling factor and coarse-graining, the best result is shown in
    \textbf{bold}.
    Model sizes are reported as the number of parameters in millions (M).
    }
    \label{tab:hbdt_detailed}
    
    \resizebox{0.92\textwidth}{!}{%
    \begin{tabular}{
        c %
        l %
        l %
        S[table-format=2.2] %
        S[table-format=1.4] %
        S[table-format=2.2] %
        S[table-format=1.4] %
        S[table-format=2.2] %
    }
    \toprule
    \textbf{Factor} & \textbf{Coarse-gr.} & \textbf{Method} & {\textbf{\#Params. (M)}} & {\textbf{SSIM↑}} & {\textbf{PSNR↑}} & {\textbf{nRMSE↓}} & {\textbf{fRMSE↓}} \\
    \midrule
    
    \multirow{27}{*}{\textbf{8$\times$}}
    & \multirow{9}{*}{Favre} & Tricubic & \text{--} & 0.9226 & 46.87 & 0.0933 & 3.89 \\
    & & RRDB & 4.3 & 0.9778 & 57.95 & 0.0307 & 1.38 \\
    & & RCAN & 5.1 &  {0.9800} &  {61.57} &  {0.0262} &  {1.27} \\
    & & EDSR & 4.7 & 0.9779 & 59.22 & 0.0286 & 1.34 \\
    & & FNO & 34.3 & 0.9264 & 51.04 & 0.0544 & 2.14 \\
    & & PhySR & 6.1 & 0.9668 & 59.38 & 0.0316 & 1.45 \\
    & & DiffSR & \multicolumn{1}{c}{7.3} & \multicolumn{1}{c}{0.9655} & \multicolumn{1}{c}{57.80} & \multicolumn{1}{c}{0.0264} & \multicolumn{1}{c}{1.37} \\
    & & SwinIR & \multicolumn{1}{c}{4.0} & \multicolumn{1}{c}{0.9672} & \multicolumn{1}{c}{55.65} & \multicolumn{1}{c}{0.0353} & \multicolumn{1}{c}{1.47} \\
    & & Ours & 3.9 & \textbf{0.9862} & \textbf{66.63} & \textbf{0.0195} & \multicolumn{1}{c}{\textbf{1.01}} \\
    \cmidrule(l){2-8}
    & \multirow{9}{*}{Spectral} & Tricubic & \text{--} & 0.8382 & 44.54 & 0.1196 & 4.76 \\
    & & RRDB & 4.3 & 0.9619 & 55.12 & 0.0432 &  {1.96} \\
    & & RCAN & 5.1 &  {0.9625} &  {57.46} &  {0.0410} & 1.97 \\
    & & EDSR & 4.7 & 0.9611 & 55.71 & 0.0428 & 1.99 \\
    & & FNO & 34.3 & 0.8936 & 47.60 & 0.0812 & 3.05 \\
    & & PhySR & 6.1 & 0.9478 & 55.79 & 0.0452 & 2.01 \\
    & & DiffSR & \multicolumn{1}{c}{7.3} & \multicolumn{1}{c}{0.9389} & \multicolumn{1}{c}{52.67} & \multicolumn{1}{c}{0.0496} & \multicolumn{1}{c}{1.79} \\
    & & SwinIR & \multicolumn{1}{c}{4.0} & \multicolumn{1}{c}{0.9483} & \multicolumn{1}{c}{52.51} & \multicolumn{1}{c}{0.0521} & \multicolumn{1}{c}{2.20} \\
    & & Ours & 3.9 & \textbf{0.9702} & \textbf{60.12} & \textbf{0.0348} & \textbf{1.68} \\
    \cmidrule(l){2-8}
    & \multirow{9}{*}{Gaussian} & Tricubic & \text{--} & 0.9211 & 46.70 & 0.0959 & 3.97 \\
    & & RRDB & 4.3 & 0.9782 & 57.54 & 0.0316 & 1.42 \\
    & & RCAN & 5.1 &  {0.9794} &  {61.31} &  {0.0275} &  {1.33} \\
    & & EDSR & 4.7 & 0.9777 & 58.80 & 0.0299 & 1.40 \\
    & & FNO & 34.3 & 0.9296 & 51.30 & 0.0535 & 2.14 \\
    & & PhySR & 6.1 & 0.9531 & 56.39 & 0.0389 & 1.73 \\
    & & DiffSR & \multicolumn{1}{c}{7.3} & \multicolumn{1}{c}{0.9586} & \multicolumn{1}{c}{57.25} & \multicolumn{1}{c}{0.0300} & \multicolumn{1}{c}{1.39} \\
    & & SwinIR & \multicolumn{1}{c}{4.0} & \multicolumn{1}{c}{0.9668} & \multicolumn{1}{c}{55.57} & \multicolumn{1}{c}{0.0354} & \multicolumn{1}{c}{1.49} \\
    & & Ours & 3.9 & \textbf{0.9864} & \textbf{66.37} & \textbf{0.0200} & \multicolumn{1}{c}{\textbf{1.03}} \\
    \midrule
    
    \multirow{27}{*}{\textbf{16$\times$}}
    & \multirow{9}{*}{Favre} & Tricubic & \text{--} & 0.8364 & 42.00 & 0.1666 & 6.87 \\
    & & RRDB & 4.4 &  {0.9037} &  {47.78} &  {0.0940} &  {4.14} \\
    & & RCAN & 6.0 & 0.8626 & 46.65 & 0.1396 & 5.03 \\
    & & EDSR & 5.1 & 0.8639 & 46.42 & 0.1301 & 5.00 \\
    & & FNO & 5.1 & 0.8393 & 43.85 & 0.1224 & 4.94 \\
    & & PhySR & 27.9 & 0.8173 & 45.37 & 0.1159 & 4.93 \\
    & & DiffSR & \multicolumn{1}{c}{7.3} & \multicolumn{1}{c}{0.9073} & \multicolumn{1}{c}{48.08} & \multicolumn{1}{c}{0.0860} & \multicolumn{1}{c}{{3.92}} \\
    & & SwinIR & \multicolumn{1}{c}{4.2} & \multicolumn{1}{c}{0.8400} & \multicolumn{1}{c}{44.19} & \multicolumn{1}{c}{0.1228} & \multicolumn{1}{c}{4.78} \\
    & & Ours & 4.1 & \textbf{0.9265} & \textbf{52.42} & \textbf{0.0739} & \multicolumn{1}{c}{\textbf{3.49}} \\
    \cmidrule(l){2-8}
    & \multirow{9}{*}{Spectral} & Tricubic & \text{--} & 0.8378 & 41.96 & 0.1671 & 6.87 \\
    & & RRDB & 4.4 &  {0.8545} &  {44.39} &  {0.1354} &  {5.77} \\
    & & RCAN & 6.0 & 0.8357 & 44.20 & 0.1438 & 6.07 \\
    & & EDSR & 5.1 & 0.8227 & 43.38 & 0.1476 & 6.06 \\
    & & FNO & 5.1 & 0.7770 & 40.87 & 0.1767 & 6.61 \\
    & & PhySR & 27.9 & 0.7724 & 43.13 & 0.1511 & 6.22 \\
    & & DiffSR & \multicolumn{1}{c}{7.3} & \multicolumn{1}{c}{0.8605} & \multicolumn{1}{c}{43.96} & \multicolumn{1}{c}{0.1366} & \multicolumn{1}{c}{{5.82}} \\
    & & SwinIR & \multicolumn{1}{c}{4.2} & \multicolumn{1}{c}{0.8085} & \multicolumn{1}{c}{42.16} & \multicolumn{1}{c}{0.1602} & \multicolumn{1}{c}{6.16} \\
    & & Ours & 4.1 & \textbf{0.8778} & \textbf{47.53} & \textbf{0.1127} & \multicolumn{1}{c}{\textbf{5.00}} \\
    \cmidrule(l){2-8}
    & \multirow{9}{*}{Gaussian} & Tricubic & \text{--} & 0.7588 & 39.97 & 0.2074 & 8.24 \\
    & & RRDB & 4.4 &  {0.9011} &  {47.25} &  {0.0985} &  {4.31} \\
    & & RCAN & 6.0 & 0.8876 & 47.33 & 0.1055 & 4.56 \\
    & & EDSR & 5.1 & 0.8532 & 45.71 & 0.1452 & 5.11 \\
    & & FNO & 5.1 & 0.8393 & 43.85 & 0.1224 & 4.94 \\
    & & PhySR & 27.9 & 0.8210 & 45.40 & 0.1165 & 4.98 \\
    & & DiffSR & \multicolumn{1}{c}{7.3} & \multicolumn{1}{c}{0.9020} & \multicolumn{1}{c}{47.62} & \multicolumn{1}{c}{0.0984} & \multicolumn{1}{c}{{4.19}} \\
    & & SwinIR & \multicolumn{1}{c}{4.2} & \multicolumn{1}{c}{0.8462} & \multicolumn{1}{c}{44.06} & \multicolumn{1}{c}{0.1242} & \multicolumn{1}{c}{4.86} \\
    & & Ours & 4.1 & \textbf{0.9268} & \textbf{52.25} & \textbf{0.0754} & \multicolumn{1}{c}{\textbf{3.55}} \\

    \bottomrule
    \end{tabular}
    } %
\end{table*}

\clearpage
\subsection{Noise robustness}
\label{ssec:noise_robustness}

Super-resolution analysis can also be viewed as a framework for reconstructing global flow fields from spatially sparse measurements, with LR flow data representing sparse or noisy sensor observations~\cite{fukami2023super}. This perspective motivates examining how reconstruction performance depends on the noise magnitude in the LR input. We evaluate this dependence using FIT at \(4\times\) upscaling with box-filtered LR inputs. After channel-wise standardization using the training-set statistics, additive Gaussian noise \(\bm{\xi}\sim\mathcal{N}(0,\sigma_\xi^2)\) is added to the LR input, where \(\sigma_\xi\) is expressed relative to the standardized channel scale. All learned methods use the same noise augmentation during training, with \(\sigma_\xi\) sampled with equal probability from \(\{0,0.01,0.03,0.05,0.10\}\). At evaluation, reconstruction performance is measured separately at each of these noise levels.

\begin{figure}[htbp]
    \centering
    \includegraphics[width=0.6\textwidth]{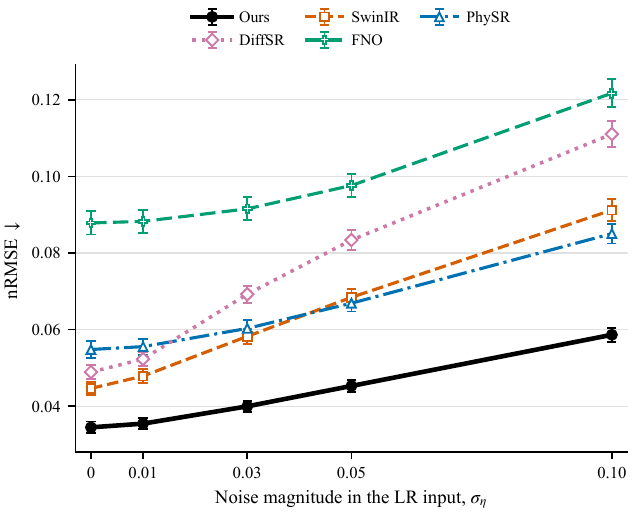}
    \caption{\textbf{Noise robustness on FIT.}
    Mean nRMSE at \(4\times\) upscaling as a function of the noise magnitude \(\sigma_\xi\) in box-filtered LR inputs. Error bars denote 95\% confidence intervals.}
    \label{fig:noise_robustness}
\end{figure}

\begin{table}[htbp]
    \centering
    \caption{\textbf{Quantitative comparison of noise robustness on FIT.}
    Mean nRMSE at \(4\times\) upscaling for box-filtered LR inputs. \(\Delta\)nRMSE denotes the absolute increase from \(\sigma_\xi=0\) to \(0.10\). The best result in each column is shown in \textbf{bold}.}
    \label{tab:noise_robustness}
    \small
    \setlength{\tabcolsep}{5pt}
    \begin{tabular}{lcccccc}
        \toprule
        \multirow{2}{*}{Method} & \multicolumn{5}{c}{Noise magnitude \(\sigma_\xi\)} & \multirow{2}{*}{\(\Delta\)nRMSE} \\
        \cmidrule(lr){2-6}
        & \(0\) & \(0.01\) & \(0.03\) & \(0.05\) & \(0.10\) & \\
        \midrule
        Ours   & \textbf{0.0345} & \textbf{0.0355} & \textbf{0.0400} & \textbf{0.0453} & \textbf{0.0587} & \textbf{0.0242} \\
        SwinIR & 0.0447 & 0.0479 & 0.0582 & 0.0684 & 0.0912 & 0.0466 \\
        PhySR  & 0.0548 & 0.0556 & 0.0603 & 0.0669 & 0.0851 & 0.0302 \\
        DiffSR & 0.0489 & 0.0523 & 0.0692 & 0.0834 & 0.1110 & 0.0621 \\
        FNO    & 0.0878 & 0.0883 & 0.0915 & 0.0976 & 0.1217 & 0.0339 \\
        \bottomrule
    \end{tabular}
\end{table}

Under this common training and evaluation protocol, our model achieves the lowest nRMSE at every examined noise magnitude and the smallest absolute increase in nRMSE from \(\sigma_\xi=0\) to \(0.10\) (Supplementary Fig.~\ref{fig:noise_robustness}; Supplementary Table~\ref{tab:noise_robustness}). Its mean nRMSE increases from \(0.0345\) to \(0.0587\), an absolute increase of \(0.0242\), whereas the increases for the other methods range from \(0.0302\) to \(0.0621\). At \(\sigma_\xi=0.10\), its nRMSE is \(0.0587\), \(31.0\%\) lower than PhySR, the second-best method at this noise magnitude.

\FloatBarrier

\subsection{Ablation study of the multiscale neural operator}
\label{ssec:mno_ablation}

We use FIT at \(4\times\) upscaling as a controlled case for attributing the two spatial operators. The full model is compared with w/o \(\mathcal{O}_s\), which removes the global Fourier operator, and w/o \(\mathcal{O}_c\), which removes the local three-dimensional convolutional operator. All three models use the same training and evaluation settings.

We evaluate the ablations as functions of the physical wavenumber \(k\) and summarize the metrics in three ranges: Low (\(k<64\)), Mid (\(64\leq k\leq144\)), and High (\(k>144\)). Velocity nRMSE measures reconstruction error across spatial scales, while vorticity MAE additionally probes velocity gradients and rotational structures. Each table entry is the mean within the corresponding range.

\begin{figure}[htbp]
    \centering
    \includegraphics[width=\textwidth]{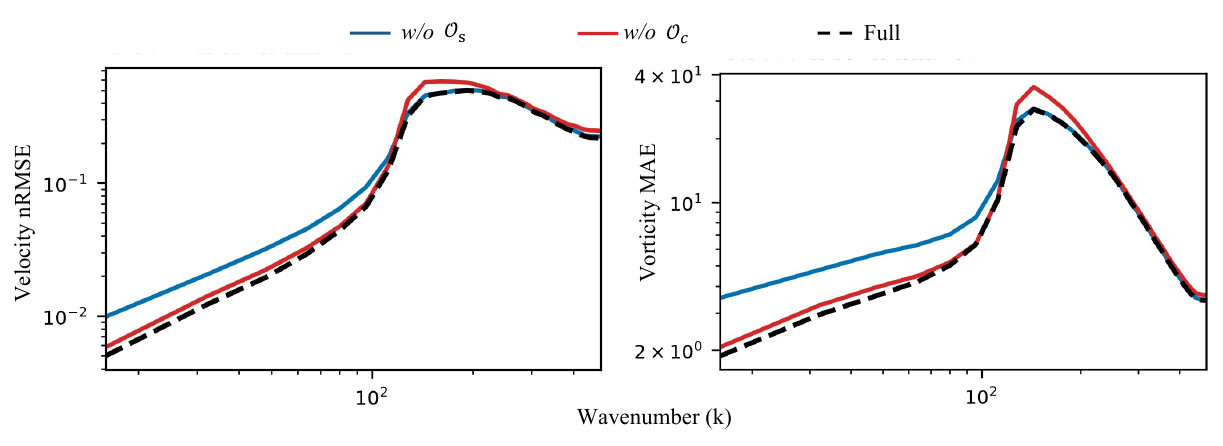}
    \caption{\textbf{Ablation study of the multiscale neural operator on FIT.}
    The full model is compared with models without the spectral operator \(\mathcal{O}_s\) or the convolutional operator \(\mathcal{O}_c\) at \(4\times\) upscaling with spectral-cutoff coarse-graining.
    \textbf{Left,} Velocity nRMSE versus physical wavenumber.
    \textbf{Right,} Vorticity MAE versus physical wavenumber.}
    \label{fig:mno_ablation}
\end{figure}

\begin{table}[htbp]
\centering
\caption{\textbf{Scale-resolved errors for the multiscale neural operator ablations.}
The comparison uses FIT at $4\times$ with spectral-cutoff coarse-graining. Entries are mean errors in the Low (\(k<64\)), Mid (\(64\leq k\leq144\)), and High (\(k>144\)) physical-wavenumber ranges. Arrows give the relative increase from the Full model for the same metric and range.}
\label{tab:mno_ablation}
{
\footnotesize
\setlength{\tabcolsep}{3pt}
\begin{tabular}{lcccccc}
\toprule
& \multicolumn{3}{c}{\textbf{Velocity nRMSE}} & \multicolumn{3}{c}{\textbf{Vorticity MAE}} \\
\cmidrule(lr){2-4}\cmidrule(lr){5-7}
\textbf{Model} & \textbf{Low} & \textbf{Mid} & \textbf{High} & \textbf{Low} & \textbf{Mid} & \textbf{High} \\
\midrule
Full & \textbf{0.0122} & \textbf{0.1725} & \textbf{0.3440} & \textbf{2.8059} & \textbf{12.6611} & \textbf{10.2234} \\
w/o \(\bm{\mathcal{O}_s}\) & \(0.0207^{\scriptscriptstyle\uparrow69.9\%}\) & \(0.1914^{\scriptscriptstyle\uparrow11.0\%}\) & \(0.3485^{\scriptscriptstyle\uparrow1.3\%}\) & \(4.6619^{\scriptscriptstyle\uparrow66.1\%}\) & \(14.3753^{\scriptscriptstyle\uparrow13.5\%}\) & \(10.3058^{\scriptscriptstyle\uparrow0.8\%}\) \\
w/o \(\bm{\mathcal{O}_c}\) & \(0.0140^{\scriptscriptstyle\uparrow14.9\%}\) & \(0.2137^{\scriptscriptstyle\uparrow23.9\%}\) & \(0.3794^{\scriptscriptstyle\uparrow10.3\%}\) & \(3.0898^{\scriptscriptstyle\uparrow10.1\%}\) & \(15.0576^{\scriptscriptstyle\uparrow18.9\%}\) & \(11.2589^{\scriptscriptstyle\uparrow10.1\%}\) \\
\bottomrule
\end{tabular}
}
\end{table}

\FloatBarrier

The component-removal effects vary systematically across the three ranges (Supplementary Fig.~\ref{fig:mno_ablation} and Supplementary Table~\ref{tab:mno_ablation}). The Low range is most sensitive to removing \(\mathcal{O}_s\): velocity and vorticity errors increase by \(69.9\%\) and \(66.1\%\), versus \(14.9\%\) and \(10.1\%\) without \(\mathcal{O}_c\). The Mid and High ranges are more sensitive to removing \(\mathcal{O}_c\), with increases of \(23.9\%/18.9\%\) and \(10.3\%/10.1\%\), respectively, versus \(11.0\%/13.5\%\) and \(1.3\%/0.8\%\) without \(\mathcal{O}_s\). Thus, in FIT, \(\mathcal{O}_s\) is most influential at the largest represented scales through global coupling, and \(\mathcal{O}_c\) at intermediate and small scales, including velocity-gradient and rotational content. Both removals affect the full spectrum, and the full model is best at every evaluated wavenumber, indicating complementary and overlapping roles.

\noindent\textbf{Comparison with an alternative multiscale design.}
We replace every multiscale neural-operator block with a three-dimensional multiscale dense residual block derived from MS-ESRGAN~\cite{yousif2021highfidelity}. The replacement combines parallel convolutions with different receptive-field sizes, dense feature connections, and residual scaling. The temporal module, reconstruction head, loss, and training and evaluation settings are held fixed. The full model and the replacement contain $3.7$ M and $3.8$ M parameters, respectively, providing a comparison at closely matched capacity.

On FIT at $4\times$, the replacement increases overall nRMSE from $0.0615$ to $0.0739$ ($20.1\%$). The proposed spatial design also yields lower mean shell-wise velocity nRMSE in each of the Low, Mid, and High ranges (Supplementary Table~\ref{tab:mno_msesrgan_ablation}) and reduces the high-wavenumber energy underestimation of the replacement (Supplementary Fig.~\ref{fig:mno_msesrgan_ablation}). Together with the component-removal results, this comparison supports the value of combining global Fourier coupling with local convolutional refinement for recovering the multiscale content of this flow.

\begin{table}[htbp]
\centering
\caption{\textbf{Comparison of Ours and MS-ESRGAN spatial designs at closely matched capacity.}
Ours denotes the proposed multiscale neural operator; MS-ESRGAN denotes the three-dimensional multiscale replacement derived from MS-ESRGAN. Overall nRMSE is evaluated on the complete FIT \(4\times\) spectral-cutoff test fields. Velocity nRMSE entries are mean shell-wise errors in the Low (\(k<64\)), Mid (\(64\leq k\leq144\)), and High (\(k>144\)) ranges. The best result in each column is shown in \textbf{bold}.}
\label{tab:mno_msesrgan_ablation}
{
\small
\setlength{\tabcolsep}{9pt}
\begin{tabular}{lcccc}
\toprule
& \textbf{Overall} & \multicolumn{3}{c}{\textbf{Velocity nRMSE}} \\
\cmidrule(lr){2-2}\cmidrule(lr){3-5}
\textbf{Spatial design} & \textbf{nRMSE} & \textbf{Low} & \textbf{Mid} & \textbf{High} \\
\midrule
Ours & \textbf{0.0615} & \textbf{0.0122} & \textbf{0.1725} & \textbf{0.3440} \\
MS-ESRGAN & 0.0739 & 0.0139 & 0.2174 & 0.3803 \\
\bottomrule
\end{tabular}
}
\end{table}

\begin{figure}[htbp]
    \centering
    \includegraphics[width=0.9\textwidth]{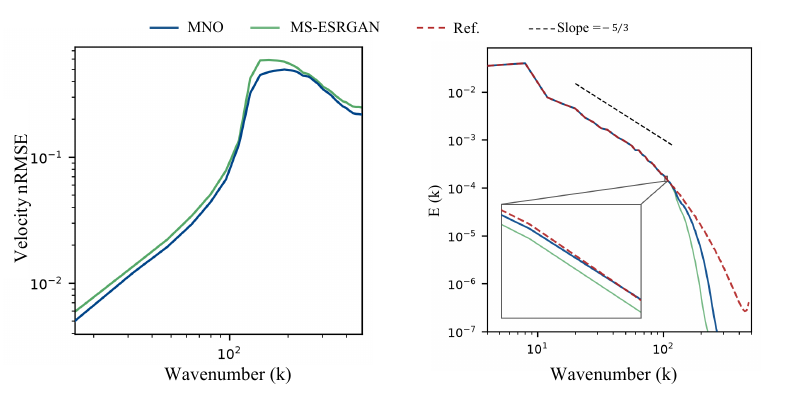}
    \vspace{-10pt}
    \caption{\textbf{Comparison of spatial modules on FIT at \(4\times\) with spectral-cutoff coarse-graining.}
    The proposed multiscale neural operator (MNO) is compared with an MS-ESRGAN-derived three-dimensional multiscale replacement (labelled MS-ESRGAN) at closely matched capacity. \textbf{Left,} shell-wise velocity nRMSE. \textbf{Right,} kinetic-energy spectra; the inset magnifies the high-wavenumber range. Ref. denotes the reference field, and the black dashed line indicates a slope of $-5/3$.}
    \label{fig:mno_msesrgan_ablation}
\end{figure}

\FloatBarrier

\clearpage
\subsection{Ablation and sensitivity study of the state space model}\label{ssec:temporal_ablation}\label{ssec:ssm_removal}\label{ssec:ssm_sensitivity}

\noindent\textbf{Contribution of state-space-model augmentation.} The full model (Ours) is compared with a model trained without state space model (SSM) augmentation (Ours (w/o SSM)) on FIT at $4\times$ and $8\times$.

Supplementary Table~\ref{tab:ssm_removal} reports the direct ablation alongside SwinIR, DiffSR, and PhySR. Relative to Ours (w/o SSM), the full model reduces nRMSE by $14.36\%$ and tRMSE by $16.05\%$ at $4\times$; the corresponding reductions at $8\times$ are $12.87\%$ and $6.52\%$. Percentages are calculated from the unrounded measurements. The full model also achieves the lowest values of both metrics among all five evaluated models at both factors. These results support the contribution of SSM augmentation to field reconstruction and recovery of reference temporal increments in the tested FIT settings.

\begin{table}[htbp]
\centering
\caption{\textbf{SSM-removal ablation and baseline comparisons on FIT at $4\times$ and $8\times$.} Inputs are box-filtered. nRMSE measures field reconstruction error and tRMSE measures error in frame-to-frame increments relative to the reference; lower values are preferred. Values are rounded to four decimal places, and the best value in each column is bold.}
\label{tab:ssm_removal}
{\small
\setlength{\tabcolsep}{5pt}
\renewcommand{\arraystretch}{1.15}
\begin{tabular}{lcccc}
\toprule
& \multicolumn{2}{c}{\textbf{FIT $4\times$}} & \multicolumn{2}{c}{\textbf{FIT $8\times$}} \\
\cmidrule(lr){2-3}\cmidrule(lr){4-5}
\textbf{Model} & \textbf{nRMSE} $\downarrow$ & \textbf{tRMSE} $\downarrow$ & \textbf{nRMSE} $\downarrow$ & \textbf{tRMSE} $\downarrow$ \\
\midrule
SwinIR & 0.0421 & 0.0052 & 0.1499 & 0.0103 \\
DiffSR & 0.0423 & 0.0095 & 0.1440 & 0.0244 \\
PhySR & 0.0476 & 0.0060 & 0.1486 & 0.0106 \\
\midrule
Ours (w/o SSM) & 0.0340 & 0.0049 & 0.1371 & 0.0101 \\
Ours & \textbf{0.0291} & \textbf{0.0041} & \textbf{0.1195} & \textbf{0.0094} \\
\bottomrule
\end{tabular}
}
\end{table}

Supplementary Fig.~\ref{fig:ssm_removal_tx} shows that Ours closely matches the reference structures and preserves their smooth evolution over time. The model without SSM and the other baselines exhibit varying degrees of local blurring, deformation, and reduced temporal smoothness. Together with the lower tRMSE, these observations support the contribution of SSM augmentation to temporal fidelity.

\begin{figure}[htbp]
\centering
\includegraphics[width=\textwidth]{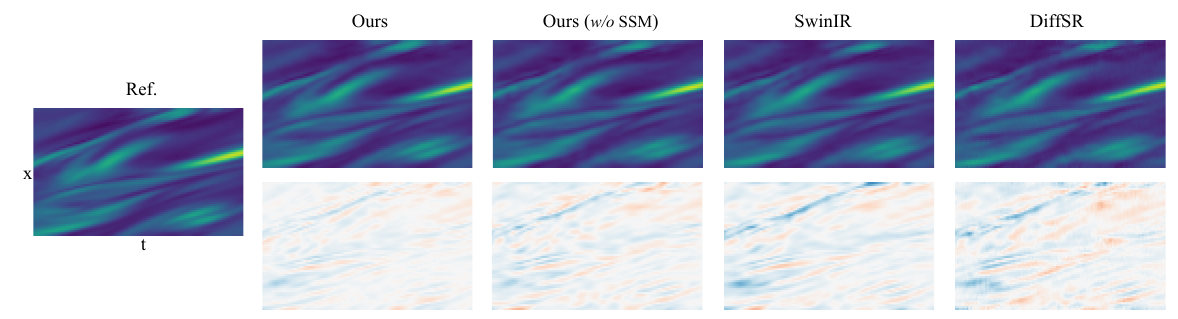}
\caption{\textbf{Time--space visualization of the SSM-removal ablation on FIT.} The left panel shows the reference; the upper row compares Ours, Ours (w/o SSM), SwinIR, and DiffSR, and the lower row shows the corresponding reconstruction errors. Time $t$ runs horizontally and position $x$ vertically.}
\label{fig:ssm_removal_tx}
\end{figure}

\FloatBarrier

\medskip
\noindent\textbf{Comparison with alternative temporal modules.} We compare the proposed state space model with two temporal alternatives, ConvLSTM~\cite{shi2015convolutional} and a causal Transformer decoder~\cite{vaswani2017attention}, on FIT at $4\times$ upscaling. Only the temporal module is replaced; the remaining architecture, data partitions, sequence length, preprocessing, loss, and training and evaluation settings are identical. Both the temporal-module and total parameter counts are closely matched, with $3.7$ M parameters for the state space model variant and $3.8$ M for each alternative. ConvLSTM represents gated convolutional recurrence, and the causal Transformer represents temporal aggregation through masked self-attention.

\begin{table}[htbp]
\centering
\caption{\textbf{Comparison of temporal models on FIT at $4\times$ with box-filtered inputs at closely matched parameter counts.}
Causal Transformer denotes the Transformer decoder baseline; parameter counts refer to the complete networks. Metrics follow Section~\ref{ssec:metrics} and Supplementary Note~\ref{ap:metrics}; arrows indicate the preferred direction, and the best value for each metric is in \textbf{bold}.}
\label{tab:temporal_ablation}
{
\small
\setlength{\tabcolsep}{4pt}
\renewcommand{\arraystretch}{1.15}
\begin{tabular}{lcccccc}
\toprule
\textbf{Temporal model} & \textbf{Params (M)} & \textbf{nRMSE} $\downarrow$ & \textbf{SSIM} $\uparrow$ & \textbf{PSNR} $\uparrow$ & \textbf{tRMSE} $\downarrow$ & \textbf{fRMSE} $\downarrow$ \\
\midrule
State space model & 3.7 & \textbf{0.0291} & \textbf{0.9834} & \textbf{55.09} & \textbf{0.0041} & \textbf{7.22} \\
ConvLSTM & 3.8 & 0.0306 & 0.9819 & 54.60 & 0.0044 & 7.59 \\
Causal Transformer & 3.8 & 0.0314 & 0.9809 & 54.36 & 0.0044 & 7.79 \\
\bottomrule
\end{tabular}
}
\end{table}

\FloatBarrier

The state space model achieves the lowest nRMSE, tRMSE, and fRMSE and the highest SSIM and PSNR (Supplementary Table~\ref{tab:temporal_ablation}). Relative to ConvLSTM and the causal Transformer, respectively, nRMSE decreases by $4.95\%$ and $7.44\%$, tRMSE by $5.10\%$ and $6.28\%$, and fRMSE by $4.87\%$ and $7.32\%$. PSNR increases by $0.49$ and $0.73$ dB. The temporal-increment improvement is therefore accompanied by more accurate spatial fields and Fourier representations.

The state space formulation in Section~\ref{ssec:ssm} maintains temporal memory through a recurrent latent state at each Eulerian location. Input-conditioned $\bm{B}_t$ and $\bm{C}_t$ control how local features enter the state and how the accumulated history is read out, and the selection gate further modulates the output. This mechanism integrates temporal context, namely flow features accumulated from preceding time steps, into the reconstruction of the current field. The joint reductions in tRMSE and nRMSE are consistent with more effective use of this temporal context, supporting improvements in both temporal fidelity and field reconstruction accuracy.

\begin{table}[htbp]
\centering
\caption{\textbf{Sensitivity of SSM reconstruction to state dimension and feature width on FIT and MHDT at $4\times$.} $d_s$ denotes state dimension and $d_f$ feature width. Params and GFLOPs count the SSM module alone. All tRMSE values are reported in units of $10^{-3}$ and rounded to two decimal places; nRMSE is unscaled.}
\label{tab:ssm_sensitivity}
{\small
\setlength{\tabcolsep}{4pt}
\renewcommand{\arraystretch}{1.15}
\begin{tabular}{rrrrcccc}
\toprule
& & & & \multicolumn{2}{c}{\textbf{FIT}} & \multicolumn{2}{c}{\textbf{MHDT}} \\
\cmidrule(lr){5-6}\cmidrule(lr){7-8}
$d_s$ & $d_f$ & \textbf{Params (M)} & \textbf{GFLOPs} & \textbf{nRMSE} $\downarrow$ & \textbf{tRMSE} $\downarrow$ & \textbf{nRMSE} $\downarrow$ & \textbf{tRMSE} $\downarrow$ \\
\midrule
8 & 64 & 0.968 & 63.468 & {0.0295} & {4.23} & {0.0613} & {0.90} \\
16 & 64 & 1.034 & 67.864 & 0.0291 & 4.14 & 0.0603 & 0.88 \\
32 & 64 & 1.166 & 76.655 & {0.0290} & {4.11} & {0.0599} & {0.87} \\
16 & 32 & 0.484 & 31.789 & {0.0305} & {4.33} & {0.0634} & {0.92} \\
16 & 128 & 2.329 & 152.899 & {0.0287} & {4.07} & {0.0593} & {0.86} \\
\bottomrule
\end{tabular}
}
\end{table}
\noindent\textbf{Sensitivity to state dimension and feature width.} The SSM has two capacity parameters: feature width $d_f$ (\texttt{dimension}) and state dimension $d_s$ (\texttt{state}), with default $(d_f,d_s)=(64,16)$. Feature width sets the number of learned channels representing local flow information at each Eulerian location, while state dimension sets the capacity of the recurrent representation through which past features influence the current reconstruction. In a channelwise selective SSM, $d_s$ counts state components per internal channel~\cite{gu2023mamba}; the learned transition and input-dependent selection determine the effective memory timescales. Together, $d_f$ and $d_s$ control the richness of the flow features represented and their temporal history, with capacity requirements depending on the spatial complexity and temporal correlations of the flow. 

We vary $d_s\in\{8,16,32\}$ at $d_f=64$ and $d_f\in\{32,64,128\}$ at $d_s=16$, giving five configurations on FIT and MHDT. All other architecture, data, loss, and training and evaluation settings are held fixed within each task. Supplementary Table~\ref{tab:ssm_sensitivity} reports the reconstruction errors, parameter counts, and GFLOPs for all five configurations.

Relative to the default $(d_f,d_s)=(64,16)$, doubling $d_s$ to 32 increases SSM parameters by approximately $12.8\%$ and GFLOPs by $13.0\%$. Doubling $d_f$ to 128 increases both to approximately $2.25$ times the default, whereas halving $d_f$ to 32 reduces both by approximately $53.2\%$.

Reducing feature width from $d_f=64$ to 32 increases nRMSE by $4.8\%$ on FIT and $5.1\%$ on MHDT, accompanied by higher tRMSE, while reducing SSM GFLOPs by $53.2\%$. This joint deterioration is consistent with a narrower feature representation limiting both instantaneous field reconstruction and the representation of evolving flow structures. Halving state dimension from $d_s=16$ to 8 produces smaller nRMSE increases of $1.4\%$ and $1.7\%$, respectively, with a larger relative increase in tRMSE on both datasets. This pattern is consistent with reduced history-representation capacity affecting the fidelity of temporal increments more strongly than instantaneous field accuracy.
Increasing either capacity yields modest improvements. Doubling $d_s$ reduces nRMSE by $0.3\%$ on FIT and $0.7\%$ on MHDT, with larger relative reductions in tRMSE, at $13.0\%$ higher SSM GFLOPs. Doubling $d_f$ improves nRMSE by $1.4\%$ and $1.7\%$, respectively, and also reduces tRMSE, while increasing SSM GFLOPs to $2.25$ times the default. Both FIT and MHDT show a similar accuracy--cost trade-off across the tested configurations.

Together, the modest gains from capacity expansion and the larger penalties from reducing feature width indicate diminishing returns near the default configuration. These trends support $(d_f,d_s)=(64,16)$ as a practical accuracy--cost compromise for the two tested $4\times$ tasks; $d_f=32$ offers a lower-cost alternative with an approximately $5\%$ accuracy penalty.

\clearpage
\subsection{Recovering multiscale spatial structure from coarse-grained data}
\label{ssec:spatial_structure_appendix}

\begin{figure}[htbp]
    \centering
    \includegraphics[width=1\textwidth]{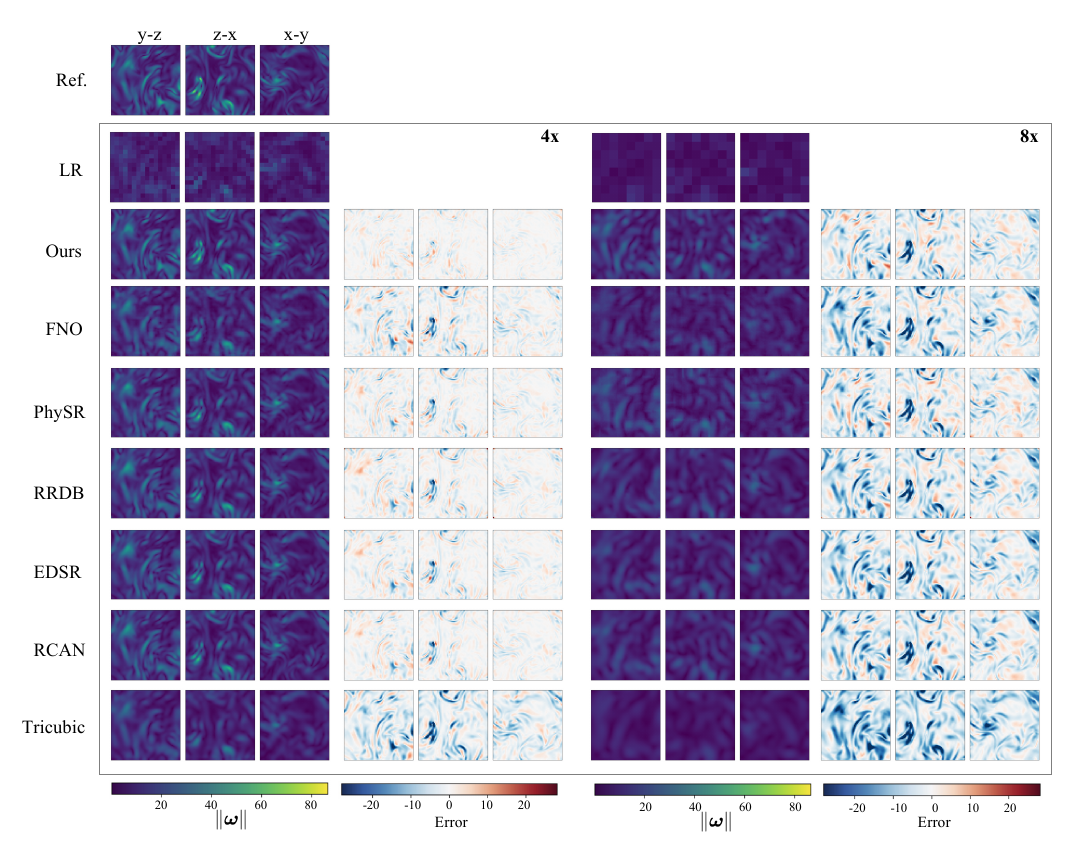}
    \caption{\textbf{{Vorticity-magnitude} reconstruction and error visualization for FIT.}
    Comparison of reconstructed vorticity magnitude $\|\bm{\omega}\|=\|\nabla\times\bm{u}\|$ and errors among different super-resolution models under box-filtered coarse-graining.
    Each column block corresponds to $4\times$ (left) and $8\times$ (right) upscaling.
    For each case, three orthogonal slices (y--z, z--x, x--y) are shown for the reference high-resolution field (Ref.), the low-resolution input (LR), the reconstructed fields from our method, FNO, PhySR, RRDB, EDSR, and RCAN, and the tricubic baseline.
    The right panels visualize the corresponding error maps (Prediction -- Reference) using a symmetric color scale.
    All samples are from held-out test data; color bars denote vorticity magnitude (left) and reconstruction error (right) under identical dynamic ranges across methods.
    }
    \label{fig:spatial_fidelity_sp1}
\end{figure}

\begin{figure}[htbp]
    \centering
    \includegraphics[width=1\textwidth]{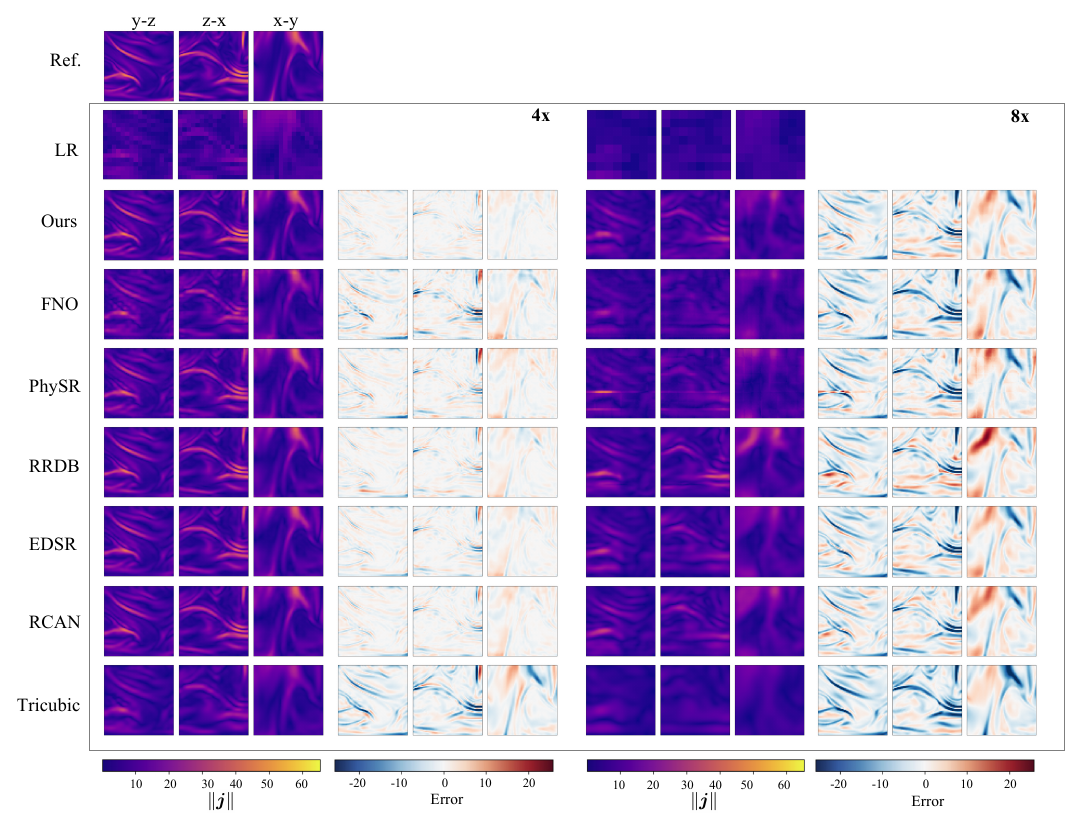}
    \caption{\textbf{Reconstruction of the magnetic field curl magnitude for MHDT.}
    Comparison of the reconstructed current magnitude $\|\bm{j}\|=\|\nabla\times\bm{b}\|$ field and corresponding voxelwise errors under the box-filtered coarse-graining.
    Columns correspond to $4\times$ (left) and $8\times$ (right) super-resolution.
    Each group shows three orthogonal slices (y--z, z--x, x--y) for the reference high-resolution field (Ref.), the low-resolution input (LR), and reconstructions by our method and baselines (FNO, PhySR, RRDB, EDSR, RCAN, and tricubic interpolation).
    The right panels visualize error fields (Prediction -- Reference) with symmetric color scaling.
    Color bars denote $\|\bm{j}\|$ (left) and reconstruction error (right).
    }
    \label{fig:spatial_fidelity_sp2}
\end{figure}

\begin{figure}[htbp]
    \centering
    \includegraphics[width=1\textwidth]{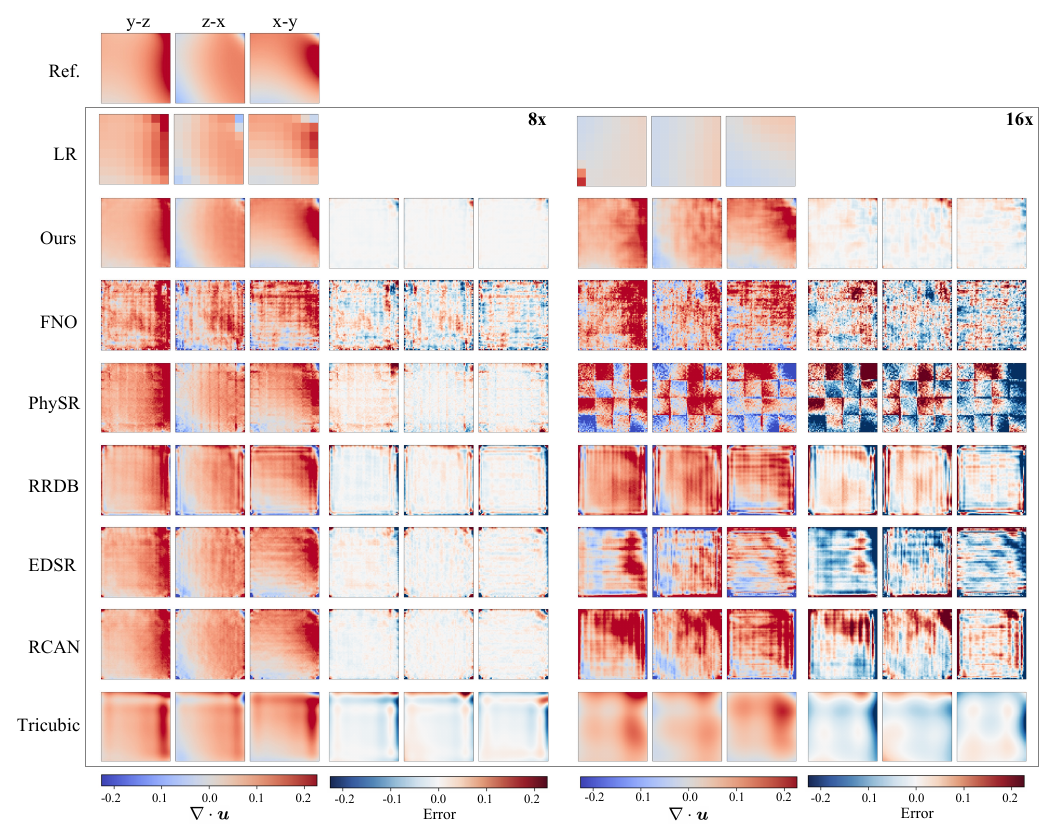}
    \caption{\textbf{Velocity divergence reconstruction and error visualization for HBDT.}
    Comparison of reconstructed and reference velocity divergence fields $\nabla\cdot\bm{u}$ under the Favre-filtered coarse-graining for $8\times$ (left) and $16\times$ (right) super-resolution reconstruction.
    Each column block presents three orthogonal slices (y--z, z--x, x--y) of the reference high-resolution field (Ref.), low-resolution input (LR), and reconstructions from our method, FNO, PhySR, RRDB, EDSR, RCAN, and tricubic interpolation.
    The right panels visualize error fields (Prediction -- Reference) with symmetric color scaling.
    Color bars indicate $\nabla\cdot\bm{u}$ (left) and corresponding reconstruction error (right).
    }
    \label{fig:spatial_fidelity_sp3}
\end{figure}

\begin{figure}[!p]
    \centering
    \includegraphics[width=1\textwidth]{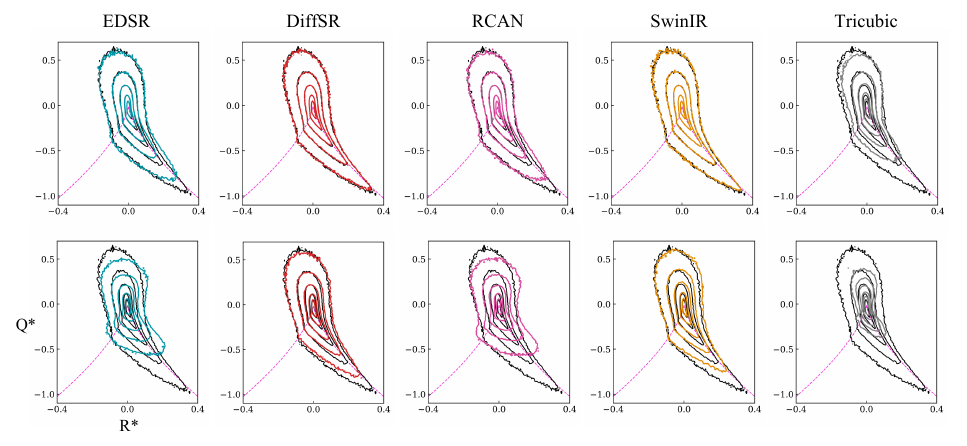}
    \caption{\textbf{Q--R joint probability distributions.} Isocontour plots of joint PDFs of normalized $Q^*$ and $R^*$ for FIT at $4\times$ (top) and $8\times$ (bottom) super-resolution reconstruction. Columns show EDSR, DiffSR, RCAN, SwinIR, and tricubic interpolation, from left to right. Black contours denote the reference; colored contours (gray for tricubic interpolation) denote the reconstructed fields. The magenta dashed curve marks $27(R^*)^2+4(Q^*)^3=0$. Evaluation details are given in Supplementary Note~\ref{ssec:spatial_structure_appendix}.}
    \label{fig:ap_qr_jpdf}
\par\medskip

    \includegraphics[width=0.88\textwidth]{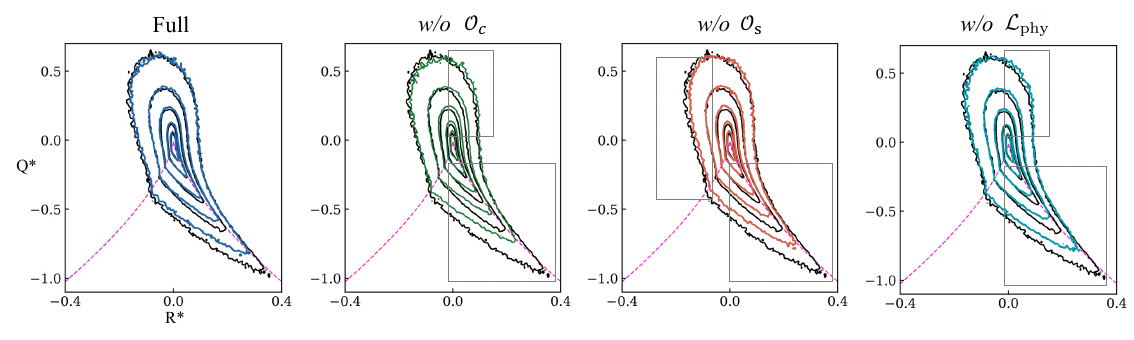}
    \caption{\textbf{Component contributions to Q--R joint distributions on FIT at $8\times$ with spectral-cutoff coarse-graining.}
    Left to right: the full model, removal of the local convolutional operator $\mathcal{O}_c$, removal of the global spectral operator $\mathcal{O}_s$, and removal of the physical loss $\mathcal{L}_{\mathrm{phy}}$, with the remaining training settings held fixed. Black contours show the reference and colored contours the corresponding reconstruction; gray rectangles highlight regions of contour displacement. The dashed curve is $27(R^*)^2+4(Q^*)^3=0$. Evaluation details are given in Supplementary Note~\ref{ssec:spatial_structure_appendix}.}
    \label{fig:ap_qr_ablation}
\end{figure}

\FloatBarrier

\noindent\textbf{Evaluation of Q--R statistics.}
For all Q--R comparisons, $Q^*=Q/\langle S_{ij}S_{ij}\rangle_{\mathrm{DNS}}$ and $R^*=R/\langle S_{ij}S_{ij}\rangle_{\mathrm{DNS}}^{3/2}$, where the average is taken over the evaluated DNS reference samples and the same scale is used for every reconstruction. The statistics are computed directly from the DNS and reconstructed velocity fields on the HR grid without additional spatial filtering; the coarse-graining specified for each experiment defines its LR inputs. Joint PDFs are estimated using unsmoothed two-dimensional histograms on common bin edges. Contours show $p/p_{\mathrm{DNS,max}}=0.003,0.01,0.03,0.1,0.3$, so the DNS peak sets common absolute density thresholds for all methods. The comparison assesses gradient configurations represented by the displayed contours; rarer events require additional tail sampling.

\noindent\textbf{Physical interpretation and component contributions in the Q--R plane.}
For incompressible flow, let $A_{ij}=\partial u_i/\partial x_j$, $\bm{S}=(\bm{A}+\bm{A}^{\mathsf T})/2$, and $\bm{\Omega}=(\bm{A}-\bm{A}^{\mathsf T})/2$. These symmetric and antisymmetric parts describe local deformation and rotation, respectively. The invariants can be written as~\cite{chong1990classification,meneveau2011lagrangian}
\begin{equation}
\begin{aligned}
 Q&=-\tfrac12\operatorname{tr}(\bm{A}^2)
   =\tfrac12\left(\|\bm{\Omega}\|_F^2-\|\bm{S}\|_F^2\right),\\
 R&=-\tfrac13\operatorname{tr}(\bm{A}^3)
   =-\tfrac13\operatorname{tr}(\bm{S}^3)
    -\tfrac14\bm{\omega}^{\mathsf T}\bm{S}\bm{\omega},
\end{aligned}
\label{eq:qr_strain_rotation}
\end{equation}
where $\bm{\omega}=\nabla\times\bm{u}$ and $\|\cdot\|_F^2$ is the sum of squared tensor entries. Thus $Q$ compares rotation and strain intensity. The first contribution to $R$ depends on the principal stretching and compression rates; the second depends on the strength and alignment of vorticity relative to strain. Positive $\bm{\omega}^{\mathsf T}\bm{S}\bm{\omega}$ describes vortex stretching and contributes negatively to $R$. The sign of $R$ therefore reflects the balance of both terms.

The joint PDF measures the frequency of local gradient configurations. Its teardrop asymmetry is associated with the coupled amplification of strain and vorticity, whose evolution is regulated by pressure and viscous effects~\cite{chevillard2006lagrangian,meneveau2011lagrangian}. The positive-$Q$ region represents rotation-dominated motion; the lower-right extension contains strong strain-dominated states. For incompressible flow, the discriminant $27R^2+4Q^3=0$ separates different velocity-gradient eigenvalue structures~\cite{chong1990classification}. Its positive-$R$ branch includes axisymmetric compression: extension in two directions and compression in the third, flattening a fluid element while conserving volume. The distribution thus captures preferred combinations of local deformation and rotation, including motions relevant to dissipation and vortex stretching.

In the FIT $8\times$ box-filtered comparison (Fig.~\ref{fig:spatial_fidelity_2}c), RRDB's displaced positive-$Q^*$ contours indicate altered combinations of rotation, strain, and their coupling, while its shortened lower-right extension indicates deficient recovery of the strongly strain-dominated states represented by the reference tail. Spatial differentiation emphasizes fine-scale velocity variations, and $Q$ and $R$ depend quadratically and cubically on gradients. Their joint statistics also depend on correlations among gradient components: an energy spectrum constrains energy at each wavenumber, while local gradient products additionally depend on phase and component relationships. Consequently, spectral agreement alone leaves room for Q--R distortions. RRDB builds spatial context through successive residual dense convolutions~\cite{wang2018esrgan}; the observed discrepancies are consistent with incomplete recovery of gradient amplitudes and their relationships under stronger coarse-graining. Q--R contours reveal the combined statistical effect; separating the contributions would require strain and vorticity-alignment statistics.

\begin{figure}[!tb]
    \centering
    \includegraphics[width=\textwidth]{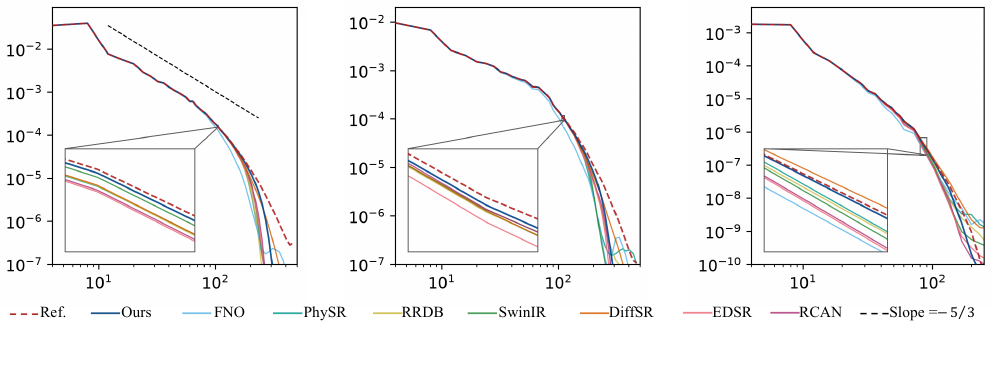}
    \caption{\textbf{Turbulent energy spectra under {box/Favre coarse-graining}.} Panels show FIT ($4\times$), MHDT ($4\times$), and HBDT ($8\times$) from left to right. Horizontal and vertical axes denote wavenumber $k$ and energy spectrum $E(k)$, respectively. Reconstructions from our method, FNO, PhySR, RRDB, SwinIR, DiffSR, EDSR, and RCAN are compared with the reference (Ref.). For MHDT, the spectrum is the sum of kinetic and magnetic contributions. Insets magnify the high-wavenumber range to highlight small-scale recovery. The black dashed line indicates a slope of $-5/3$.}
    \label{fig:energy_spectrum}
\end{figure}

Supplementary Fig.~\ref{fig:ap_qr_ablation} compares the full model with the three removals on FIT at $8\times$ with spectral-cutoff coarse-graining, keeping the remaining training settings fixed. This experiment examines the mechanisms within the proposed model and complements the $8\times$ box-filtered baseline comparison in Fig.~\ref{fig:spatial_fidelity_2}c. The highlighted regions reveal distinct changes in the Q--R distribution. Removing the local operator $\mathcal{O}_c$ or the physical loss $\mathcal{L}_{\mathrm{phy}}$ shifts the right flank of the rotation-dominated contours ($Q^*>0$) towards larger $R^*$. Removing the spectral operator $\mathcal{O}_s$ contracts the negative-$R^*$ side of the distribution across both rotation- and strain-dominated regions. All three removals also degrade the strain-dominated lower-right branch ($Q^*<0$, $R^*>0$), with several intermediate contours retracting relative to the reference and showing larger departures than in the full model. The outermost branch remains extended after removing $\mathcal{O}_c$, so the loss of extension depends on the contour level. Together, these changes indicate that the components contribute to recovering both the characteristic strain-dominated extension and the distribution of strain--vorticity configurations, with different effects across the Q--R plane. The full model retains residual deviations in the outer contours.

The local operator $\mathcal{O}_c$ refines neighboring velocity differences, which determine both strain and rotation. The spectral operator $\mathcal{O}_s$ couples spatially separated features~\cite{guibas2022efficient}, supplying context for the recovered local field. Their contributions are consistent with the separate FIT $4\times$ spectral-cutoff velocity/vorticity ablation in Supplementary Note~\ref{ssec:mno_ablation}. The physical loss encourages local volume conservation through its divergence penalty and couples temporal changes, advection, pressure gradients, and viscosity through its scale-selective momentum term (Eqs.~\eqref{eq:loss_p} and~\eqref{eq:scale_selective_momentum}). These constraints regularize relationships among recovered derivatives. The contour changes therefore support the roles of local refinement, global context, and physical regularization in recovering strain--rotation statistics. Each acts on the velocity field and can affect both strain and rotation; the component removals establish their contributions within the proposed model under the tested spectral-cutoff setting.

\noindent\textbf{Spectral ranges and plotting conventions.}
Energy spectra are plotted against physical wavenumber \(k\). The FIT and MHDT reference simulations use phase-shift dealiasing with \(2\sqrt{2}/3\) spectral truncation, giving a maximum retained wavenumber of approximately \(482\). For HBDT, the \(1024^3\) grid on \([0,2\pi)^3\) gives an axial Nyquist wavenumber of \(512\). We display the HBDT spectra up to approximately \(k=256\), as the reconstructed spectra develop nonphysical distortions at higher wavenumbers. The same display range is used for the reference and all compared methods.

\begin{figure}[htbp]
    \centering
    \includegraphics[width=0.65\textwidth]{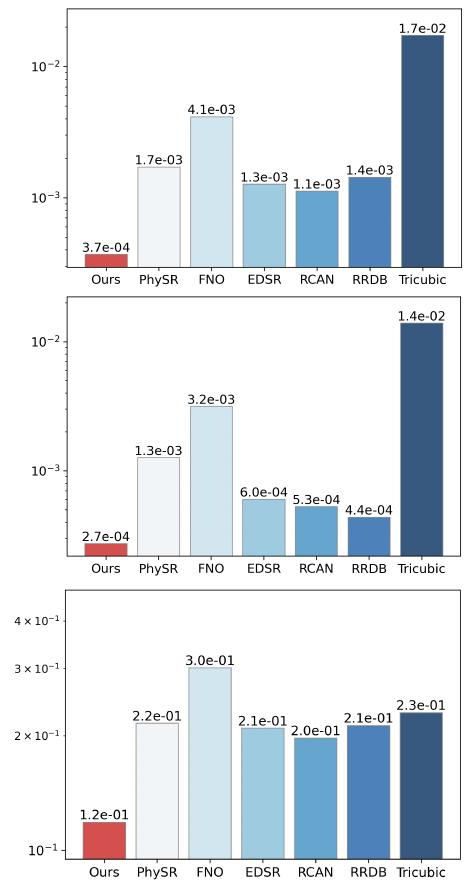}
    \caption{\textbf{Quantitative distributional similarity across turbulent regimes.} Shown are the Jensen--Shannon (JS) divergences between the reconstructed and reference PDFs of three canonical turbulent diagnostics: \textbf{top}, vorticity magnitude $\|\bm{\omega}\|$ in FIT; \textbf{middle}, current magnitude $\|\bm{j}\|$ in MHDT; \textbf{bottom}, velocity divergence $\nabla\cdot\bm{u}$ in HBDT; these correspond to Fig.~\ref{fig:spatial_fidelity_2}b.}
    \label{fig:jsd}
\end{figure}

\clearpage
\subsection{Atmospheric downscaling on ERA5} \label{ap:era5}
Cross-level structure in atmospheric winds is assessed via bulk vertical wind shear. In atmospheric dynamics, bulk shear, commonly expressed as the wind-vector difference between two pressure levels, is associated with dynamically active features such as jet structures and frontal zones and summarizes how horizontal winds vary with height~\cite{Barton2025soil}. Based on this, two shear-based metrics are introduced to evaluate both amplitude accuracy and the spatial organization of strong cross-level shear in the downscaled fields.
Let $\bm{V}=(u,v)$ denote the horizontal wind field. The magnitude of the wind-vector difference between two pressure levels $p_a$ and $p_b$ is defined as
\begin{equation} \label{eq:s_def_compact}
S := \left\|\bm{V}(\cdot,p_a)-\bm{V}(\cdot,p_b)\right\|_2,
\end{equation}
where the norm is taken over the wind-vector components at each horizontal grid point. Here, $S$ has units of m/s and serves as a bulk-difference proxy of shear magnitude. Following common practice in diagnosing the vertical distribution of environmental shear~\cite{jacob2024tropical}, the wind-vector difference between 850 and 500~hPa is used as a representative shallow-to-midtropospheric shear measure; $\hat{S}$ is computed analogously from $\hat{\bm{V}}$.
The resulting vertical  metrics are denoted as vRMSE and vCoh. The amplitude error of cross-level shear is quantified by
\begin{equation}
\label{eq:vrmse_compact}
\mathrm{vRMSE} := \sqrt{\frac{1}{|\mathcal{D}|}\sum\nolimits_{i \in \mathcal{D}} (\hat{S}_i - S_i)^2},
\end{equation}
where $\mathcal{D}$ denotes valid horizontal grid points. To emphasize the spatial organization of dynamically active, high-shear regions, we define a vertical coherence score as the Pearson correlation restricted to strong-shear points:
\begin{equation}
\label{eq:vcoh_compact}
\mathrm{vCoh} := \mathrm{Corr}_{\Omega_q}\!(\hat{S},S),
\qquad
\Omega_q := \left\{i\in\mathcal{D}: S_i > Q_q(S)\right\},
\end{equation}
where $Q_q(S)$ is the $q$-th quantile of $S$ over $\mathcal{D}$ and $\mathrm{Corr}_{\Omega_q}$ denotes the Pearson correlation computed on $\Omega_q$. In practice, $q=0.9$ is used, focusing on the top 10\% strongest-shear points (defined by the reference), to assess whether strong-shear structures are reconstructed in the correct locations. Metrics are computed per time step and averaged over the evaluation set.

\begin{table}[htbp]
\centering
\caption{
\textbf{Cross-level wind diagnostics for ERA5 downscaling.}
Performance is evaluated using vCoh and vRMSE from Supplementary Eqs.~\eqref{eq:s_def_compact}--\eqref{eq:vcoh_compact}. Higher vCoh and lower vRMSE indicate closer agreement in the spatial organization and amplitude, respectively, of the 850--500~hPa bulk-shear field. The best result is in \textbf{bold}.
}
\label{tab:vertical}
\begin{tabular}{clllll}
\toprule
Factors & Metrics & Ours & SwinIR & FNO & Bicubic \\
\midrule
\multirow{2}{*}{$4\times$} 
& vCoh  & \textbf{0.9925} &  {0.9906} & 0.9871 & 0.9743 \\
& vRMSE & \textbf{0.4672} &  {0.5167} & 0.5925 & 0.8482 \\
\midrule
\multirow{2}{*}{$8\times$} 
& vCoh  & \textbf{0.9533} &  {0.9442} & 0.9259 & 0.8745 \\
& vRMSE & \textbf{1.1340} &  {1.2582} & 1.4174 & 1.8646 \\
\bottomrule
\end{tabular}
\end{table}

\FloatBarrier
\subsection{Training-seed uncertainty}
\label{ssec:seed_uncertainty}
Training-seed uncertainty is quantified using independent seeds 42, 43, and 44 ($n=3$). The analysis covers all eight learned methods on FIT at $4\times$ upscaling with box-filtered inputs and Ours, FNO, and SwinIR on ERA5 at both $4\times$ and $8\times$. The predefined data partitions, model-specific hyperparameters, training budgets, and evaluation protocols are held fixed across seeds, so the resulting variation reflects stochastic model training. Supplementary Tables~\ref{tab:seed_fit} and~\ref{tab:seed_era5} report the mean and sample standard deviation across the three runs. A standard deviation displayed as $0.0000$ is below the four-decimal reporting precision.

The proposed method achieves the best mean for every reported metric in all three settings. On FIT, its nRMSE is $0.0292\,{\scriptscriptstyle\pm 0.0001}$, compared with $0.0410\,{\scriptscriptstyle\pm 0.0006}$ for RCAN, the closest baseline for this metric; its tRMSE is $0.0042\,{\scriptscriptstyle\pm 0.0000}$, compared with $0.0053\,{\scriptscriptstyle\pm 0.0000}$ for SwinIR. On ERA5, its MSE is $0.0762\,{\scriptscriptstyle\pm 0.0011}$ at $4\times$ and $0.4451\,{\scriptscriptstyle\pm 0.0007}$ at $8\times$, compared with $0.0947\,{\scriptscriptstyle\pm 0.0005}$ and $0.5333\,{\scriptscriptstyle\pm 0.0006}$, respectively, for SwinIR. The sample standard deviations are small relative to these between-method differences, demonstrating stable performance under training randomness.

\begin{table*}[htbp]
\centering
\caption{\textbf{Training-seed uncertainty on FIT at $4\times$ with box-filtered inputs.} Values are mean $\pm$ sample standard deviation over independent training seeds 42, 43, and 44 ($n=3$). Arrows indicate the preferred direction. A standard deviation displayed as $0.0000$ is below the reporting precision. The best mean is shown in \textbf{bold}.}
\label{tab:seed_fit}
\resizebox{0.85\textwidth}{!}{%
\begin{tabular}{lccccc}
\toprule
Model & SSIM $\uparrow$ & nRMSE $\downarrow$ & PSNR $\uparrow$ & tRMSE $\downarrow$ & fRMSE $\downarrow$ \\
\midrule
Ours     & \(\mathbf{0.9835}\,{\scriptscriptstyle\pm 0.0001}\) & \(\mathbf{0.0292}\,{\scriptscriptstyle\pm 0.0001}\) & \(\mathbf{55.0727}\,{\scriptscriptstyle\pm 0.0178}\) & \(\mathbf{0.0042}\,{\scriptscriptstyle\pm 0.0000}\) & \(\mathbf{7.2528}\,{\scriptscriptstyle\pm 0.0225}\) \\
FNO      & \(0.9163\,{\scriptscriptstyle\pm 0.0040}\) & \(0.0704\,{\scriptscriptstyle\pm 0.0022}\) & \(46.9694\,{\scriptscriptstyle\pm 0.3089}\) & \(0.0072\,{\scriptscriptstyle\pm 0.0004}\) & \(16.7646\,{\scriptscriptstyle\pm 0.4352}\) \\
PhySR    & \(0.9585\,{\scriptscriptstyle\pm 0.0001}\) & \(0.0477\,{\scriptscriptstyle\pm 0.0001}\) & \(50.6396\,{\scriptscriptstyle\pm 0.0175}\) & \(0.0062\,{\scriptscriptstyle\pm 0.0002}\) & \(11.6351\,{\scriptscriptstyle\pm 0.0106}\) \\
RCAN     & \(0.9674\,{\scriptscriptstyle\pm 0.0007}\) & \(0.0410\,{\scriptscriptstyle\pm 0.0006}\) & \(52.0477\,{\scriptscriptstyle\pm 0.1220}\) & \(0.0064\,{\scriptscriptstyle\pm 0.0001}\) & \(10.0879\,{\scriptscriptstyle\pm 0.1354}\) \\
EDSR     & \(0.9657\,{\scriptscriptstyle\pm 0.0002}\) & \(0.0422\,{\scriptscriptstyle\pm 0.0001}\) & \(51.7950\,{\scriptscriptstyle\pm 0.0221}\) & \(0.0065\,{\scriptscriptstyle\pm 0.0001}\) & \(10.3804\,{\scriptscriptstyle\pm 0.0256}\) \\
RRDB     & \(0.9641\,{\scriptscriptstyle\pm 0.0028}\) & \(0.0434\,{\scriptscriptstyle\pm 0.0022}\) & \(51.4815\,{\scriptscriptstyle\pm 0.5152}\) & \(0.0068\,{\scriptscriptstyle\pm 0.0004}\) & \(10.6929\,{\scriptscriptstyle\pm 0.5478}\) \\
SwinIR & \(0.9685\,{\scriptscriptstyle\pm 0.0004}\) & \(0.0424\,{\scriptscriptstyle\pm 0.0003}\) & \(51.3854\,{\scriptscriptstyle\pm 0.0713}\) & \(0.0053\,{\scriptscriptstyle\pm 0.0000}\) & \(10.2953\,{\scriptscriptstyle\pm 0.0671}\) \\
DiffSR   & \(0.9652\,{\scriptscriptstyle\pm 0.0004}\) & \(0.0421\,{\scriptscriptstyle\pm 0.0003}\) & \(51.6693\,{\scriptscriptstyle\pm 0.0687}\) & \(0.0096\,{\scriptscriptstyle\pm 0.0003}\) & \(10.2910\,{\scriptscriptstyle\pm 0.0774}\) \\
\bottomrule
\end{tabular}%
}
\end{table*}

\begin{table*}[htbp]
\centering
\caption{\textbf{Training-seed uncertainty on ERA5.} Values are mean $\pm$ sample standard deviation over independent training seeds 42, 43, and 44 ($n=3$). The best mean within each factor is shown in \textbf{bold}.}
\label{tab:seed_era5}
\resizebox{0.85\textwidth}{!}{%
\begin{tabular}{clccccc}
\toprule
Factor & Model & MAE $\downarrow$ & MSE $\downarrow$ & tRMSE $\downarrow$ & vCoh $\uparrow$ & vRMSE $\downarrow$ \\
\midrule
\multirow{3}{*}{\(4\times\)}
& Ours & \(\mathbf{0.1202}\,{\scriptscriptstyle\pm 0.0007}\) & \(\mathbf{0.0762}\,{\scriptscriptstyle\pm 0.0011}\) & \(\mathbf{0.3146}\,{\scriptscriptstyle\pm 0.0015}\) & \(\mathbf{0.9925}\,{\scriptscriptstyle\pm 0.0001}\) & \(\mathbf{0.4625}\,{\scriptscriptstyle\pm 0.0039}\) \\
& FNO  & \(0.1692\,{\scriptscriptstyle\pm 0.0012}\) & \(0.1255\,{\scriptscriptstyle\pm 0.0011}\) & \(0.4038\,{\scriptscriptstyle\pm 0.0015}\) & \(0.9870\,{\scriptscriptstyle\pm 0.0001}\) & \(0.5942\,{\scriptscriptstyle\pm 0.0012}\) \\
& SwinIR & \(0.1336\,{\scriptscriptstyle\pm 0.0004}\) & \(0.0947\,{\scriptscriptstyle\pm 0.0005}\) & \(0.3433\,{\scriptscriptstyle\pm 0.0009}\) & \(0.9907\,{\scriptscriptstyle\pm 0.0001}\) & \(0.5162\,{\scriptscriptstyle\pm 0.0013}\) \\
\midrule
\multirow{3}{*}{\(8\times\)}
& Ours & \(\mathbf{0.3491}\,{\scriptscriptstyle\pm 0.0002}\) & \(\mathbf{0.4451}\,{\scriptscriptstyle\pm 0.0007}\) & \(\mathbf{0.8041}\,{\scriptscriptstyle\pm 0.0007}\) & \(\mathbf{0.9529}\,{\scriptscriptstyle\pm 0.0012}\) & \(\mathbf{1.1363}\,{\scriptscriptstyle\pm 0.0031}\) \\
& FNO  & \(0.4404\,{\scriptscriptstyle\pm 0.0021}\) & \(0.6794\,{\scriptscriptstyle\pm 0.0029}\) & \(1.0002\,{\scriptscriptstyle\pm 0.0011}\) & \(0.9250\,{\scriptscriptstyle\pm 0.0006}\) & \(1.4211\,{\scriptscriptstyle\pm 0.0026}\) \\
& SwinIR & \(0.3782\,{\scriptscriptstyle\pm 0.0004}\) & \(0.5333\,{\scriptscriptstyle\pm 0.0006}\) & \(0.8817\,{\scriptscriptstyle\pm 0.0008}\) & \(0.9442\,{\scriptscriptstyle\pm 0.0004}\) & \(1.2576\,{\scriptscriptstyle\pm 0.0002}\) \\
\bottomrule
\end{tabular}%
}
\end{table*}
\color{black}
\FloatBarrier

\subsection{Generalizing to unknown coarse graining}
\label{ap:generalzation}
The DNS simulation and LES--DNS pairing for the decaying-turbulence experiment are described in Supplementary Notes~\ref{ap:dit_dns} and~\ref{ap:les_dns_pairing}.

Although all three expert models can in principle be included in the MoE, we find that the box and spectral experts already exhibit  strong generalization performance (Supplementary Table~\ref{tab:generalization_results}), and thus adopt a two-expert MoE as our main configuration.
To understand this behavior, we analyze reconstruction error fields under tricubic downsampling and compute Pearson correlations between experts (Fig.~\ref{fig:generalization}f).
The box and spectral experts show near-zero correlation, indicating largely independent and therefore complementary error modes.
In contrast, the Gaussian expert exhibits a moderate positive correlation (approximately $0.5$) with both the box and spectral experts, suggesting partially redundant error characteristics.
This observation is consistent with the distinct filtering characteristics of these coarse-graining strategies~\cite{pope2000turbulent} and suggests that expert diversity, rather than simply increasing the number of experts, is key to robust generalization.
\begin{table*}[htbp]
    \centering
\caption{
\textbf{Full quantitative generalization results on FIT at $4\times$ upscaling.}
We evaluate three unseen test-time coarse-graining settings: tricubic downsampling, Lanczos filtering, and uniform subsampling with $5\%$ additive Gaussian noise.
We compare three single-expert models (trained with box, Gaussian, or spectral-cutoff coarse-graining) and two MoE variants (2 experts: box+spectral; 3 experts: box+Gaussian+spectral).
Higher SSIM/PSNR and lower nRMSE/fRMSE indicate better performance; best results are \textbf{bold}.
}
\label{tab:generalization_results}
\resizebox{0.8\textwidth}{!}{%
\begin{tabular}{
        l %
        l %
        S[table-format=1.4] %
        S[table-format=2.2] %
        S[table-format=1.4] %
        S[table-format=2.2] %
}
\toprule
\textbf{Coarse-graining} & \textbf{Expert} & {\textbf{SSIM↑}} & {\textbf{PSNR↑}} & {\textbf{nRMSE↓}} & {\textbf{fRMSE↓}} \\
\midrule
\multirow{5}{*}{Tricubic}
    & Spectral     & 0.8679 & 43.38 & 0.1048 & 25.27 \\
    & Box      & {0.9077} & {45.02} & {0.0879} & {21.44} \\
    & Gaussian    & 0.8897 & 43.82 & 0.1001 & 24.08 \\
    & MoE (2 experts)    & \textbf{0.9406} & \textbf{47.19} & \textbf{0.0673} & \textbf{16.45} \\
    & MoE (3 experts)      &  {0.9333} &  {46.48} &  {0.0731} &  {17.72} \\
\midrule
\multirow{5}{*}{Lanczos}
    & Spectral     & 0.8581 & 42.78 & 0.1130 & 27.04 \\
    & Box      & {0.8926} & {44.32} & {0.0978} & {23.60} \\
    & Gaussian    & 0.8735 & 43.11 & 0.1106 & 26.39 \\  
    & MoE (2 experts)    & \textbf{0.9300} & \textbf{46.37} & \textbf{0.0757} & \textbf{18.20} \\
    & MoE (3 experts)      &  {0.9212} &  {45.65} &  {0.0822} &  {19.65} \\
\midrule
\multirow{5}{*}{Uniform (5\% noise)}
    & Spectral     & {0.8274} & {43.01} & {0.1093} & {25.48} \\
    & Box      & 0.6774 & 39.25 & 0.1723 & 39.98 \\
    & Gaussian    & 0.7757 & 42.39 & 0.1208 & 27.68 \\
    & MoE (2 experts)    & \textbf{0.8460} &  {43.91} &  {0.0991} &  {22.85} \\
    & MoE (3 experts)      &  {0.8420} & \textbf{44.32} & \textbf{0.0972} & \textbf{22.27} \\
\bottomrule
\end{tabular}
}
\end{table*}

\FloatBarrier
\end{supplementaryinformation}
\clearpage 

\end{document}